\documentclass[11pt]{article}
\pdfoutput=1
\usepackage{gensymb}
\usepackage{jheppub} \usepackage{braket,slashed,bm}
\usepackage{booktabs}
\usepackage{array,multirow}
\usepackage{amsmath}
\usepackage[normalem]{ulem}
\usepackage[T1]{fontenc}
\usepackage{mathrsfs}
\usepackage{tikz}
\usepackage{stmaryrd}
\usepackage{subcaption}
\usetikzlibrary{arrows,decorations.pathmorphing,backgrounds,positioning,fit,petri,automata,shadows,calendar,mindmap,
decorations.markings,calc}
\usepackage{lscape}
\usepackage{hyperref}
\usepackage{graphicx,wrapfig}
\usepackage{floatrow}
\newfloatcommand{capbtabbox}{table}[][\FBwidth]
\def\bea{\begin{eqnarray}}
\def\eea{\end{eqnarray}}
\def\that{{\hat \theta}}

\usepackage{braket,slashed,bm}
\def\nn{\nonumber}
\def\hyp{\mathsf{y}}
\title{The $Z$ decay width in the SMEFT: $y_t$ and $\lambda$ corrections at one loop.}
\author{
Christine Hartmann$^a$, William Shepherd$^{a,b}$ and Michael Trott$^a$ \\
$^a$Niels Bohr International Academy,
University of Copenhagen, Blegdamsvej 17, DK-2100 Copenhagen, Denmark\\
$^b$Institut f\"{u}r Physik, Johannes-Gutenberg-Universit\"{a}t Mainz, Staudingerweg 7, D-55128 Mainz, Germany }
\abstract{We calculate one loop $y_t$ and $\lambda$  dependent corrections to $\bar{\Gamma}_Z,\bar{R}_f^0$ and the partial $Z$ widths due to dimension six operators in the Standard Model Effective Field Theory (SMEFT), including finite terms. We assume $\rm CP$ symmetry and a $\rm U(3)^5$  symmetry
in the UV matching onto the dimension six operators, dominantly broken by the Standard Model Yukawa matrices. Corrections to these observables are predicted using the input parameters $\{\hat{\alpha}_{ew}, \hat{M}_Z, \hat{G}_F, \hat{m}_t, \hat{m}_h\}$ extracted with one loop corrections in the same limit.  We show that at one loop the number of SMEFT parameters contributing to the precise LEPI pseudo-observables
exceeds the number of measurements. As a result the SMEFT parameters contributing to LEP data are formally unbounded when the size of loop corrections
are reached until other data is considered in a global analysis. The size of these loop effects is generically a correction of order $\sim\%$ to leading effects in the SMEFT, but we find multiple large numerical coefficients in our calculation at this order.
We use a $\rm \overline{MS}$ scheme, modified  for the SMEFT, for renormalization. Some subtleties involving novel evanescent scheme dependence
present in this result are explained.}
\begin{document}
\maketitle

\section{Introduction}
The vast LHC data set already reported, and the more expansive data set expected to be reported during the full LHC
experimental run, is unprecedented. Such a treasure trove of data on the interactions of the Standard Model (SM) fields around the electroweak scale $(\bar{v}_T)$
enables a paradigm shift in what is possible and
advisable in Effective Field Theory (EFT) studies of beyond the SM physics.
With such a data set, it is possible to systematically study the SM as a real EFT. The idea is that the SM Lagrangian is just the leading order terms
of a more complete theory at higher energy scales. This hypothesis is adopted by embedding the SM in an expansion including
higher dimensional operators, and promoting it to the Standard Model Effective Field Theory (SMEFT).

In the SMEFT, it is assumed that  $\rm SU(2)_L \times U(1)_Y$ is spontaneously broken to $\rm U(1)_{em}$ by the vacuum expectation value ($\langle H^\dagger H \rangle  \equiv \bar{v}_T^2/2$) of the Higgs field. The minimum of the potential is determined including the effect of
operators in $\mathcal{L}_6$.
The observed scalar is assumed to be  $J^P = 0^+$ and embedded in a doublet of $\rm SU(2)_L$, and a mass gap to the scale(s) of new physics (referred to as $\sim \Lambda$) leads to an expansion parameter $\bar{v}_T^2/\Lambda^2 < 1$. The SMEFT follows from these assumptions, and is the sum of the SM Lagrangian and a series of $\rm SU(3)_C \times SU(2)_L \times U(1)_Y$ invariant higher dimensional operators built out of SM fields
\bea
\mathcal{L}_{SMEFT} = \mathcal{L}_{SM} + \mathcal{L}^{(5)} + \mathcal{L}^{(6)} + \mathcal{L}^{(7)} + ..., \quad \quad \mathcal{L}^{(k)}= \sum_{i = 1}^{n_k} \frac{C_i^{(k)}}{\Lambda^{k-4}} Q_i^{(k)} \hspace{0.25cm} \text{ for $k > 4$.}
\eea
The number of non redundant operators in $\mathcal{L}^{(5)}$, $\mathcal{L}^{(6)}$, $\mathcal{L}^{(7)}$ and $\mathcal{L}^{(8)}$ is known \cite{Buchmuller:1985jz,Grzadkowski:2010es,Weinberg:1979sa,Abbott:1980zj,Lehman:2014jma,Lehman:2015coa,Henning:2015alf} and a general algorithm to determine operator bases at higher orders has been established in Refs.~\cite{Lehman:2015via,Lehman:2015coa,Henning:2015daa,Henning:2015alf}, making the SMEFT, in principle, defined to all orders in the expansion in local operators. In this work, we use a naive power counting in mass dimension so that the operators  $Q_i^{(k)}$ will be suppressed by $k-4$ powers of the cutoff scale $\Lambda$, where the $C_i^{(k)}$ are the Wilson coefficients. We generally absorb the cut off scale into the Wilson coefficients as a notational choice unless otherwise noted.

Lower energy data ($E \ll \bar{v}_T$) on flavour changing processes and dipole moments already place strong
constraints on possible deviations from the SM. The SMEFT formalism is most interesting when, in spite of these constraints, deviations captured by this formalism could possibly be observed in the LHC experimental program.
For this reason, it is of interest to examine the effect of higher dimensional operators in the limit that $\rm U(3)^5$ flavour symmetry and CP symmetry is assumed to be (at least approximately) present in a $\sim \, {\rm TeV}$ scale physics sector. In this paper, we calculate in this limit in the SMEFT.

The measurements of the properties of the $Z$ boson gathered at LEPI and LEPII, a subset of Electroweak Precision Data (EWPD), is important to incorporate in the SMEFT formalism. Due to the high precision and accuracy of some of these measurements, it is very important to inform expectations of possible deviations that can be measured
at LHC consistent with these LEP results. For these expectations to be of most value, it is necessary to project these constraints into the SMEFT formalism consistently. This is particularly the case when drawing strong model independent conclusions, that can potentially impact the {\it experimental reporting of LHC data} interpreted in the SMEFT. In some recent literature very strong claims are made that certain SMEFT parameters can be set to zero in reporting LHC data, in a model independent manner, if they impact LEP measurements. This is argued to be a valid conclusion due to model independent interpretations of experimental constraints.  However, the neglect of dimension eight operator effects and loop corrections involving higher dimensional operators
if the cut off scale of the theory is in the $\sim {\rm TeV}$ mass range is problematic if bounds on the SMEFT parameters are pushed beyond the $\sim \%$ level due to such experimental constraints. This is exactly what is required to robustly justify neglecting such effects in LHC data reporting.  Naively neglecting such effects leads to conclusions that are not truly model independent, as an implicit assumption of a large value of the parameters $\Lambda/\sqrt{C_i} \gtrsim 3 \, {\rm TeV}$ is effectively adopted without justification by experiment.
See the detailed discussion on this point in Refs.~\cite{Berthier:2015oma,Berthier:2015gja,Berthier:2016tkq,Passarino:2016pzb}. For results and discussion (with supporting calculations) characterizing various NLO corrections in the SMEFT see Refs.~\cite{Ghezzi:2015vva,Passarino:2012cb,Passarino:2016pzb,David:2015waa,Englert:2014cva,Bylund:2016phk,Zhang:2016omx,Zhang:2013xya,Maltoni:2016yxb,Gauld:2015lmb,Gauld:2016kuu,Gorbahn:2016uoy,Bizon:2016wgr,Degrassi:2016wml,Passarino:2016owu}.

To form a more model independent and consistent picture of the constraint of LEPI measurements in the SMEFT, it is necessary to calculate
the effects of higher dimensional operators at one loop on EWPD. This is the purpose of this paper. We report one loop results for the $Z$ total width ($\bar{\Gamma}_Z$), partial widths ($\bar{\Gamma}_{Z\rightarrow \bar{\psi} \psi}$) and ratios of partial widths ($\bar{R}_{\psi}^0$) in the limit where one loop $y_t,\lambda$ dependent corrections are retained.\footnote{The notation $\bar{X}$ indicates a theoretical prediction of an observable $X$ in the canonically normalized SMEFT. Also note that $\psi =\{\ell,u,d,\nu\}$.}
Our results show that a number of SMEFT parameters are introduced at one loop that are not present at tree level in the SMEFT modification of the LEPI pseudo-observables.
The number of parameters present exceeds the number of the precise LEPI pseudo-observables. As a result, the SMEFT parameters present in EWPD become formally
unbounded by LEPI data alone when the typical size of one loop corrections involving these new parameters
is reached. The size of this effect depends on {\it a priori} unknown Wilson coefficients, but is not
robustly below the $\sim \%$ level in the SMEFT \cite{Berthier:2015oma,Berthier:2015gja,Berthier:2016tkq,Passarino:2016pzb}.

The prediction of $\bar{\Gamma}_Z$ in the SM, or the SMEFT, is a multi-scale problem. Measured input observables are used
to specify Lagrangian parameters, which are in turn used to predict the partial and total widths. The scales
present are hierarchical\footnote{The notation $\hat{X}$ indicates a measured value of an observable $X$ in the canonically normalized SMEFT.
In particular this notation is used to indicate experimentally extracted masses.}
\bea
0 \ll \hat{m}_\mu^2 \ll \hat{m}_Z^2 \sim \hat{m}_h^2 \sim \hat{m}_t^2.
\eea
Following historical conventions the input value of $\hat{\alpha}$ is run up to the scale $\hat{M}_Z$ in this work (the running of $\hat{G}_F$ is further suppressed).
To project experimental constraints onto Wilson coefficients defined at $\Lambda$,
we renormalize the theory at the scale $\mu \simeq \Lambda$. This introduces logarithmic terms in the prediction of
the observables of interest, and, crucially, Wilson coefficients that are not present at tree level. A re-definition of the parameters present can be performed to absorb a subset of the  logarithmic corrections at the measurement scale into effective lower scale parameters. Although such an approach is consistent with historical uses of EFT's,
we do not perform such a re-definition here for the following reasons. First, these
analyses aim to infer a consistent set of constraints on the parameters in the SMEFT at the matching scale \cite{Freitas:2016iwx}. Second, the matching scale cannot be far separated from the electroweak scale ($\Lambda \gg v$ as opposed to $\Lambda > v$)
without introducing decoupling to a degree that these studies are not of interest. As such, non-logarithmic finite terms are not guaranteed to be significantly sub-dominant in the one loop results. In some cases, and in some schemes, the logs can dominate for classes of perturbative corrections \cite{Gauld:2015lmb}.
In other cases, the logs are not very dominant numerically \cite{Hartmann:2015oia,Hartmann:2015aia,Ghezzi:2015vva,Maltoni:2016yxb,Gauld:2016kuu}.
In either case, the logs would be reintroduced when mapping experimental constraints to the high scale theory.  See Refs.~\cite{Hartmann:2015oia,Hartmann:2015aia,Ghezzi:2015vva,Passarino:2012cb,Passarino:2016pzb,David:2015waa,Englert:2014cva,Bylund:2016phk,Zhang:2016omx,Zhang:2013xya,Maltoni:2016yxb,Gauld:2015lmb,Gauld:2016kuu,Gorbahn:2016uoy,Bizon:2016wgr} for related discussion on NLO SMEFT results. Our results explicitly show that not all of the new parameters
present in the observables, we consider, can be trivially reabsorbed into the tree level Wilson coefficients modifying EWPD, due to the multi-scale nature of the problem.
The number of parameters present in LEP data at one loop in the SMEFT exceeds the number of measurements.
This is the main reason the results of the (partial) explicit calculation already do not support an idea that parameters that affect LEPI data can be trivially set to zero
in a truly model independent fashion in the SMEFT when reporting LHC data. This point is expected to be even further reinforced when the full one loop corrections to EWPD are known in the SMEFT.

The outline of this paper is as follows.
As the technical details of this calculation contain a number of novel features, we include an extensive discussion on how the results were developed.  The interplay of the counter terms for the SM fields and the pure dimension six operator counter terms that underly the cancelation
of the divergences present in this calculation is discussed and demonstrated at length in Sections \ref{renorms},\ref{divcancel}.
Another interesting aspect of the results is the appearance of a subtle evanescent scheme dependence for dimension six operators.
The cancelation of this scheme dependence is discussed and demonstrated in Section \ref{evanescentappendix}.
We further develop
a $\rm \overline{MS}$ scheme for perturbation theory for SMEFT loop corrections introduced in Refs.~\cite{Hartmann:2015oia,Hartmann:2015aia} in this paper suitable for the SMEFT.
This development includes (partial) one loop results for the input parameters $\{\hat{\alpha},\hat{G}_F,\hat{m}_Z \}$ extracted from experimental measurements in the SMEFT
in Section \ref{inputcorr}. Explicit amplitude results are also reported, including finite terms, in Section \ref{naiveresults}.
We then report results for
$\bar{\Gamma}_{Z}$, $\bar{R}_f^0$ and the $Z$ partial widths in terms of these input parameters in this limit, expressed in terms of effective couplings
to one loop in the SMEFT, mirroring past leading order SMEFT work, in Section \ref{fullresults}.  Finally, we conclude in Section \ref{conclude}.

\section{Renormalization and notational conventions}\label{renorms}
The bare and renormalized spin one fields/couplings ($F/c = \{A,B,W,G\}/\{e, g_1, g_2, g_3\}$) are related as
\begin{align}
F_\mu^{(r)} &= \frac{1}{\sqrt{Z_{F}}} \, F^{(0)}_\mu,  & c^{(r)} &= \frac{1}{Z_c} \mu^{- \epsilon} c^{(0)}.
\end{align}
The factor $\mu^{- \epsilon}$ is included in the coupling relation to render the renormalized coupling dimensionless \cite{Manohar:2000dt}
and we use $(0)$ superscripts for bare fields and $(r)$ superscripts for renormalized quantities.
In addition, the scalar field renormalization is defined as
$\sqrt{Z_s} \, S^{(r)} = S^{(0)}$ for  $S = \{h, \phi^0,\phi^\pm\}$.
All the divergence subtractions are defined in the modified $\rm \overline{MS}$ scheme for $d = 4 - 2 \, \epsilon$ dimensions (as in Refs.~\cite{Hartmann:2015oia,Hartmann:2015aia}). This
$\rm \overline{MS}$ scheme utilizes the Background Field method (BFM) \cite{DeWitt:1967ub,Abbott:1981ke} to define the divergence subtractions.\footnote{In this sense it is
even further "modified" from some other $\rm \overline{MS}$ conventions in the literature.}
In addition we implement on-shell renormalization conditions to define the external states wavefunctions, and a tadpole prescription to define the vev. This introduces significant technical advantages and simplifications, that primarily follow from the use of the BFM. In the BFM, the fields are split into classical and quantum components.
A gauge fixing term breaks the gauge
invariance of the quantum fields while maintaining the gauge invariance of the classical background fields.
As a result of the BFM, for a specific choice of operator normalization,
cancelations occur between insertions of the divergent renormalization factors of the SM gauge fields and couplings. This follows from the unbroken Ward identities of the classical background fields of the theory.

We use the convention that the Higgs doublet is defined as
\bea \label{Hexpansion}
 H = \frac{1}{\sqrt{2}}\left( \begin{array}{c}
\sqrt{2} i \phi^+ \\
 h + \bar{v}_T + \Delta v + i\phi_0 \end{array}  \right),
 \eea
in $R_\xi$ gauge (with background field gauge fixing), with $\phi^\pm$ and $\phi_0$ the goldstone bosons. In this normalization, the physical Higgs mass in the SM is
$m_h^2 = 2 \, \lambda \, (\bar{v}_T + \Delta v)^2$. Finite tadpole counter terms,
indicated by the introduction of $\Delta v$, are introduced to fix the one point function of the $H$ field. This induces gauge dependent terms into the definition of the vev and $\Delta v$
is formally of one loop order.\footnote{Note that we have switched our notation to $\Delta v$ from $\delta v$ in some closely related previous works \cite{Alonso:2013hga,Berthier:2015oma,Berthier:2015gja,Berthier:2016tkq,Bjorn:2016zlr}. This is to consistently reserve $\delta$ notation
for a correction at leading order in the tree level power counting corrections in the SMEFT (i.e. suppressed by $1/\Lambda^2$).}

We use the notation that the $\bar{g}_i$ are the canonically normalized gauge couplings of the SM, including the effects of $\mathcal{L}_6$, as defined in Ref.~\cite{Alonso:2013hga}. The sign convention on the covariant derivative is
$D_\mu = \partial_\mu + i \, g_3 \, A^A_\mu \, T^A + i \, g_2 \, W^a_\mu \, \sigma^a/2 + i g_1 \, B_\mu \, \hyp_i$ with $\hyp_i$ the $U(1)_Y$ hypercharge generator for the state $i$ that
$D$ acts on. Here $\sigma_a$ is the Pauli matrix.
The yukawa coupling $y_t$ is defined with the convention that the Lagrangian mass term at leading order is $m_t = y_t \, \bar{v}_T/\sqrt{2}$ in the SM. $\epsilon^\mu_Z$ is the polarization vector of the massive $Z$ boson. We introduce the notation $C_Z = i \, \sqrt{\bar{g}_1^2+ \bar{g}_2^2} \, \bar{v}_T^2/2$ to characterize the coupling of the $Z$ boson from SMEFT corrections. The SM weak mixing angle is defined as $s_\theta^2 = \bar{g}_1^2/(\bar{g}_1^2+ \bar{g}_2^2)$ at leading order and $\mathcal{Q}_i$ is the electric charge of a state $i$ in units of $e$.  The SM electric charge is defined as $e = g_1 \, g_2/ \sqrt{g_1^2 + g_2^2}$.
The chiral fermion fields are summed over the $p,r,s,t$ flavour indices on the fields
\bea
U^p_{L/R} &=& \{u,c\}_{L/R},\quad
D^r_{L/R} = \{d,s,b\}_{L/R}, \quad
\ell^s_{L/R} = \{e,\mu,\tau\}_{L/R} \quad
\nu^t_{L} = \{\nu_e,\nu_\mu,\nu_\tau\}.
\eea
In addition, the fermion indices are at times made explicit.
The left handed projector is defined with the convention $P_{L} = (1-\gamma_5)/2$. We discuss a subtlety involving the definition of $\gamma_5$ in $d$ dimensions in these
results in Appendix \ref{evanescentappendix}.

We use hat superscripts for measured quantities. For the measured values of the input parameters we use $\{\hat{\alpha}_{ew}, \hat{M}_Z, \hat{G}_F, \hat{m}_t, \hat{m}_h\}$.
Quantities related at tree level to these input parameters are also labeled with hat superscripts. We generally use $\delta$ to indicate a correction to a SM prediction due to a power counting correction of order $1/\Lambda^2$, and $\Delta$ to indicate a correction to a SM prediction that is at least one loop in the perturbative expansion. Terms that are labeled as
$\Delta$ shifts can also include power counting corrections. We retain some terms of order $1/\Lambda^4$ in the intermediate results below for ease of presentation;
such higher order terms are consistently dropped in the final numerical results.

\subsection{SM counter terms}
For the Electroweak (EW) terms, we choose to use define the basic counter term subtractions in the modified $\rm \overline{MS}$ scheme for the $W^\pm,Z^0$ fields, the gauge couplings $g_1,g_2$, the vev and the scalar fields.
The vev is re-normalized with the inclusion of $\bar{v}_T = \bar{v}_T^{(0)}/\sqrt{Z_v}$.
We will use the subscript "div" when only the divergent part of $(\sqrt{Z_v} + \frac{\Delta v}{\bar{v}_T})_{div}$ is used.
All other counter terms in the EW sector are then derived quantities at one loop. The wavefunction renormalization counter terms
are only introduced for the background fields  \cite{DeWitt:1967ub,Abbott:1981ke}.
One finds the relations among the SM counter terms in the BFM \cite{Denner:1994xt}
\begin{align} \label{BFgoodness}
\sqrt{Z_A} Z_e &= 1, &
Z_h &= Z_{\phi_\pm} = Z_{\phi_0}, &
\sqrt{Z_B} Z_{g_1} &= 1, &
\sqrt{Z_W} Z_{g_2} &= 1, &
\sqrt{Z_G} Z_{g_3} &= 1.
\end{align}
See Refs.~\cite{Einhorn:1988tc,Denner:1994xt,Sperling:2013eva,Hartmann:2015oia} for more discussion.
 In the scalar sector, the BFM gives the relationship
$ (\sqrt{Z_v}+ \frac{\Delta v}{\bar{v}_T})_{div} = \sqrt{Z_h}$ and the Higgs wavefunction renormalization is given by
\begin{align}\label{remainingSMsubtract}
Z_h &= 1  + \frac{(3 + \xi)\, (\bar{g}_1^2+ 3 \, \bar{g}_2^2)}{64 \, \pi^2 \, \epsilon}  - \frac{N_c \, y_t^2}{16 \, \pi^2 \, \epsilon}.
\end{align}

\begin{figure}
\hspace{-2.5cm}
\begin{tikzpicture}
\draw [decorate,decoration=snake](-1.1,0) -- (-0.5,0);
\draw [decorate,decoration=snake](0.5,0) -- (1.1,0);

\draw  [->][thick]  (270:0.5) -- + (0:-0.05) ;
\draw  [->][thick]  (90:0.5)  -- + (0:0.05) ;

\draw  (0,0) circle (0.50);

\draw (0,-1.25) node [align=center] {(a)};

\node [above][ultra thick] at (0,0.5) {$t$};
\node [above][ultra thick] at (0,-1.1) {$b$};
\node [left][ultra thick] at (-0.5,-0.3) {$W^+$};
\node [right][ultra thick] at (0.5,-0.3) {$W^-$};

\end{tikzpicture}
\begin{tikzpicture}
\hspace{0.5cm}

\draw  [decorate,decoration=snake] (-1,0) -- (1,0);

\draw [thick] [dashed] (0,0.45) circle (0.50);

\draw (0,-1) node [align=center] {(b)};

\node [above][ultra thick] at (0,1) {$h$};
\node [left][ultra thick] at (-0.3,-0.3) {$W^+$};
\node [right][ultra thick] at (0.3,-0.3) {$W^-$};

\end{tikzpicture}
\begin{tikzpicture}
\hspace{1cm}
\draw [decorate,decoration=snake](-1.1,-0.3) -- (-0.5,-0.3);
\draw [decorate,decoration=snake](0.5,-0.3) -- (1.1,-0.3);

\draw  (0,-0.3) [dashed] circle (0.50);

\draw (0,-1.6) node [align=center] {(c)};

\node [above][ultra thick] at (0,+0.1) {$h$};
\node [above][ultra thick] at (0,-0.8) {$\phi_\pm$};
\node [left][ultra thick] at (-0.5,-0.6) {$W^+$};
\node [right][ultra thick] at (0.5,-0.6) {$W^-$};
\end{tikzpicture}
\begin{tikzpicture}
\hspace{1.5cm}
\draw [decorate,decoration=snake](-1.1,0) -- (-0.5,0);
\draw [decorate,decoration=snake](0.5,0) -- (1.1,0);

\draw  [->][thick]  (90:0.5)  -- + (0:-0.05) ;
\draw  [->][thick]  (270:0.5)  -- + (0:0.05) ;

\draw  (0,0) circle (0.50);

\draw (0,-1.25) node [align=center] {(d)};

\node [above][ultra thick] at (0,0.5) {$t$};
\node [above][ultra thick] at (0,-1.1) {$t$};
\node [left][ultra thick] at (-0.5,-0.3) {$A$};
\node [right][ultra thick] at (0.5,-0.3) {$A$};

\end{tikzpicture}
\begin{tikzpicture}
\hspace{2cm}
\draw [decorate,decoration=snake](-1.1,0) -- (-0.5,0);
\draw [decorate,decoration=snake](0.5,0) -- (1.1,0);

\draw  [->][thick]  (90:0.5)  -- + (0:-0.05) ;
\draw  [->][thick]  (270:0.5)  -- + (0:0.05) ;

\draw  (0,0) circle (0.50);

\draw (0,-1.25) node [align=center] {(e)};

\node [above][ultra thick] at (0,0.5) {$t$};
\node [above][ultra thick] at (0,-1.1) {$t$};
\node [left][ultra thick] at (-0.5,-0.3) {$Z^0$};
\node [right][ultra thick] at (0.5,-0.3) {$Z^0$};

\end{tikzpicture}

\caption{Subset of diagrams determining the SM counter terms for $Z_{g_1},Z_{g_2},Z_W,Z_{Z^0},Z_e,Z_A$ in this calculation.}
\label{WSMrenorm1}
\end{figure}
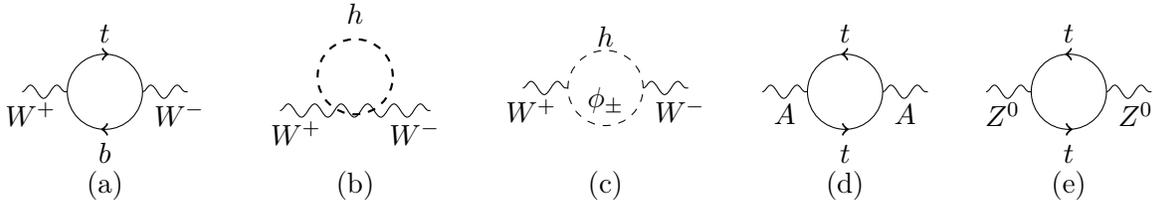
The explicit form of the renormalization constants remaining
for the SM (in our partial result) are directly determined from the diagrams in Fig.~\ref{WSMrenorm1}. The $\lambda$ pole cancels between Fig.~\ref{WSMrenorm1} $b)$ and $c)$ and the remaining $y_t^2$ pole in  Fig.~\ref{WSMrenorm1} $a)$ is canceled by the vev counter term. One then identifies in the vanishing gauge coupling limit at one loop
\bea
Z_{g_2} =  Z_{W} =1.
\eea
The counter term for $Z_{A}$ can be directly determined from Fig.~\ref{WSMrenorm1} d) and is trivially $Z_{A} = 1$ in the vanishing gauge coupling limit.
This determines $Z_e = 1$ and hence $Z_{g_1} =1$ from the tree level relation defining the electric coupling.
Directly calculating  Fig.~\ref{WSMrenorm1} $e)$ and the corresponding $Z^0$ diagrams for the topologies shown in $b)$, $c)$ determines $Z_{Z^0} = 1$ in the vanishing gauge coupling limit. Again, the $\lambda$ pole cancels between the $b)$, $c)$ topologies for the $Z^0$, and the $y_t^2$ pole cancels with the insertion of the vev renormalization factor.

The fermion fields are renormalized with the
divergent counter term introduced as $\psi_{L/R}^{(r)} =  \psi_{L/R}^{(0)}/ \sqrt{Z^\psi_{L/R}}$. In this calculation, an important
divergent fermion renormalisation factor is $b_L^{(r)} = b_L^{(0)}/ \sqrt{Z^b_L}$.
The large $y_t^2$ contribution to this fermion field re-normalization factor is given by
\bea\label{ZbeqnSM}
Z_L^b = 1 - \frac{y_t^2}{32 \, \pi^2 \, \epsilon}.
\eea

\subsection{SMEFT counter terms}
\begin{figure}
\begin{tikzpicture}
\hspace{-1.5cm}
\draw    [very thick](-1,0) -- (1,0);
\draw [dashed] (0.5,0) arc [radius=0.45, start angle=0, end angle= 190];
\draw (0,-1) node [align=center] {(a)};
\node [above][ultra thick] at (0,0.6) {$\phi^{\pm}$};
\node [left][ultra thick] at (-0.6,-0.3) {$b$};
\end{tikzpicture}
\begin{tikzpicture}
\hspace{1.5cm}
\filldraw (-0.6,-0.1) rectangle (-0.4,0.1);
\draw    [very thick](-1,0) -- (1,0);
\draw [dashed] (0.5,0) arc [radius=0.45, start angle=0, end angle= 190];
\draw (0,-1) node [align=center] {(b)};
\node [above][ultra thick] at (0,0.6) {$\phi^{\pm}$};
\node [left][ultra thick] at (-0.6,-0.3) {$b$};
\end{tikzpicture}
\caption{Fermion self energy diagrams where $\phi^\pm$ are the Goldstone fields. Diagram (a) is the SM contribution, while diagram (b) contains the insertion of the Effective Lagrangian including $\mathcal{L}_6$ corrections, indicated with a black square.}
\label{wavefunction}
\end{figure}
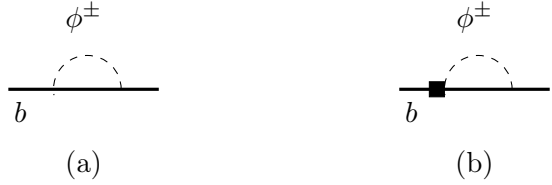

An interesting result of the complete renormalization of $\mathcal{L}_{SMEFT}$ up to $\mathcal{L}_6$ \cite{Grojean:2013kd,Jenkins:2013zja,Jenkins:2013wua,Alonso:2013hga,Alonso:2014zka} is the modification of the SM counter terms, due to the introduced cut-off scale.
In short, the SM with a cut-off scale -- the SMEFT -- is a different field theory than the SM alone, even when only considering the running of the usual SM Lagrangian parameters.   A straightforward modification of the SM counter terms of this form is given for the fermion wavefunction renormalization as:
\bea\label{Zbeqn}
(Z_L^{b})_{SMEFT} = 1 - \frac{y_t^2}{32 \, \pi^2 \, \epsilon} + \frac{y_t^2 \, \bar{v}_T^2}{16 \, \pi^2 \, \epsilon} \, C_{\substack{Hq\\bb}}^{(3)}.
\eea
The first term in Eqn.~\ref{Zbeqn} is the SM contribution, and the second term comes from the insertion of the Effective Lagrangian in the second diagram of Fig.~\ref{wavefunction}.
More subtle is the fact that the gauge couplings also run with a dependence on parameters in $\mathcal{L}_6$  \cite{Jenkins:2013zja}. For the $\{U(1), SU(2), SU(3)\}$ gauge fields $\{B,W,G\}$
the $\mathcal{L}_4$ Effective Lagrangian has the extra contribution to the SM counter terms \cite{Jenkins:2013zja}
\begin{align}
\Delta Z_{g_1} &=  - \frac{\lambda \, \bar{v}_T^2}{4 \, \pi^2 \, \epsilon} \,  g_1^2 \, C_{HB}, &
\Delta Z_{g_2} &= - \frac{\lambda \, \bar{v}_T^2}{4 \, \pi^2 \, \epsilon} \,  g_2^2 \,  C_{HW}, &
\Delta Z_{g_3} &=  - \frac{\lambda \, \bar{v}_T^2}{4 \, \pi^2 \, \epsilon} \,  g_3^2 \, C_{HG}.
\end{align}
In the BFM, this leads to $\mathcal{L}_6$ parameter dependence in the field strengths renormalization due to Eqn.~\ref{BFgoodness}
\begin{align}
\Delta Z_{B} &=  \frac{2 \, \lambda \, \bar{v}_T^2}{4 \, \pi^2 \, \epsilon} \, g_1^2 \,C_{HB}, &
\Delta Z_{W} &=   \frac{2 \, \lambda \, \bar{v}_T^2}{4 \, \pi^2 \, \epsilon} \, g_2^2 \, C_{HW}, &
\Delta Z_{G} &=  \frac{2 \, \lambda \, \bar{v}_T^2}{4 \, \pi^2 \, \epsilon} \, g_3^2 \, C_{HG}.
\end{align}
This dependence is utilized in Section \ref{MZextract}. In addition $Z_h$ is also modified as
\bea
\Delta Z_h = \frac{m_h^2}{16 \, \pi^2 \, \epsilon} \left[-5 \, C_{HD} + 14 \, C_{H \Box} - \frac{6}{\lambda} \, C_H + \frac{N_c}{2 \, \lambda} (C_{\substack{uH\\ 33}} \, y^2_t + h.c.)
- 4 \, \frac{N_c}{\lambda} \, C_{\substack{Hq\\ 33}}^{(3)} \, y_t^2 \right],
\eea
in the BFM. This result is extracted from the modification of the running of $\lambda$ and $m_h^2$ reported in Ref.~\cite{Jenkins:2013zja}.
Finally, the SMEFT operator counter term matrices themselves are introduced with the convention
\bea
Q^{(0)}_{i} = Z_{i,j} \, Q^{(r)}_{j},
\eea
where $Q^{(0)}$ is the bare operator, and  $Q^{(r)}$ is the renormalized operator. We choose to renormalize the theory at the scale $\mu^2 = \Lambda^2$ so that the constraints derived can be directly related to the matching scale to infer properties on the
underlying theory generating the SMEFT \cite{Freitas:2016iwx}. We use the Warsaw basis for the $\mathcal{L}_6$ Effective Lagrangian
\cite{Grzadkowski:2010es}. The reasons for this are extensive. First, no other construction in the literature prior to 2010 is a well defined, i.e. fully reduced by the Equations of Motion (EoM) operator basis for $\mathcal{L}_6$\footnote{The Warsaw basis builds upon the results reported in Refs.~\cite{Leung:1984ni,Buchmuller:1985jz,AguilarSaavedra:2010zi}.}. Second, the Warsaw basis systematically removes derivative operators, following the long established convention in EFT analyses, which directly enables recent NLO work in the SMEFT.
This is not an accidental choice, but is a very well known technical advantage among EFT practitioners used in many EFT calculations. More recently, the benefit of the Warsaw basis for constructing helicity arguments for the SMEFT was also discussed in Ref.~\cite{Cheung:2015aba},
and then in Refs.~\cite{Azatov:2016sqh,Falkowski:2016cxu}.

Due to the presence of the mixing angles of the SM rotating the vector fields to their mass eigenstates, and the SMEFT modifications of these mixing angles,
defining the large $y_t$ and $\lambda$ limit of $\bar{\Gamma}_Z$ at one loop in the SMEFT is subtle. When considering class 4, and 6 operators of the Warsaw basis,
the choice to scale the Wilson coefficients of operators in $\mathcal{L}_6$ by gauge couplings when a field strength is present, or not, can be made.
This choice can effect the terms retained in the large $y_t$ and $\lambda$ limit of interest in this paper. Our convention is to define the operator
normalization as
\begin{align}\label{opnorm}
Q_{\substack{uW\\ rs}} &= g_2 \,  \bar{q}_{r,A} \sigma^{\mu\nu} \, u_s \, \tau_{AB}^I \,  \tilde{H}_{B} \, W_{\mu \, \nu}^I, \quad
& Q_{\substack{uB\\ rs}} &= g_1 \,\bar{q}_{r,A} \sigma^{\mu\nu} \, u_s  \,  \tilde{H}_{A} \, B_{\mu \, \nu},  \nonumber \\
 Q_{HWB} &= g_1 \, g_2 \, H_A^\dagger \,  \tau_{AB}^I H_B \,  B_{\mu \, \nu} \, W_I^{\mu \, \nu}, \quad
&Q_{HB} &= g_1^2 \, H^\dagger \, H \, B_{\mu \, \nu} B^{\mu \, \nu}, \nonumber \\
  Q_{HW} &= g_2^2 \, H^\dagger \, H \, W_A^{\mu \, \nu} W^A_{\mu \, \nu}, \quad
&Q_{HG} &= g_3^2 \, H^\dagger \, H \, G_{\mu \, \nu} G^{\mu \, \nu},
\end{align}
where $\tilde{H}_{A} = \epsilon_{AB} \, H^{\dagger, B}$ where $\epsilon_{12} = 1$ and $\epsilon_{AB} = - \epsilon_{BA}$. Here $prst..$
are flavour indices and the upper case Roman letters $AB..$ are $\rm SU(2)_L$ indices.
For these operators, we directly define the terms that are included below.

\section{The SMEFT decay amplitude at one loop}\label{divcancel}
We define the naive amplitudes leading the decay width $\bar{\Gamma}_Z = \sum_{\psi_i} \bar{\Gamma}_{Z \rightarrow \bar{\psi}_i \, \psi_i}$, with all fermionic final states $\psi_i$ summed over that are kinematically allowed, schematically as
\bea\label{schematic}
\mathcal{A}_{Z\bar{\psi}_i \, \psi_i} = \mathcal{A}_{SM} + \delta \mathcal{A}_{SMEFT} + \Delta \mathcal{A}_{SMEFT}.
\eea
Here the leading term $\mathcal{A}_{SM}$ is the SM amplitude and the corresponding $\bar{\Gamma}_Z$ that follows from this expression
is completely known at one loop, and partially at higher loop orders in the SM. Higher order terms that are unknown are reviewed in Ref.~\cite{Freitas:2016sty} and include
missing bosonic two-loop contributions and three and four loop terms of $\mathcal{O}(\alpha_{ew}^3,\alpha_{ew}^2 \, \alpha_s,\alpha_{ew} \, \alpha_s^2,\alpha_{ew} \, \alpha_s^3)$. The estimated size of the contributions to $\bar{\Gamma}_Z$ from these missing higher order corrections
in the SM are $\sim 0.5 \, {\rm MeV}$ \cite{Awramik:2006uz,Freitas:2016sty,Freitas:2014hra}. This is smaller {\it in the SM}
than the experimental error quoted \cite{Agashe:2014kda}, and it deserves to be emphasized that the SM passes the hypothesis test
of being consistent globally with EWPD and, in particular, the inferred value of $\hat{\Gamma}_Z$ from LEP measurements \cite{ALEPH:2005ab}.

To interpret measurements of $\hat{\Gamma}_Z$ in the SMEFT, one first includes the tree level (LO) corrections to the SM predictions suppressed by $\mathcal{O}(1/\Lambda^2)$ in Eqn.~\ref{schematic}, due to local contact operators modifying the amplitudes contributing directly to $\bar{\Gamma}_Z$. These corrections are denoted as $\delta \mathcal{A}_{SMEFT}$ and are referred to as
the "naive LO contributions" in this paper.  At $\mathcal{O}(1/\Lambda^2)$ further corrections are present in the predicted value of $\bar{\Gamma}_Z$ in the SMEFT, due to the mapping of the Lagrangian parameters to the measurements defining the input parameters. The full set of LO corrections due to $\mathcal{L}_6$ in the SMEFT are now well understood \cite{Kennedy:1988sn,Altarelli:1990zd,Altarelli:1991fk,Golden:1990ig,Holdom:1990tc,Peskin:1990zt,Peskin:1991sw,Grinstein:1991cd,Maksymyk:1993zm,Georgi:1991ci,Burgess:1993vc,Hagiwara:1986vm,Han:2004az,Alonso:2013hga,Berthier:2015oma,Berthier:2015gja,Berthier:2016tkq,Bjorn:2016zlr}. Here we are concerned with the corrections due to $\mathcal{L}_6$ at one loop in this decay.
We first discuss the "naive one loop amplitude" corrections of this form - denoted $\Delta \mathcal{A}_{SMEFT}$ in Eqn.~\ref{schematic}. Subsequently the one loop improvement of measurements used to define the input parameters\footnote{Hence the numerical values of the SMEFT Lagrangian couplings.} and interference terms of $\delta \mathcal{A}_{SMEFT}$ with the one loop contributions to the SM are discussed
in Section~\ref{inputcorr}. We assume a narrow width approximation in an $e^+ \, e^- \rightarrow \bar{\psi} \, \psi$ process  used to measure the width, to treat the $Z$ effectively as a factorized initial state when making a theoretical
prediction in the SMEFT. This approximation should be relaxed once the full one loop results are known to fully benefit from the calculations reported here.

The cancelation of the divergences in the naive amplitude is an important guide to determining the full NLO result.
These cancelations occur as follows for the naive amplitudes, calculating in the broken phase of the theory.
\subsection{Class $8$ contributions to $\Delta \mathcal{A}_{SMEFT}$}\label{class8section}
\begin{figure}[t]
\begin{tikzpicture}
[
decoration={
	markings,
	mark=at position 0.55 with {\arrow[scale=1.5]{stealth'}};
}]

\hspace{-3cm}
\draw (0,0) circle (0.75);
\filldraw (0.6,-0.1) rectangle (0.8,0.1);

\draw  [postaction=decorate] (0:0.75) -- + (30:1.30) ;
\draw [postaction=decorate] (-20:2.0) -- +(150:1.40);
\draw  [->][ultra thick]  (90:0.75)  -- + (0:-0.1) ;
\draw  [->][ultra thick]  (270:0.75)  -- + (0:0.01) ;


\draw [decorate,decoration=snake] (-1.7,0) -- (-0.75,0);

\node [left][ultra thick] at (-1.7,0) {$Z$};
\node [above][ultra thick] at (0,0.8) {$t$};
\node [below][ultra thick] at (0,-0.8) {$t$};
\node [right][ultra thick] at (1.80,0.7) {$\bar{\psi}$};
\node [right][ultra thick] at (1.80,-0.7) {$\psi$};

\draw (0,-1.7) node [align=center] {(a)};
\end{tikzpicture}
\begin{tikzpicture}
[
decoration={
	markings,
	mark=at position 0.55 with {\arrow[scale=1.5]{stealth'}};
}]
\hspace{-1.8cm}

\draw  [postaction=decorate] (0,0) -- (1,0.6) ;
\draw  [postaction=decorate] (1,-0.6)  -- (0,0) ;


\draw [postaction=decorate] (-1.15,0.6)  -- (-0.1,0);
\draw [postaction=decorate] (-0.1,0) -- (-0.4,-0,2) ;

\draw (0,-1.7) node [align=center] {(b)};

\node [above][ultra thick] at (-0.6,0.5) {$t$};
\node [below][ultra thick] at (1.1,-0.5) {$\psi$};
\node [above][ultra thick] at (1.1,0.5) {$\bar{\psi}$};
\node [below][ultra thick] at (-0.6,-0.5) {$t$};

\hspace{5cm}

\draw  [postaction=decorate] (0,0.1) -- (1,0.6) ;
\draw  [postaction=decorate] (-1,0.6)  -- (0,0.1) ;
\draw  [postaction=decorate] (1,-0.6) --  (0,-0.1) ;
\draw  [postaction=decorate] (0,-0.1) -- (-1,-0.6);




\draw (0,-1.7) node [align=center] {(c)};

\node [above][ultra thick] at (-0.5,0.5) {$t$};
\node [below][ultra thick] at (0.5,-0.5) {$b$};
\node [above][ultra thick] at (0.5,0.5) {$\bar{b}$};
\node [below][ultra thick] at (-0.5,-0.5) {$t$};

\end{tikzpicture}
\caption{Direct contributions due to four fermion $\mathcal{L}_6$ operators to $\Delta \mathcal{A}_{SMEFT}^{C8}$. The two distinct contractions are illustrated with sub-Figures
(b), (c). In the figure, $\psi$ correspond to all up, down, charged lepton and neutrino final states that are kinematically allowed.\label{4fermion}}
\end{figure}
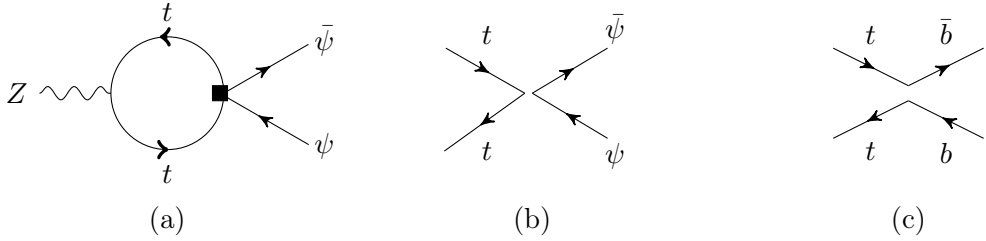

Four fermion operators are labelled as "Class $8$" in the Warsaw basis \cite{Grzadkowski:2010es} used
for $\mathcal{L}_6$. The Class 8 operators that contribute to $\Delta \mathcal{A}_{SMEFT}$, consistent with our assumptions, are
\bea
O_{C8} = \{Q_{qq}^{(1)},Q_{qq}^{(3)},Q_{\ell q}^{(1)},Q_{\ell q}^{(3)},Q_{uu},Q_{eu},Q_{lu},Q_{qe},Q_{ud}^{(1)},Q_{qu}^{(1)},Q_{qd}^{(1)}\}.
\eea
The $pqrt$ flavour subscripts on each operator are suppressed. The operators are precisely defined in Table \ref{op59} for completeness.
The two distinct contractions in the Feynman diagrams are illustrated in Fig.\ref{4fermion}.
We define the notation
\bea
C_\epsilon = \frac{y_t^2 \, C_Z}{16 \, \pi^2 \, \epsilon}.
\eea
to report results calculating in dimensional regularization with $d = 4 - 2 \, \epsilon$ (and a universally anti-commuting $\gamma^5$). We find the divergent result in the $\rm \overline{MS}$ scheme
\bea\label{4Famp}
i \, \Delta \mathcal{A}_\epsilon^{C8} &=& - C_\epsilon \, N_C \, \bar{U}_L^{p} \, \slashed{\epsilon}_Z \, U_L^p
 \, \left[-C_{\substack{qu\\ pp33}}^{(1)} + C_{\substack{qq\\ 33pp}}^{(1)} + C_{\substack{qq\\ pp33}}^{(1)} + C_{\substack{qq\\ 33pp}}^{(3)} + C_{\substack{qq\\ pp33}}^{(3)}\right], \nonumber \\
&-& C_\epsilon \, N_C \,  \bar{U}_R^{p} \,  \slashed{\epsilon}_Z \, U_R^p \,  \left[C_{\substack{qu\\ 33pp}}^{(1)} - C_{\substack{uu\\ 33pp}} - C_{\substack{uu\\ pp33}}\right],\\
&+&  C_\epsilon \,  \bar{D}_L^{r} \,  \slashed{\epsilon}_Z \,  D_L^r \, \left[N_C \left(C_{\substack{qu\\ rr33}}^{(1)} - C_{\substack{qq\\ 33rr}}^{(1)} - C_{\substack{qq\\ rr33}}^{(1)} + C_{\substack{qq\\ 33rr}}^{(3)}
+ C_{\substack{qq\\ rr33}}^{(3)}\right) - 4 \, C_{\substack{qq\\ rr33}}^{(3)} \delta_{r3} \right], \nonumber \\
&+& C_\epsilon \, N_C \,  \bar{D}_R^{r} \,  \slashed{\epsilon}_Z \, D_R^r \, \left[-C_{\substack{qd\\ 33rr}}^{(1)} + C_{\substack{ud\\ 33rr}}^{(1)}\right] - C_\epsilon \, N_C \,  \bar{\nu}_L^{t} \,  \slashed{\epsilon}_Z \, \nu_L^t \, \left[C_{\substack{lq\\ tt33}}^{(1)}+ C_{\substack{lq\\ tt33}}^{(3)} - C_{\substack{lu\\ tt33}}\right], \nonumber \\
&-&  C_\epsilon \, N_C \,  \bar{\ell}_L^{s} \,  \slashed{\epsilon}_Z \, \ell_L^s \,  \left[C^{(1)}_{\substack{lq\\ ss33}} - C_{\substack{lu\\ ss33}} - C_{\substack{lq\\ ss33}}^{(3)} \right]
- C_\epsilon \, N_C \,  \bar{\ell}_R^{s} \,  \slashed{\epsilon}_Z \,  \ell_R^s \,  \left[C_{\substack{qe\\ 33ss}} - C_{\substack{eu\\ ss33}}\right]. \nonumber
\eea
The poles in Eqn.~\ref{4Famp} are directly canceled by the insertion of the counter term matrix $Z_{C7}$ determined in Ref.\cite{Jenkins:2013wua} for
the operators
\bea
O_{C7} = \{Q_{\substack{Hq\\pr}}^{(1)}, \, Q_{\substack{Hq\\pr}}^{(3)}, \, Q_{\substack{H\ell\\pr}}^{(1)},\, Q_{\substack{H\ell\\pr}}^{(3)},\, Q_{\substack{He \\pr}},\, Q_{\substack{Hu \\pr}},\, Q_{\substack{Hd \\pr}}\}.
\eea
The flavour sum in this list is suppressed.
The tree level expansion of the operators $O_{C7}$ contributing to the $Z$ decay amplitude gives
\bea\label{Hiamp}
i \, \delta \mathcal{A}^{C7} &=&C_Z \, \bar{U}_L^{p} \, \slashed{\epsilon}_Z \, U_L^p \, \left[C_{\substack{Hq\\pp}}^{(1)} - C_{\substack{Hq\\pp}}^{(3)} \right] + C_Z \,  \bar{U}_R^{p} \,  \slashed{\epsilon}_Z \, U_R^p \,  \left[C_{\substack{Hu\\pp}}\right], \\
&+& C_Z \,  \bar{D}_L^{r} \,  \slashed{\epsilon}_Z \,  D_L^r \, \left[C_{\substack{Hq\\rr}}^{(1)} + C_{\substack{Hq\\rr}}^{(3)} \right] + C_Z \, \bar{D}_R^{r} \,  \slashed{\epsilon}_Z \, D_R^r \, \left[C_{\substack{Hd\\rr}}\right], \nonumber \\
&+& C_Z \,  \bar{\ell}_L^{s} \,  \slashed{\epsilon}_Z \, \ell_L^s \,  \left[C_{\substack{H\ell\\ss}}^{(1)} + C_{\substack{H\ell\\ss}}^{(3)}  \right] +
C_Z \, \bar{\ell}_R^{s} \,  \slashed{\epsilon}_Z \,  \ell_R^s \,  \left[C_{\substack{He\\ss}} \right] + C_Z \,  \bar{\nu}_L^{t} \,  \slashed{\epsilon}_Z \, \nu_L^t \, \left[C_{\substack{H\ell\\tt}}^{(1)}-C_{\substack{H\ell\\tt}}^{(3)} \right]. \nonumber
\eea
The counter term matrix for the Wilson operators expands out to cancel the poles in Eqn.~\ref{Hiamp}
as $i \, Z_{C7} \, \delta \mathcal{A}^{C7} \rightarrow i \, \delta \mathcal{A}^{C7} - i \, \Delta \mathcal{A}^{C8}_\epsilon$.

\subsection{Class $7$ contributions to $\Delta \mathcal{A}_{SMEFT}$}
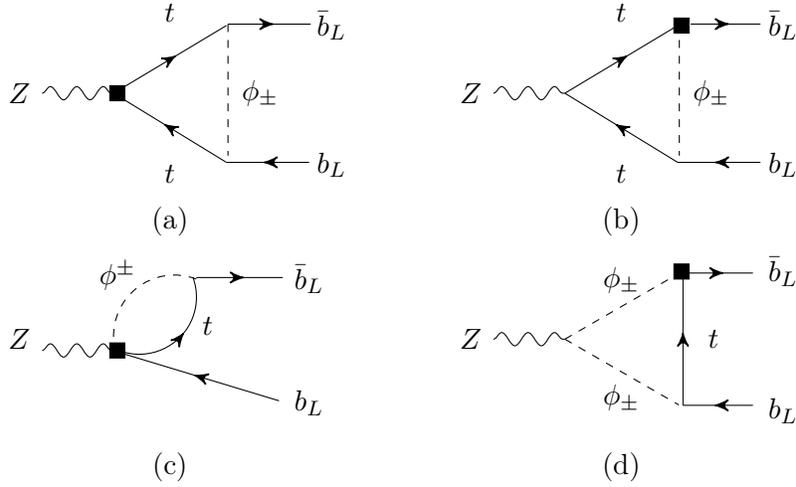
\begin{figure}[h]
\begin{tikzpicture}
[
decoration={
	markings,
	mark=at position 0.55 with {\arrow[scale=1.5]{stealth'}};
}]
\hspace{-3cm}

\draw  [decorate,decoration=snake]  (-1.7,0) -- (-0.75,0);

\draw [postaction=decorate] (-0.75,0)-- (0.75,0.9) ;
\draw [postaction=decorate] (0.75,-0.92) -- (-0.75,0);
\draw [postaction=decorate] (1.85,-0.92) -- (0.75,-0.92);
\draw [postaction=decorate] (50:1.2) -- +(0:1.1) ;

\draw [dashed](50:1.2) -- +(270:1.75) ;

\filldraw (-0.6,-0.1) rectangle (-0.8,0.1);

\draw (0,-1.7) node [align=center] {(a)};

\node [left][ultra thick] at (-1.7,0) {$Z$};
\node [above][ultra thick] at (0,0.8) {$t$};
\node [below][ultra thick] at (0,-0.8) {$t$};
\node [right][ultra thick] at (0.8,0) {$\phi_{\pm}$};
\node [right][ultra thick] at (1.8,0.9) {$\bar{b}_L$};
\node [right][ultra thick] at (1.8,-0.95) {$b_L$};

\hspace{6cm}

\draw  [decorate,decoration=snake]  (-1.7,0) -- (-0.75,0);

\draw [postaction=decorate] (-0.75,0)-- (0.75,0.9) ;
\draw [postaction=decorate] (0.75,-0.92) -- (-0.75,0);
\draw [postaction=decorate] (1.85,-0.92) -- (0.75,-0.92);
\draw [postaction=decorate] (0.92,0.92) -- (1.85,0.92) ;

\filldraw (0.7,0.8) rectangle (0.9,1);

\draw [dashed](50:1.2) -- +(270:1.75) ;

\draw (0,-1.7) node [align=center] {(b)};

\node [left][ultra thick] at (-1.7,0) {$Z$};
\node [above][ultra thick] at (0,0.8) {$t$};
\node [below][ultra thick] at (0,-0.8) {$t$};
\node [right][ultra thick] at (0.8,0) {$\phi_{\pm}$};
\node [right][ultra thick] at (1.8,0.9) {$\bar{b}_L$};
\node [right][ultra thick] at (1.8,-0.95) {$b_L$};

\end{tikzpicture}

\begin{tikzpicture}
[
decoration={
	markings,
	mark=at position 0.55 with {\arrow[scale=1.5]{stealth'}};
}]
\hspace{-3cm}

\draw [decorate,decoration=snake](-1.7,-0.15) --(-0.7,-0.15);

\draw [dashed] (0.35,0.8) arc [radius=0.81, start angle=69, end angle= 182];
\draw [postaction=decorate] (-0.68,-0.15) arc [radius=0.76, start angle=249, end angle= 380];
\draw [postaction=decorate] (0.35,0.82) -- (1.5,0.82) ;
\draw [postaction=decorate] (1.45,-0.82) -- (-0.6,-0.2) ;


\filldraw (-0.6,-0.25) rectangle (-0.8,-0.05);

\draw (0,-1.7) node [align=center] {(c)};

\node [left][ultra thick] at (-1.7,0) {$Z$};
\node [above][ultra thick] at (-0.7,0.5) {$\phi^{\pm}$};
\node [below][ultra thick] at (0.5,0.45) {$t$};
\node [right][ultra thick] at (1.5,0.8) {$\bar{b}_L$};
\node [right][ultra thick] at (1.5,-0.85) {$b_L$};

\hspace{6cm}

\draw  [decorate,decoration=snake]  (-1.7,0) -- (-0.75,0);

\draw[dashed] (180:0.75) -- +(30:1.75) ;

 \draw [dashed](180:0.75) -- +(330:1.75) ;

\filldraw (0.7,0.8) rectangle (0.9,1);

\draw [postaction=decorate] (0.88,0.88) -- (1.75,0.88) ;
\draw [postaction=decorate] (1.75, -0.88)  -- (0.82,-0.88);
\draw [postaction=decorate] (0.82,-0.88)  -- (0.82,0.88);

\draw (0,-1.7) node [align=center] {(d)};

\node [left][ultra thick] at (-1.7,0) {$Z$};
\node [above][ultra thick] at (0,0.45) {$\phi_{\pm}$};
\node [below][ultra thick] at (0,-0.45) {$\phi_{\pm}$};
\node [right][ultra thick] at (1,0) {$t$};
\node [right][ultra thick] at (1.8,0.9) {$\bar{b}_L$};
\node [right][ultra thick] at (1.8,-0.95) {$b_L$};

\end{tikzpicture}
\caption{One loop diagrams contributing to $Z \rightarrow \bar{b}_L \, b_L$ decay through Class 7 operators.  The insertion
of the SMEFT Effective Lagrangian in the diagram is indicated with a black square. Diagrams b), c) and d) also have mirror diagrams which are not shown.}
\label{class7}
\end{figure}

A subset of Class 7 operators also contribute directly to the $Z \rightarrow \bar{b}_L \, b_L$ decay amplitude in the diagrams shown in Fig.~\ref{class7}.
These diagrams give the result
\bea
i \, \Delta \mathcal{A}_{\epsilon}^{C7} = \frac{C_\epsilon}{2} \, \bar{b}_L \,  \slashed{\epsilon}_Z \,  b_L \left[C_{\substack{Hu\\33}} + C_{\substack{Hq\\33}}^{(3)} (6 + 4 s_\theta^2 \, (1- \mathcal{Q}_u))\right].
\eea
For $b_L$ final states another contribution is present from $Z_b$ which appears in the LSZ formula \cite{Lehmann:1954rq} as $ \langle \epsilon_Z^\alpha | \bar{b}_L \,  \gamma_\alpha \,  b_L \, | \bar{b}_L \, b_L \rangle \, Z_b$.
This results in  a contribution to the $Z \rightarrow \bar{b}_L \, b_L$ matrix element
\bea
 \langle \epsilon_Z^\alpha | \bar{b}_L \,  \gamma_\alpha \,  b_L \, | \bar{b}_L \, b_L \rangle \, Z_b = \frac{C_\epsilon}{2} \, \bar{b}_L \,  \slashed{\epsilon}_Z \,  b_L \, \left[C_{\substack{Hq\\33}}^{(3)} (1 - 4 s_\theta^2 \, (1- \mathcal{Q}_u)) -C_{\substack{Hq\\33}}^{(1)} \right].
\eea
A final contribution for this class of operators results from the renormalization of the vev in the SMEFT. The decay amplitude from the LO  insertion of the Class 7 operators is proportional to $\bar{v}_T^2$. This leads to a tadpole contribution
\bea
i \, \delta \mathcal{A}^{C7} \, (\sqrt{Z_v}+ \frac{\Delta v}{\bar{v}_T})_{div}^2,
\eea
which results in a net contribution of the form $- N_c \,C_\epsilon \, i \delta \mathcal{A}^{C7}/C_Z$.
The sum of these Class 7 contributions to the matrix element of $Z$ decay is given by
\bea
\langle \epsilon_Z^\alpha | \bar{b}_L \,  \gamma_\alpha \,  b_L \, | \bar{b}_L \, b_L \rangle \, Z_b \, (\sqrt{Z_v}+ \frac{\Delta v}{\bar{v}_T})_{div}^2 = - \frac{N_c \, y_t^2}{16 \, \pi^2 \, \epsilon} \, i \, \delta\mathcal{A}^{C7} +
\frac{C_\epsilon}{2} \, \bar{b}_L \,  \slashed{\epsilon}_Z \,  b_L \, \left[7 \, C_{\substack{Hq\\33}}^{(3)} -C_{\substack{Hq\\33}}^{(1)} + C_{\substack{Hu\\33}} \right]. \nonumber \\
\eea
This expression directly cancels the $y_t^2$ dependence of the counter term matrices reported for the Class 7 operators in Ref.~\cite{Jenkins:2013wua}.

\subsection{Class 4 and 6 contributions to $\delta \mathcal{A}_{SMEFT}$ and $\Delta \mathcal{A}_{SMEFT}$}\label{class46}

The contributions of the "Class 6" dipole operators to $\bar{\Gamma}_Z$ are intimately related to the contributions
of the "Class 4" operators of Ref.~\cite{Grzadkowski:2010es}.
Due to our assumption of an approximate $\rm U(3)^5$ symmetry in $\mathcal{L}_6$, dominantly broken by the $\rm SM$ Yukawa matrices to satisfy flavour constraints,
the flavour indices of the Class 6 operators are contracted with the $\rm SM$ Yukawas. This leads to flavour violation in an MFV pattern \cite{Chivukula:1987py,Hall:1990ac,D'Ambrosio:2002ex,Feldmann:2008ja,Kagan:2009bn,Feldmann:2009dc}.
We neglect all fermion masses except the top quark, leaving the operators $O_6 = \{Q_{uG},Q_{uW},Q_{uB} \}$
to consider.  For the naive amplitude in Eqn.~\ref{schematic}, there is no contribution to $i \, \delta \mathcal{A}_{SMEFT}$ due to $O_6$ in the MFV limit considered.

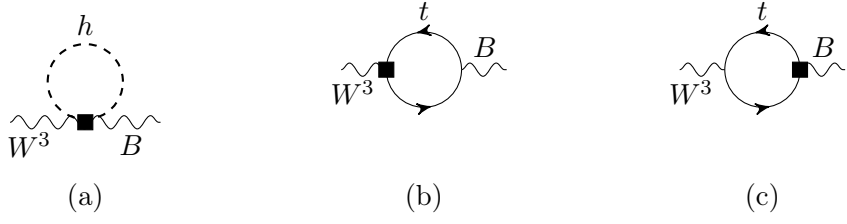
\begin{figure}
\hspace{-6cm}
\begin{tikzpicture}

\filldraw (-0.1,-0.1) rectangle (0.1,0.1);

\draw  [decorate,decoration=snake] (-1,0) -- (1,0);

\draw [thick] [dashed] (0,0.54) circle (0.50);

\draw (0,-1) node [align=center] {(a)};

\node [above][ultra thick] at (0,1) {$h$};
\node [left][ultra thick] at (-0.3,-0.3) {$W^3$};
\node [right][ultra thick] at (0.3,-0.3) {$B$};

\end{tikzpicture}
\begin{tikzpicture}
[
decoration={
	markings,
	mark=at position 0.56 with {\arrow[scale=1.5]{stealth'}};
}]
\hspace{2cm}
\draw [decorate,decoration=snake](-1.1,0) -- (-0.5,0);
\draw [decorate,decoration=snake](0.5,0) -- (1.1,0);

\draw [postaction=decorate] (0.5,0) arc [radius=0.5, start angle=0, end angle= 180];
\draw [postaction=decorate] (-0.5,0) arc [radius=0.5, start angle=180, end angle= 360];


\draw (0,-1.7) node [align=center] {(b)};

\filldraw (-0.4,-0.1) rectangle (-0.6,0.1);

\node [above][ultra thick] at (0,0.5) {$t$};
\node [left][ultra thick] at (-0.5,-0.3) {$W^3$};
\node [right][ultra thick] at (0.5,0.3) {$B$};

\hspace{4.5cm}

\draw [decorate,decoration=snake](-1.1,0) -- (-0.5,0);
\draw [decorate,decoration=snake](0.5,0) -- (1.1,0);

\draw [postaction=decorate] (0.5,0) arc [radius=0.5, start angle=0, end angle= 180];
\draw [postaction=decorate] (-0.5,0) arc [radius=0.5, start angle=180, end angle= 360];

\filldraw (0.4,-0.1) rectangle (0.6,0.1);

\draw (0,-1.7) node [align=center] {(c)};

\node [above][ultra thick] at (0,0.5) {$t$};
\node [left][ultra thick] at (-0.5,-0.3) {$W^3$};
\node [right][ultra thick] at (0.5,0.3) {$B$};

\end{tikzpicture}
\caption{The contribution of $Q_{HWB}, Q_{uW},Q_{uB}$ to the un-diagonalized kinetic term $W^3_{\mu \, \nu} \, B^{\mu \, \nu}$.
The insertion of the Effective Lagrangian is indicated with a black box.}
\label{CHWB_loop}
\end{figure}

The operator $Q_{HWB}$ contributes to $\delta \mathcal{A}_{SMEFT}$ at tree level.
Working in the canonically normalized SMEFT the operator $\mathcal{Q}_{HWB}$ contributes to $\delta \mathcal{A}_{SMEFT}$ in the following manner \cite{Alonso:2013hga,Berthier:2015oma}
\bea\label{A0CHWB}
i \, \delta \mathcal{A}^{HWB} =  i \, \sqrt{\bar{g}_1^2+ \bar{g}_2^2} \, \bar{\Psi}_i \, \slashed{\epsilon}_Z \, \Psi_i \, \mathcal{Q}_{\Psi_s} \left(\frac{\bar{g}_1^2 \, \bar{g}_2^2 \, (\bar{g}_2^2 - \bar{g}_1^2)}{(\bar{g}_1^2+ \bar{g}_2^2)^2} \right) \, \bar{v}^2_T \, C_{HWB},
\eea
where $\Psi_i = \{u,c,d,s,b,e,\mu,\tau\}$ and $\mathcal{Q}_{\Psi_i} = \{2/3,2/3,-1/3,-1/3,-1/3,-1,-1,-1\}$. The counter terms are
introduced into Eqn~\ref{A0CHWB} following the convention
\bea\label{counter termsCHWB}
i \, \delta  \mathcal{A}^{HWB} \rightarrow i \, \delta \mathcal{A}^{HWB}\left(1+  (\sqrt{Z_v}+ \frac{\Delta v}{\bar{v}_T})_{div}^2\right)  \, \left( 1+   \frac{1}{32 \, \pi^2 \, \epsilon} \left(2 \, N_c \, y_t^2 + 4 \, \lambda \right)\right),
\eea
where the relevant RGE entries of the SMEFT were reported in Refs.~\cite{Jenkins:2013zja,Jenkins:2013wua}.
The counter term $\propto N_c \,y_t^2$ in this expression is exactly canceled by the vev counter term. The counter term $\propto \lambda$ in Eqn.~\ref{counter termsCHWB} is directly canceled by the one loop contribution to the un-diagonalized kinetic term
$W^3_{\mu \, \nu} \, B^{\mu \, \nu}$ shown in Fig.~\ref{CHWB_loop} (a), which gives a divergent contribution
\bea
\frac{\bar{g}_1 \, \bar{g}_2 \, \lambda \, \bar{v}_T^2}{16 \, \pi^2 \, \epsilon} \, C_{HWB} \, W^3_{\mu \, \nu} \, B^{\mu \, \nu}.
\eea
\begin{figure}[bt]
\begin{tikzpicture}
[
decoration={
	markings,
	mark=at position 0.56 with {\arrow[scale=1.5]{stealth'}};
}]
\hspace{-5cm}
\draw  [decorate,decoration=snake]  (-1.7,0) -- (-0.75,0);
\draw [postaction=decorate] (0.75,0.9) -- (1.9,0.9);
\draw [postaction=decorate] (1.9,-0.9) --(0.75,-0.9);
\draw [postaction=decorate] (0.75,-0.9) -- (-0.75,0);
\draw [postaction=decorate] (-0.75,0) -- (0.75,0.9);

\draw [dashed](50:1.2) -- +(270:1.75) ;

\filldraw (-0.6,-0.1) rectangle (-0.8,0.1);

\draw (0,-1.7) node [align=center] {(a)};

\node [left][ultra thick] at (-1.7,0) {$Z$};
\node [above][ultra thick] at (0,0.6) {$t$};
\node [below][ultra thick] at (0,-0.6) {$t$};
\node [right][ultra thick] at (1,0) {$\phi_{\pm}$};
\node [right][ultra thick] at (1.8,0.9) {$\bar{b}_L$};
\node [right][ultra thick] at (1.8,-0.95) {$b_L$};

\hspace{5cm}

\draw  [decorate,decoration=snake]  (-1.7,0) -- (-0.75,0);
\draw [postaction=decorate] (1.9,-0.9) --(0.75,-0.9);
\draw [postaction=decorate] (0.75,0.9) -- (1.9,0.9);
\draw [postaction=decorate] (0.75,-0.9) -- (0.75,0.9);

\draw [dashed](180:0.75) -- +(30:1.75) ;
\draw [dashed](180:0.75) -- +(330:1.75) ;

\filldraw (-0.6,-0.1) rectangle (-0.8,0.1);

\draw (0,-1.7) node [align=center] {(b)};

\node [left][ultra thick] at (-1.7,0) {$Z$};
\node [above][ultra thick] at (0,0.6) {$\phi_{\pm}$};
\node [below][ultra thick] at (0,-0.6) {$\phi_{\pm}$};
\node [right][ultra thick] at (0.8,0) {$t$};
\node [right][ultra thick] at (1.8,0.9) {$\bar{b}_L$};
\node [right][ultra thick] at (1.8,-0.95) {$b_L$};

\hspace{5cm}

\draw  [decorate,decoration=snake]  (-1.7,0) -- (-0.75,0);
\draw [dashed](180:0.75) -- +(30:1.75) ;
\draw  [decorate,decoration=snake] (180:0.75) -- +(330:1.75) ;

\draw [postaction=decorate] (1.9,-0.9) --(0.75,-0.9);
\draw [postaction=decorate] (0.75,0.9) -- (1.9,0.9);
\draw [postaction=decorate] (0.75,-0.9) -- (0.75,0.9);

\filldraw (-0.6,-0.1) rectangle (-0.8,0.1);

\draw (0,-1.7) node [align=center] {(c)};

\node [left][ultra thick] at (-1.7,0) {$Z$};
\node [above][ultra thick] at (0,0.6) {$\phi_{\pm}$};
\node [below][ultra thick] at (0,-0.6) {$W_{\pm}$};
\node [right][ultra thick] at (0.8,0) {$t$};
\node [right][ultra thick] at (1.8,0.9) {$\bar{b}_L$};
\node [right][ultra thick] at (1.8,-0.95) {$b_L$};

\end{tikzpicture}
\caption{One loop diagrams contributing to $Z \rightarrow \bar{b}_L \, b_L$ decay due to $\mathcal{Q}_{HWB}$. The insertion
of the Effective Lagrangian in the diagram is indicated with a black square.\label{CHWB_loop2}}
\end{figure}
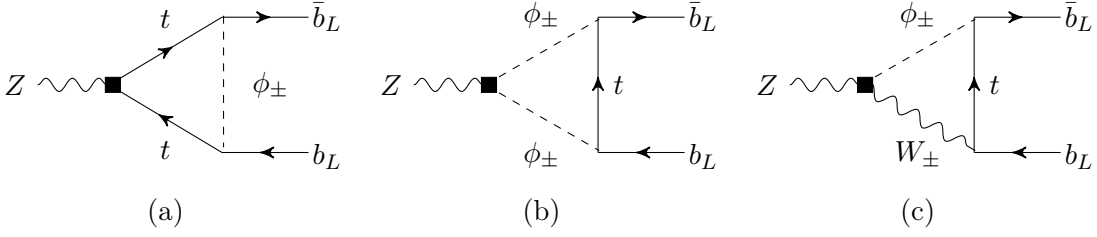
The operator $Q_{HWB}$ also contributes directly to $\Delta \mathcal{A}_{SMEFT}$ through the diagrams shown in
Fig.~\ref{CHWB_loop2}.  These diagrams directly give the result
\bea\label{CHWBresult}
i \, \Delta \mathcal{A}_\epsilon^{HWB} = C_\epsilon \, \bar{b}_L \,  \slashed{\epsilon}_Z \,  b_L \, \left(\mathcal{Q}_u -1\right) \left[\frac{\bar{g}_1^2 \, \bar{g}_2^2 (\bar{g}_2^2 - \bar{g}_1^2)}{(\bar{g}_1^2 + \bar{g}_2^2)^2}\right] \, C_{HWB}.
\eea
The wavefunction renormalization of the $b$ quarks $Z_b$ given in Eqn.~\ref{Zbeqn} cancels this $\epsilon$ pole, and is introduced as
\bea
i \, \delta \mathcal{A}^{HWB} \rightarrow  i \, \delta \mathcal{A}^{HWB} \, Z_b.
\eea
A final term from the Class 6 operators is due to the un-diagonalized one loop
two point function $\Pi_{W^3 \, B}$, as shown in Fig.~\ref{CHWB_loop} (b,c). The contribution of $\mathcal{Q}_{HWB}$ to this kinetic term at tree level is given by
\bea
-\frac{\bar{v}_T^2 \, \bar{g}_1 \, \bar{g}_2}{2} \, C_{HWB}^{(0)} W_3^{\mu \, \nu} \, B_{\mu \, \nu},
\eea
which leads to the counter term of the $Q_{HWB}$ operator giving the contributions
\bea\label{QHWBdipoles}
-\frac{\bar{v}_T^2 \, \bar{g}_1 \, \bar{g}_2}{2} \, W_3^{\mu \, \nu} \, B_{\mu \, \nu} \frac{N_c \, y_t^2}{32 \, \pi^2 \, \epsilon} \, \left[
(C^{(r)}_{\substack{uB\\ 33}}+  C^{(r) \, \star}_{\substack{uB \\ 33}})  + 2  \,  (\hyp_q + \hyp_u) \, \left( C^{(r)}_{\substack{uW \\ 33}} +  C^{(r), \star}_{\substack{uW\\ 33}} \right) \right],
\eea
where $\hyp_q = 1/6$ and $\hyp_u =2/3$. Note that in the limit with no CP violating BSM phases in $\mathcal{L}_6$:
$C^{(r)}_{\substack{uB\\ 33}} = C^{(r), \star}_{\substack{uB\\ 33}}$ and $C^{(r)}_{\substack{uW \\ 33}} = C^{(r), \star}_{\substack{uW\\ 33}}$.
These poles are directly canceled by the $\epsilon$ divergences in Fig.~\ref{CHWB_loop} (b,c).

\subsection{Naive amplitude one loop finite amplitude}\label{naiveresults}

The massless limit considered in the final states of the $Z$ decay, encourages expressing the finite term results in terms
of the chiral fields of the SM, as was done in Section \ref{class8section}.
The finite terms for the Class $8$ operators are
\bea
i \,\Delta \mathcal{A}^{C8}_{U,D} &=& \Delta_8^{y_t} \bar{U}_L^{p} \, \slashed{\epsilon}_Z \, U_L^p  \left[C_{\substack{qq\\ 33pp}}^{(1)} + C_{\substack{qq\\ pp33}}^{(1)} + C_{\substack{qq\\ 33pp}}^{(3)} + C_{\substack{qq\\ pp33}}^{(3)} -C_{\substack{qu\\ pp33}}^{(1)} \right], \nn \\
&+& \Delta_8^{y_t} \bar{U}_R^{p} \, \slashed{\epsilon}_Z \, U_R^p \left[C_{\substack{qu\\ 33pp}}^{(1)} - C_{\substack{uu\\ 33pp}} - C_{\substack{uu\\ pp33}}\right], \nn \\
&+& \Delta_8^{y_t} \bar{D}_L^{r} \, \slashed{\epsilon}_Z \, D_L^r  \left[C_{\substack{qq\\ 33rr}}^{(1)} + C_{\substack{qq\\ rr33}}^{(1)} - C_{\substack{qq\\ 33rr}}^{(3)}
- C_{\substack{qq\\ rr33}}^{(3)} - C_{\substack{qu\\ rr33}}^{(1)} \right],  \\
&+& \Delta_8^{y_t} \bar{D}_L^{r} \, \slashed{\epsilon}_Z \, D_L^r \left[\frac{4 \, \delta_{r3}}{N_c} \, C_{\substack{qq\\ 3333}}^{(3)} \left(- \log^{-1} \left(\frac{\Lambda^2}{\hat{m}_t^2} \right) + 1 \right)\right]
+ \Delta_8^{y_t} \bar{D}_R^{r} \, \slashed{\epsilon}_Z \, D_R^r \left[C_{\substack{qd\\ 33rr}}^{(1)}  - C_{\substack{ud\\ 33rr}}^{(1)}  \right], \nn \\
&+&
\Delta_8^{y_t} \bar{\ell}_L^{s} \, \slashed{\epsilon}_Z \, \ell_L^s  \left[C_{\substack{\ell q\\ ss33}}^{(1)} - C_{\substack{\ell q\\ ss33}}^{(3)} - C_{\substack{\ell u\\ ss33}} \right]
+ \Delta_8^{y_t} \bar{\ell}_R^{s} \, \slashed{\epsilon}_Z \, \ell_R^s  \left[-C_{\substack{eu\\ ss33}} + C_{\substack{q e \\ 33ss}} \right], \nn \\
&+&\Delta_8^{y_t} \bar{\nu}_L^{t} \, \slashed{\epsilon}_Z \, \nu_L^t  \left[C_{\substack{\ell q\\ tt33}}^{(1)} + C_{\substack{\ell q\\ tt33}}^{(3)} - C_{\substack{\ell u\\ tt33}} \right], \nn
\eea
where
\bea
\Delta_8^{y_t}= - \frac{C_Z \, y_t^2 \, N_c}{16 \, \pi^2} \log \left[\frac{\Lambda^2}{\hat{m}_t^2} \right].
\eea
$\mathcal{Q}_{HWB}$ contributes directly to $Z \rightarrow \bar{b}_L \, b_L$ at one loop with the finite terms
\bea
i \, \Delta \mathcal{A}^{C4} &=&  \frac{C_Z \, y_t^2}{16 \, \pi^2} \, \bar{e}^2 (\bar{c}_\theta^2 - \bar{s}_\theta^2) C_{HWB} \,(\mathcal{Q}_u -1) \left[\frac{3}{2} +
 \log \left[\frac{\Lambda^2}{\hat{m}_t^2} \right]\right] \bar{b}_L \, \slashed{\epsilon}_Z \, b_L,
\eea
 and through one-loop Higgs tadpole terms as
\bea
i \, \Delta \mathcal{A}^{C4}_{\Delta v} &=& i\delta \mathcal{A}^{HWB}\frac{\Delta V^2}{\bar{v}_T^2}.
\eea
The vev shift at one loop ($\Delta V^2$) will be defined and addressed in detail in section \ref{MZextract}.\footnote{Note that $\Delta V \neq \Delta v$.}
The Class 7 operators contribute directly to $Z \rightarrow \bar{b}_L \, b_L$ at one loop with the finite terms
\bea
i \, \Delta \, \mathcal{A}^{C7} &=&  \frac{C_Z \, y_t^2}{16 \, \pi^2} \, \left[\left(-\frac{1}{4} + \frac{1}{2} \log \left[\frac{\Lambda^2}{\hat{m}_t^2} \right]\right) C_{Hu} + C_{Hq}^{(1)} \right] \bar{b}_L \, \slashed{\epsilon}_Z \, b_L, \nn \\
&+& \frac{C_Z \, y_t^2}{16 \, \pi^2} \, C_{Hq}^{(3)}  \, \left[\frac{1}{2} - s_{\hat{\theta}}^2 \, \mathcal{Q}_b +  (3 - 2 \mathcal{Q}_b \, s_{\hat{\theta}}^2) \, \log \left[\frac{\Lambda^2}{\hat{m}_t^2} \right] \right] \bar{b}_L \, \slashed{\epsilon}_Z \, b_L,
\eea
and through Higgs tadpoles as
\bea
i \, \Delta \mathcal{A}^{C7}_{\Delta v} &=& i\delta \mathcal{A}^{C7}\frac{\Delta V^2}{\bar{v}_T^2}.
\eea

\section{Input parameters, and corrections, in the SMEFT}\label{inputcorr}
Any prediction of $\bar{\Gamma}_Z$ in the SM or the SMEFT depends upon an input parameter set used to define
the numerical values of Lagrangian parameters.\footnote{For discussions on input parameter choices and perturbative corrections to the input parameter set determinations in the SM,
see Refs.~\cite{Wells:2005vk,Bardin:1999ak,Freitas:2016sty,Agashe:2014kda}.}
The input parameter set we use is given in Table~\ref{inputs}.
Standard Model one loop corrections to the input parameters used to define $\mathcal{A}_{SM}$ in Eqn.~\ref{schematic} interfere with $\delta \mathcal{A}_{SMEFT}$. There are also one loop contributions of $\mathcal{L}_6$
interactions to the mapping of the input parameter set to the Lagrangian terms, which are not included in the "naive LO amplitude" defined above. These contributions can lead to shifts the same order as the interference of the $\mathcal{A}_{SM} \times \Delta \mathcal{A}_{SMEFT}$ amplitudes, for the input parameters $\{\hat{\alpha}_{ew},\hat{m}_Z,\hat{G}_F\}$.
We discuss each of these input parameters in turn in this section, characterizing the corrections that we include in our predicted value of $\bar{\Gamma}_Z$ in the SMEFT.
Conversely, one loop corrections to the extraction of the input parameters $\hat{m}_t$ and $\hat{m}_h$, and the mapping of these input parameters to the Lagrangian terms, are neglected, as these are two loop effects for $\bar{\Gamma}_Z$ and the related quantities reported in this paper. A class of tree level shifts to the measured pole masses in the SMEFT is unobservable in the quantities predicted, as the results are in terms of these pole masses.
\begin{center}
\begin{table}[bt]
\centering
\tabcolsep 8pt
\begin{tabular}{|c|c|c|}
\hline
Input parameters & Value & Ref. \\ \hline
$\hat{\alpha}_{ew}$ & $1/137.035 999139(31) $ &  \cite{Agashe:2014kda,Mohr:2012tt,Mohr:2015ccw} \\
$\hat{m}_Z$ & $91.1875 \pm 0.0021$ & \cite{Z-Pole,Agashe:2014kda,Mohr:2012tt,Mohr:2015ccw} \\
$\hat{G}_F$ & $1.1663787(6) \times 10^{-5} $ &  \cite{Agashe:2014kda,Mohr:2012tt} \\
$\hat{m}_t$&$173.21 \pm 0.51 \pm 0.71$& \cite{Agashe:2014kda}\\
$\hat{m}_h$ & $125.09 \pm 0.21 \pm 0.11$ & \cite{Aad:2015zhl} \\
\hline
\end{tabular}
\caption{Current experimental best estimates of the input parameter set. We use hat superscripts
to indicate when a parameter is a measured value, consistent with the notation in Refs.~\cite{Berthier:2015oma,Berthier:2015gja,Berthier:2016tkq}\label{inputs}.
We interpret the measured values of $\hat{m}_t$, $\hat{m}_h$ to correspond to the $\rm \overline{MS}$ pole mass. Quantities with units are expressed in $\rm GeV$
to the appropriate power.}
\end{table}
\end{center}
\subsection{$\hat{\alpha}_{ew}$ extractions}
Extractions of $\hat{\alpha}_{ew}$ are dominated by $p^2 \rightarrow 0$ measurements
determined by probing the Coulomb potential of a charged particle, for example in a measurement of $g-2$ for the electron or muon.
The low scale measurement extracts a different parameter in the SMEFT, compared to the SM, due to contributions to the magnetic moments of the leptons from local contact operators. These contributions are discussed in Ref.~\cite{Alonso:2013hga}, but are not relevant here.
In the chosen $\rm U(3)^5$ limit
these corrections are proportional to light quark and lepton masses which are neglected in this work. As a result the low scale matching of $\alpha$ in the
SMEFT proceeds in the same manner as in the SM with the mapping
\bea
- i \, \left[\frac{4 \, \pi \, \hat{\alpha}(q^2)}{q^2}\right]_{q^2 \rightarrow 0}
\equiv \frac{- i \, (\bar{e}_0 + \Delta R_{\bar{e}})^2}{q^2} \left[1 + \frac{\Sigma^{AA}(q^2)}{q^2} \right]_{q^2 \rightarrow 0}
\eea
In this expression, the one loop renormalization of the electric charge $\bar{e}$ is introduced as $\bar{e} = \bar{e}_0 + \Delta R_{\bar{e}}$, where $\Delta R_{\bar{e}}$ is formally of one loop order and is fixed by renormalization conditions. The renormalization of $\bar{e}$ in the Lagrangian is related to the two-point functions as
\bea
\frac{\Delta R_{\bar{e}}}{\bar{e}_0} = \frac{1}{2} \frac{\partial \Sigma_T^{AA}(p^2)}{\partial p^2} \vert_{p^2 \rightarrow 0}
\eea
in the BFM, as a class of one loop corrections to $\Sigma_T^{AZ}(0)$ vanish in this case.\footnote{The subtleties of the BFM in the SMEFT and tadpole contributions are extensive and will be discussed in more detail in a future publication.}
Here $\Sigma_T^{AA},\Sigma_T^{AZ}$ are the transverse components of
the two point functions, consistent with the notation in Ref.~\cite{Denner:1991kt}. Our limit of retaining the one loop contributions proportional to $\lambda,y_t$, in the $\rm U(3)^5$ symmetry limit, leads to a vanishing $\Delta R_{\bar{e}}$, in agreement with the results in Ref.~\cite{Gauld:2015lmb}.

The finite terms in the low scale matching that are the largest effect are due to the vacuum polarization of the photon in the $q^2 \rightarrow 0$ limit. Following Ref.~\cite{Wells:2005vk} we rearrange this unknown term into the form
\bea
\left[\frac{\Sigma^{AA}(q^2)}{q^2} \right]_{q^2 \rightarrow 0} = {\rm Re} \frac{\Sigma^{AA}(m_Z^2)}{m_Z^2} - \left[\frac{{\rm Re} \Sigma^{AA}(m_Z^2)}{m_Z^2}  - \left[\frac{\Sigma^{AA}(q^2)}{q^2} \right]_{q^2 \rightarrow 0}\right]
\eea
and introduce the notation
\bea
\nabla \alpha = \left[\frac{{\rm Re} \Sigma^{AA}(m_Z^2)}{m_Z^2}  - \left[\frac{\Sigma^{AA}(q^2)}{q^2} \right]_{q^2 \rightarrow 0}\right].
\eea
As is standard, we decompose this quantity into perturbative and non-perturabative contributions as
 $\nabla \alpha = \nabla \alpha_\ell + \nabla \alpha_t + \nabla \alpha_{had} + \nabla \alpha_{\rm \overline{MS} -os}$.The estimates
for these quantities are $\nabla \alpha_\ell  = 0.03150$,  $\nabla \alpha_t = -0.0007$, $\nabla \alpha_{had} \approx 0.02764$ and $\nabla \alpha_{\rm \overline{MS} -os} \approx 0.0072$
\cite{Wells:2005vk,Agashe:2014kda,Baikov:2012zm,PhysRevD.22.971} where the last term is the correction due to adjusting the value of $\nabla \alpha_{had}$ from the on-shell to $\rm \overline{MS}$ scheme in use here. Note that we neglect any SMEFT corrections to the non-perturbative parameters, assuming them to be subdominant
to other theoretical uncertainties.

\subsubsection{$\alpha_{ew}$ running and hadronic corrections}
As $\hat{\alpha}_{ew}$ is extracted in the Thompson limit ($p^2 \rightarrow 0$), the corresponding input value is run up to the
scale $\mu^2 \sim m_Z^2$.  It is consistent to only retain one loop contributions proportional to $\lambda,y_t$, in the $\rm U(3)^5$ symmetry limit,
neglecting this running effect in this calculation. The effect of the hadronic contributions to the vacuum polarization is due to sub-leading gauge
coupling dependence that is generally neglected in this paper. However, we treat this numerically significant correction as leading order due
to it being defined by significant non-perturbative effects. This nonperturbative correction to the input parameter in the SM is a numerically significant contribution leading to
\cite{Agashe:2014kda,Mohr:2015ccw}
\bea
1/\tilde{\alpha}_{ew}(p^2 \sim \hat{m}_Z^2) = 127.940 \pm 0.014, \quad {\rm while} \quad 1/\hat{\alpha}_{ew}(p^2 \rightarrow 0) = 137.035 999139(31). \nonumber \\
\eea
Neglecting this modification\footnote{Our notation uses $\tilde{\alpha}_{ew}$ instead of $\hat{\alpha}_{ew}$ for $\rm \overline{MS}$ quantities, which differs from Ref.\cite{Agashe:2014kda}. This is to avoid different uses of the tilde superscript when comparing to the LO results in Refs.~\cite{Berthier:2015oma,Berthier:2015gja}.}
would necessitate introducing a large theoretical error to this input parameter, $\mathcal{O}(10\%)$.
To avoid this, we retain the (gauge coupling dependent) hadronic contributions to the vacuum polarization while neglecting the SM running.
This is done by adjusting the result for $\alpha(m_Z^2)_{ew}$ by
\bea
\tilde{\alpha}(m_Z^2)_{ew} = \frac{\hat{\alpha}(p^2\rightarrow 0)_{ew}}{1 - \nabla \alpha}
\eea
In the remainder of this paper we use the notation with a $\sim$ superscript to signify that this vacuum polarization correction is included.
Numerically, this results in the central value $1/\tilde{\alpha}(m_Z^2)_{ew} \simeq 128.041$.
This is an approximation to the full result, that would be appropriate to determine when the complete one loop of $\bar{\Gamma}_Z$ in the SMEFT is known.
Numerically, this is a reasonable approximation to the result quoted in the PDG of $1/\tilde{\alpha}(m_Z^2)_{ew} = 127.940 \pm 0.014$
using higher order results in perturbation theory, in the SM.

The gauge couplings and $\alpha$ run differently in the SMEFT compared to the SM.
These corrections are due to $C_{HW}$ and $C_{HB}$ dependent terms introduced in the RGEs \cite{Jenkins:2013zja}.
In this calculation we retain corrections proportional to $\lambda$ and $y_t$. If gauge coupling
dependence is due to a normalization choice on the operators, we retain such effects. This is the case for this running with the normalization given in Eqn.~\ref{opnorm}.
Using the results of
Ref.~\cite{Jenkins:2013zja} as an approximation to this running\footnote{In the case of the SM running and the SMEFT correction, formally one should use
a series of EFT's to relate the $p^2 \rightarrow 0$ limit and the higher scale $p^2=m_Z^2$ parameter. We approximate this running without constructing the series of EFT's to relate the low and high scale. This approximation should be improved upon once the full one loop result of $\bar{\Gamma}_Z$ is known for the SMEFT. Note that we use $p^2 \simeq 0.01 ~\rm GeV^2$ for the low measurement scale value.}, one finds the $\lambda$ dependent one loop correction to the SM value
\bea
\frac{1}{4 \, \pi} \frac{\bar{g}_1^2 \, \bar{g}_2^2}{(\bar{g}_1^2 + \bar{g}_2^2)} \left[1 + \delta \alpha + \Delta \alpha \right]  \approx 1/128.041,
\eea
which introduces dependence on $C_{HWB}$ and $C_{HB} + C_{HW}$ into precise EWPD measurements as
\bea
\delta \alpha &=& - \sqrt{2} \, \frac{4 \pi \, \tilde{\alpha}}{\hat{G}_F} C_{HWB}^{(r)},  \\
\Delta \alpha &=& -\sqrt{2} \, \frac{4 \pi \, \tilde{\alpha}}{\hat{G}_F} C_{HWB}^{(r)}\left(\Delta V^2+\frac{\Delta G_F}{\hat{G}_F}\right)+\frac{\tilde{\alpha}}{\pi} \,  \hat{m}_h^2 \, \left(C_{HB}^{(r)} + C_{HW}^{(r)} \right) \log \left[\frac{\hat{m}_Z^2}{p^2} \right], \nn \\
&\simeq&  -\sqrt{2} \, \frac{4 \pi \, \tilde{\alpha}}{\hat{G}_F} C_{HWB}^{(r)}\left(\Delta V^2+\frac{\Delta G_F}{\hat{G}_F}\right) + 0.03 \, \hat{m}_h^2 \,  \left(C_{HB}^{(r)} + C_{HW}^{(r)} \right).
\eea
If neglected, this SMEFT correction to the running can be dominant over the SM theoretical error when using the input
parameter $\tilde{\alpha}(m_Z^2)_{ew}$ to define numerical values of the gauge couplings \cite{Passarino:2016pzb},
depending on the {\it{a priori}} unknown value of $C_{HB} + C_{HW}$. $\Delta \alpha$  is formally counted as one loop and $1/\Lambda^2$ suppressed when constructing the result for $\bar{\Gamma}_Z$ and related quantities.

\subsection{$\hat{m}_Z$ extractions}\label{MZextract}
$\hat{m}_Z$ is extracted at LEP in a pole scan for the position of the $Z$ resonance peak \cite{ALEPH:2005ab}. The SMEFT LO redefinition of the Lagrangian parameter that corresponds to $m_Z$ is well known
at this time \cite{Kennedy:1988sn,Altarelli:1990zd,Altarelli:1991fk,Golden:1990ig,Holdom:1990tc,Peskin:1990zt,Peskin:1991sw,Grinstein:1991cd,Maksymyk:1993zm,Georgi:1991ci,Burgess:1993vc,Hagiwara:1986vm,Han:2004az,Ciuchini:2013pca,Alonso:2013hga,Berthier:2015oma,Berthier:2015gja,Berthier:2016tkq}
and one finds \cite{Alonso:2013hga}
\bea
\bar{m}_Z^2 = \frac{\bar{v}_T^2}{4} (\bar{g}_1^2 + \bar{g}_2^2) +  \frac{\bar{v}_T^4}{8} \, C_{HD} \, (\bar{g}_1^2 + \bar{g}_2^2) +   \frac{\bar{v}_T^4}{2} \, \bar{g}_1^2 \, \bar{g}_2^2 \, C_{HWB}.
\eea
With the $\mathcal{L}_6$ operators normalized as in Eqn.~\ref{opnorm}.
The operators of interest when extending this result to one loop are $Q_{HD}$, $Q_{HWB}$, $Q_{HW}$ and $Q_{HB}$ and the relevant Effective Lagangian terms are
\bea
{\cal L}_{SMEFT} &=& - \frac{1}{2} W^+_{\mu \nu} \, W_-^{\mu \nu} - \frac{1}{4} W^3_{\mu \nu} \, W_3^{\mu \nu} - \frac{1}{4} B_{\mu \nu} \, B^{\mu \nu} -
\frac{1}{4} G_{\mu \nu} \, G^{\mu \nu} +
 \frac12 \bar{v}_T^2 \, g_3^2 \, C_{HG}^{(0)} \,G_{\mu \nu}^A \, G^{A\mu \nu}, \nn \\
&+& \frac12 \, \bar{v}_T^2 \, g_2^2 \, C_{HW}^{(0)}  W_{\mu \nu}^I W^{I \mu \nu}+ \frac12 \bar{v}_T^2 \, g_1^2 \, C_{HB}^{(0)}  B_{\mu \nu} B^{\mu \nu} -\frac{1}{2} \bar{v}_T^2 \, C_{HWB}^{(0)} \, g_1 \, g_2 \,  W_{\mu \nu}^3 B^{\mu \nu}, \nn \\
&+& \frac14 \, g_2^2 \, \bar{v}_T^2  \, W_\mu^+ W^{-\,\mu} + \frac18 \bar{v}_T^2  (g_2 W^3_\mu -g_1 B_\mu)^2  + \frac1{16} \, \bar{v}_T^4 \, C_{H D}^{(0)} \, (g_2 W^3_\mu -g_1 B_\mu)^2 + \cdots
\eea
The counter terms for the Wilson coefficients of the operators $Q_{HD}$, $Q_{HWB}$, $Q_{HW}$, $Q_{HB}$ and $Q_{HG}$ enter the calculation --denoting the renormalized operators with $(r)$ superscripts -- as
\bea\label{bareguageops}
\mathcal{L}_6^{(0)} &=& Z_{SM} \, Z_{i,j} \, C^{(r)}_{i} \, Q^{(r)}_{j} + \cdots, \nn \\
&=& C^{(r)}_{HWB} \, (1+ \frac{2 \, \lambda+ N_c \, y_t^2}{16 \, \pi^2 \, \epsilon}) \, Z_{SM} \, Q_{HWB}^{(r)} +  C^{(r)}_{HD} \, (1+ \frac{6 \, \lambda + 2 \, N_c \, y_t^2}{16 \, \pi^2 \, \epsilon}) \, Z_{SM} \,  Q_{HD}^{(r)},  \nonumber \\
 &+& (1+ \frac{6 \, \lambda + N_c \, y_t^2}{16 \, \pi^2 \, \epsilon}) \, Z_{SM} (C^{(r)}_{HG} \,  Q_{HG}^{(r)} + C^{(r)}_{HB}  \, Q_{HB}^{(r)} + C^{(r)}_{HW}  \, Q_{HW}^{(r)} ) + \cdots
\eea
We normalize each of the operators with the gauge coupling of the field strength(s) present in the operator as in Eqn.~\ref{opnorm}. As a result, due to the BFM, the renormalization of the gauge coupling
and gauge field strengths ($F = \{W,B,G\}$) cancel for the $\mathcal{L}_6$ terms. The remaining $Z_{SM}$ factors are the same for the operators $Q_{HW}$, $Q_{HB}$ and $Q_{HG}$.

For a matrix element with no external Higgs state, the cancelation of the divergences occurs as follows.
Expanding the operators about the vev results in two insertions of $\bar{v}_T$.
The resulting $y_t^2$ dependent divergence cancels directly using the expression for $(\sqrt{Z_v} + \Delta v/\bar{v}_T)_{div}$. Similarly, for $Q_{HD}$, four insertions
of $\bar{v}_T$ leads to the $y_t^2$ divergence canceling directly. The cancelation of the divergent $\lambda$ dependence
for $Q_{HWB}$ follows from the diagram in Fig.~\ref{CHWB_loop} as discussed in Section \ref{class46}. Also, note that the mixing with the dipole operators
cancels for this operator, as shown  in Eqn.~\ref{QHWBdipoles}.

The $\lambda$ divergence cancels for $Q_{HB}$, $Q_{HG}$ and $Q_{HW}$ in an interesting manner.
Expanding the operator $Q_{HG}$ at one loop  generates a divergence from a closed Higgs loop. A final contribution comes from expanding out the Yang-Mills field strength counter term in the SMEFT. The contributions are illustrated in Fig.~\ref{YMcancelation}. The divergence cancels as
\bea
(1+ \frac{6 \, \lambda}{16 \, \pi^2 \, \epsilon}  - \frac{2 \, \lambda}{16 \, \pi^2 \, \epsilon} - \frac{4 \, \lambda}{16 \, \pi^2 \, \epsilon}) \, \frac{\bar{v}_T^2}{2} \, \bar{g}_3^2 \, G^a_{\mu \, \nu} \, G_a^{\mu \, \nu} \, C^{(r)}_{HG}.
\eea
\begin{figure}
\begin{tikzpicture}
\hspace{-2cm}
\draw     [decorate,decoration=snake]  (-1,0) -- (1,0);
\draw [thick] [dashed] (0,0.5) circle (0.50);
\node [above][ultra thick] at (0,1) {$h$};
\filldraw (-0.1,-0.1) rectangle (0.1,0.1);
\end{tikzpicture}
\begin{tikzpicture}
\draw     [decorate,decoration=snake]  (-1,0) -- (1,0);
\filldraw (0,0) circle(3pt);
\node [above][ultra thick] at (0,0.3) {$Z_F$};
\end{tikzpicture}
\begin{tikzpicture}
\hspace{2cm}
\draw     [decorate,decoration=snake]  (-1,0) -- (1,0);
\filldraw (0,0) circle(3pt);
\filldraw (-0.1,-0.1) rectangle (0.1,0.1);
\node [above][ultra thick] at (0,0.3) {$Z_{HF}$};
\end{tikzpicture}
\caption{Cancelation of the $\lambda$ divergence for the  $Q_{HB}$ $Q_{HG}$ and $Q_{HW}$ operators. The effects of $\mathcal{L}_6$ operators
and counter terms are indicated with a black square, while the insertion of the field strength counter term for the Yang-Mills $\mathcal{L}_4$ operator
in the SMEFT is indicated with a
black circle.}
\label{YMcancelation}
\end{figure}
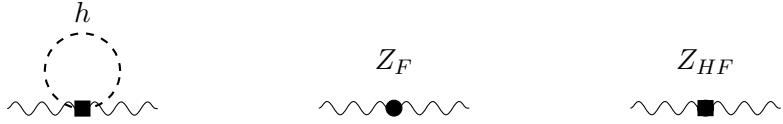
\begin{figure}
\begin{tikzpicture}
\hspace{-1.5cm}
\draw     [decorate,decoration=snake]  (-1,0) -- (1,0);
\filldraw (0,0) circle(3pt);
\filldraw (-0.1,-0.1) rectangle (0.1,0.1);
\node [above][ultra thick] at (0,0.3) {$Z_{HD}$};
\draw (0,-0.6) node [align=center] {(a)};
\end{tikzpicture}
\begin{tikzpicture}
\hspace{-0.9cm}
\draw     [decorate,decoration=snake]  (-1,1.5) -- (1,1.5);
\draw [thick] [dashed] (0,2) circle (0.50);
\node [above][ultra thick] at (0,2.5) {$h$};
\filldraw (-0.1,1.4) rectangle (0.1,1.6);
\draw (0,0.9) node [align=center] {(b)};
\end{tikzpicture}
\begin{tikzpicture}
\hspace{-0.4cm}
\draw     [decorate,decoration=snake]  (-1,1.5) -- (1,1.5);
\draw [thick] [dashed] (0,2) circle (0.50);
\node [above][ultra thick] at (0,2.7) {$h$};
\filldraw (-0.1,2.4) rectangle (0.1,2.6);
\draw (0,0.9) node [align=center] {(c)};
\end{tikzpicture}
\begin{tikzpicture}
\hspace{0.6cm}
\draw     [decorate,decoration=snake]  (-1.5,-0.5) -- (-0.51,-0.5);
\draw     [decorate,decoration=snake]  (0.5,-0.5) -- (1.5,-0.5);
\draw [thick] [dashed] (0,-0.5) circle (0.50);
\node [above][ultra thick] at (0,-0.1) {$\phi_0$};
\node [above][ultra thick] at (0,-0.8) {$h$};
\filldraw (-0.1,-0.9) rectangle (0.1,-1.1);
\draw (0,-1.8) node [align=center] {(d)};
\end{tikzpicture}
\begin{tikzpicture}
\hspace{0.8cm}
\draw     [decorate,decoration=snake]  (-1.5,-0.5) -- (-0.51,-0.5);
\draw     [decorate,decoration=snake]  (0.5,-0.5) -- (1.5,-0.5);
\draw [thick] [dashed] (0,-0.5) circle (0.50);
\node [above][ultra thick] at (0,-0.1) {$\phi_0$};
\node [above][ultra thick] at (0,-1
) {$h$};
\filldraw (-0.6,-0.6) rectangle (-0.4,-0.4);
\draw (0,-1.8) node [align=center] {(e)};
\end{tikzpicture}
\caption{Cancelation of the $\lambda$ divergence for the  $Q_{HD}$ operator for a matrix element with no external Higgs field.
The effects of $\mathcal{L}_6$ operators and counter terms are indicated with a black square. Note diagram (e) has a mirror graph
and corresponds to three different cases of external gauge fields.}
\label{QHDcancelation}
\end{figure}
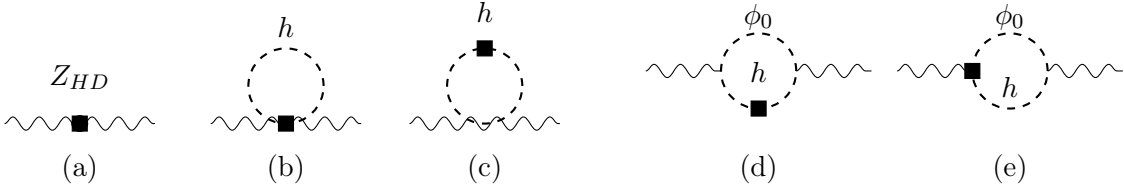
A cancelation of this form also occurs for the operators $Q_{HW}$, $Q_{HB}$.
The $\lambda$ divergence for the operator $Q_{HD}$ cancels through the combination of the operator counter term, and the one loop diagrams
shown in Fig.~\ref{QHDcancelation}. The contributions combine as
\bea
(1+ \frac{6 \, \lambda}{16 \, \pi^2 \, \epsilon}  - \frac{10 \, \lambda}{16 \, \pi^2 \, \epsilon} + \frac{2 \, \lambda}{16 \, \pi^2 \, \epsilon} + \frac{4 \, \lambda}{16 \, \pi^2 \, \epsilon}
- \frac{2 \, \lambda}{16 \, \pi^2 \, \epsilon}) \, \frac{\bar{v}_T^4}{16} \,(\bar{g}_1\, B^\mu - \bar{g}_2 \, W_3^\mu)^2 \, C_{HD}^{(r)},
\eea
where the order of the $1/\epsilon$ terms reflects the order of the sub-diagrams in Fig.~\ref{QHDcancelation}.
Using the guidance of how the divergences cancel in our chosen scheme, we can promote the tree level discussion in Ref.~\cite{Alonso:2013hga}
to a one loop treatment of $m_Z$ in the SMEFT. We choose to redefine the gauge fields and couplings as
\begin{align}
G_\mu^A &= \mathcal{G}_\mu^A \left[1 + C_{HG} \, (\bar{v}_T^2 + \Delta V^2)  \right], & \quad
\bar{g}_3 &= g_3 \left[1 + C_{HG} \, (\bar{v}_T^2 + \Delta V^2) \right], \\
W^I_\mu  &=  \mathcal{W}^I_\mu \left[1 + C_{HW} \, (\bar{v}_T^2 + \Delta V^2)  \right], & \quad
\bar{g}_2 &= g_2 \left[1 + C_{HW} \, (\bar{v}_T^2 + \Delta V^2) \right], \\
B_\mu  &=  \mathcal{B}_\mu \left[1 + C_{HB} \, (\bar{v}_T^2 + \Delta V^2)  \right]. & \quad
\bar{g}_1 &= g_1 \left[1 + C_{HB} \, (\bar{v}_T^2 + \Delta V^2) \right].
\end{align}
This redefinition is performed so that $g_3 G_\mu^A=\bar{g}_3 \mathcal{G}_\mu^A$, etc. are unchanged including $\Delta V$ corrections.
Here $\Delta V$ is defined as
\bea
\Delta V^2= 2 \, \bar{v}_T \, \Delta v   - \hat{m}_h^2 \, \Delta_1, \quad  \quad 16 \, \pi^2 \, \Delta_1 = 1 + \log \left(\frac{\Lambda^2}{\hat{m}_h^2} \right).
\eea
The electroweak terms that remain are now
\begin{align}
{\cal L} &=  - \frac{1}{2} \mathcal{W}^+_{\mu \nu} \, \mathcal{W}_-^{\mu \nu} - \frac{1}{4} \mathcal{W}^3_{\mu \nu} \, \mathcal{W}_3^{\mu \nu} - \frac{1}{4} \mathcal{B}_{\mu \nu} \, \mathcal{B}^{\mu \nu}  - \frac{1}{2} \left((\bar{v}_T^2+ \Delta V^2) \, C^{(r)}_{HWB}   + \Delta_{HWB}^{y_t}\right) \, \bar{g}_1 \, \bar{g}_2  \,  \mathcal{W}_{\mu \nu}^3 \, \mathcal{B}^{\mu \nu} \nonumber \\
&\hspace{0.5cm}+\left(\frac14  \bar{g}_2^2 \, (\bar{v}_T^2+ \Delta V^2)   +  \Delta m_W^2\right)  \, \mathcal{W}_\mu^+ \mathcal{W}^{-\,\mu}
+ \frac18 (\bar{v}_T^2+ \Delta \bar{v}_Z^2) \,  (\bar{g}_2 \, \mathcal{W}^3_\mu - \bar{g}_1 \, \mathcal{B}_\mu)^2, \nn \\
&\hspace{0.5cm}+ \frac1{16} (\bar{v}_T^4 + \Delta \bar{v}_T^4) \, C^{(r)}_{H D} \, (\bar{g}_2 \mathcal{W}^3_\mu - \bar{g}_1 \mathcal{B}_\mu)^2,
\end{align}
with
\bea
\Delta m_W^2 &=&  \frac{N_c \, \bar{g}_2^2}{32 \, \pi^2} \, \hat{m}_t^2 \,  \left(\frac{1}{2}+ \log \left(\frac{\Lambda^2}{\hat{m}_t^2}\right) \right)
+ \frac{\bar{g}_2^2}{64 \, \pi^2} \, \hat{m}_h^2 \, \left(\frac{3}{2}+ \log \left(\frac{\Lambda^2}{\hat{m}_h^2}\right) \right), \\
\Delta \bar{v}_Z^2 &=& \Delta V^2 + \frac{N_c \, g_{A,t}^2}{\pi^2} \, \hat{m}_t^2 + \frac{\hat{m}_h^2}{8 \, \pi^2}  \, \left(\frac{3}{2}+ \log \left(\frac{\Lambda^2}{\hat{m}_h^2}\right) \right),\\
\Delta \bar{v}_T^4 &=& \bar{v}_T^4 \, \sqrt{2} \, \hat{G}_F \, \left(\frac{4}{2^{1/4}} \,  \frac{\Delta v}{\sqrt{\hat{G}_F}} + \frac{\hat{m}_h^2}{32 \, \pi^2}  - 3  \, \hat{m}_h^2 \, \Delta_1 \right), \\
\Delta_{HWB}^{yt} &=& - \frac{\hat{m}_t^2 N_c}{8 \, \pi^2} \, \left[
C^{(r)}_{\substack{uB\\ tt}} + 2  \, (\hyp_q + \hyp_u) \,  C^{(r)}_{\substack{uW \\ tt}} \right] \, \log \left[\frac{\Lambda^2}{\hat{m}_t^2} \right].
\eea
On shell relations are taken in defining the $p^2$ dependence in  the logarithms, leading to the neglect of sub-leading gauge coupling dependence. We have also neglected a class of higher dimensional operator effects in the two point functions due to massive fermion loops which require
a calculation including sub-leading gauge dependence at one loop dependence to specify. The discussion in Ref.~\cite{Alonso:2013hga} is then changed by promoting $\bar{v}_T^2 \rightarrow \bar{v}_T^2+ \Delta V^2$ resulting in
\bea
\hat{m}_Z^2 &=& \frac{\bar{v}_T^2}{4} (\bar{g}_1^2 + \bar{g}_2^2) + \delta m_Z^2 + \Delta m_Z^2,
\eea
where the leading SMEFT corrections are given by \cite{Berthier:2015oma}
\bea\label{deltaMz}
\frac{\delta m_Z^2}{\hat{m}_Z^2} &=&  \frac{1}{2 \, \sqrt{2} \, \hat{G}_F} \, C^{(r)}_{HD} + \frac{4 \, \sqrt{2} \, \pi \, \tilde{\alpha}}{\hat{G}_F} C^{(r)}_{HWB},
\eea
and the one loop corrections are given by
\bea\label{DeltaMz}
\Delta m_Z^2 = \left[\Delta \bar{v}_Z^2 + \frac{\Delta \bar{v}_T^4}{2} \, C_{HD}^{(r)} \right] \, \frac{(\bar{g}_1^2 + \bar{g}_2^2)}{4}
+\frac{\bar{g}_1^2 \, \bar{g}_2^2}{2} \, \bar{v}_T^2  \,\left[( \Delta \bar{v}_Z^2 + \Delta V^2) \, C_{HWB}^{(r)} + \Delta_{HWB}^{y_t}\right].
\eea
Expressed in terms of input parameters
\bea\label{deltaMz}
\frac{\Delta m_Z^2}{\hat{m}_Z^2} &=& \sqrt{2} \, \Delta \bar{v}_Z^2 \,  \hat{G}_F \left[1 - \sqrt{2} \,\delta G_F - \frac{\delta m_Z^2}{\hat{m}_Z^2}\right]
+  C^{(r)}_{HD} \left(\frac{4}{2^{1/4}} \,  \frac{\Delta v}{\sqrt{\hat{G}_F}} + \frac{\hat{m}_h^2}{32 \, \pi^2}  - 3  \, \hat{m}_h^2 \, \Delta_1 \right), \nn \\
&+& 8 \, \pi \tilde{\alpha} \, \left[( \Delta \bar{v}_Z^2 +\Delta V^2) \, C_{HWB}^{(r)} + \Delta_{HWB}^{yt} \right].
\eea
Finally note $\Delta V^2$, $\bar{g}_2^2$ are expressed in terms of input parameters as
\bea
\Delta V^2 = \frac{2^{3/4} \, \Delta v}{\sqrt{\hat{G}_F}} \, \left(1 + \frac{\delta G_F}{\sqrt{2}} \right)  - \hat{m}_h^2 \, \Delta_1,  \quad \quad
\bar{g}_2^2 = \frac{4 \, \pi \, \tilde{\alpha}}{s_{\hat{\theta}}^2} \left[1 + \frac{\delta s_{\theta}^2}{s^2_{\hat{\theta}}} + 4 \, \hat{m}_W^2 \, C_{HWB}^{(r)} \right].
\eea
The leading order SMEFT expression for  $\bar{g}_2^2$ above was reported in Ref.~\cite{Berthier:2015oma} and $s_{\hat{\theta}}$ and $\delta s_{\theta}$ are defined in Section \ref{Deltasin}.
The expression for $\delta G_F$ is given in Section \ref{GFextract}.
We use these results to map a measurement of $\hat{m}_Z$ extracted from the LEP pole scan to the Lagrangian parameters.
In the experimental extraction of $\hat{m}_Z$ radiative corrections due to SM photon emission are present and subtracted, defining this pseudo-observable. Electroweak emissions
are modified in the SMEFT, but we assume that this effect is subdominant to the corrections retained.
These corrections are proportional to the sub-leading gauge coupling coupling dependence and the corresponding derivative operator corrections scale as $\sim p^2/\Lambda^2$ with $p^2$ a scale dominated by soft emissions.

\subsubsection{$\Delta s_\theta^2$}\label{Deltasin}
The rotation angle to take the fields to mass eigenstates in the canonically normalized SMEFT is given by
\bea
s_{\bar{\theta}}^2 = \frac{\bar{g}_1^2}{\bar{g}_1^2+ \bar{g}_2^2} + \frac{\bar{g}_1^2 \, \bar{g}_2^2 \, (\bar{g}_2^2 - \bar{g}_1^2)}{(\bar{g}_1^2+\bar{g}_2^2)^2} \left((\bar{v}_T^2+ \Delta V^2) \, C^{(r)}_{HWB}   + \Delta_{HWB}^{y_t}\right).
\eea
Expressing this mixing angle in terms of the input parameters is required to define the predictions of the observables
of interest. The effective "measured mixing angle" is inferred in terms of the input parameters
\bea\label{sinequation}
s_{\hat{\theta}}^2 = \frac{1}{2} - \frac{1}{2}\sqrt{1 - \frac{4 \, \pi \tilde{\alpha}_{ew}}{\sqrt{2} \, \hat{G}_F \, \hat{m}_Z^2}}.
\eea
The short hand notation used to capture the shifts due to the SMEFT relating $s_{\hat{\theta}}^2$ to $s_{\bar{\theta}}^2$
is introduced as
\bea
s_{\hat{\theta}}^2 = s_{\bar{\theta}}^2 + \delta s_{\theta}^2  + \Delta  s_{\theta}^2.
\eea
The leading order term is known\footnote{Note the normalization
change for $C_{HWB}$ when comparing to Ref~\cite{Berthier:2015oma}.}
to be \cite{Berthier:2015oma}
\bea
\delta s_\theta^2 =  - \frac{s^2_\that \, c^2_\that}{(1 - 2 s^2_\that)} \left[\frac{\delta m_Z^2}{\hat{m}_Z^2} + \sqrt{2} \, \delta G_F
+ 4 \, (1- 2 \, s^2_\that c^2_\that) \, \hat{m}_Z^2 \, C_{HWB} \right].
\eea
The one loop correction is given by
\bea
\Delta s_\theta^2 &=&  - \frac{s^2_\that  c^2_\that}{(1 - 2 s^2_\that)} \left[\sqrt{2} \, \Delta G_F \left(1 - \delta \alpha + \frac{\delta \,m_Z^2}{\hat{m}_Z^2}\right)
-\Delta \alpha + \frac{\Delta m_Z^2}{\hat{m}_Z^2} \left(1 - \delta \alpha + \sqrt{2}  \delta G_F \right)\right], \nn \\
&\,& \hspace{0.05cm} - s^2_\that  c^2_\that
\, 4 \, \sqrt{2} \, (1-2 s^2_\that) \, \hat{m}_Z^2 \, \hat{G}_F \, \left[\left(\Delta V^2 + \frac{\Delta G_F}{\hat{G}_F} \right)\, C^{(r)}_{HWB} + \Delta_{HWB}^{y_t}\right], \nn \\
&\,& \hspace{0.05cm} -  \frac{s^4_\that  c^4_\that}{(1 - 2 s^2_\that)} \, 16 \, \hat{m}_Z^2 \, \left[\frac{\Delta m_Z^2}{\hat{m}_Z^2} + \sqrt{2} \, \Delta G_F  \right] \, C^{(r)}_{HWB}.
\eea

\subsection{$\hat{G}_F$ extractions}\label{GFextract}

\begin{figure}[t!]
\begin{tikzpicture}
[
decoration={
	markings,
	mark=at position 0.65 with {\arrow[scale=1.5]{stealth'}};
}]
\hspace{-1cm}
\draw [postaction=decorate]  (0,1) -- (0.6,1.8) ;
\draw [postaction=decorate] (0.6,0.2) -- (0,1) ;
\draw [postaction=decorate] (-0.4,1.5) -- (-0.35,1.5) ;
\draw [postaction=decorate] (-0.55,0.5) -- (-0.65,0.5) ;
\draw [thick,] (-0.5,1) circle (0.50);
\node [above][ultra thick] at (-0.5,1.5) {$b$};
\node [above][ultra thick] at (-0.5,0) {$t$};
\filldraw (-0.1,0.9) rectangle (0.1,1.1);
\draw     [decorate,decoration=snake]  (-2.4,1) -- (-1,1);
\draw [postaction=decorate] (-2.4,1) -- (-3,1.8) ;
\draw [postaction=decorate] (-3,0.2) -- (-2.4,1) ;
\draw (-1.2,-0.4) node [align=center] {(a)};
\end{tikzpicture}
\begin{tikzpicture}
[
decoration={
	markings,
	mark=at position 0.65 with {\arrow[scale=1.5]{stealth'}};
}]
\draw [postaction=decorate] (0,1) -- (0.6,1.8) ;
\draw [postaction=decorate] (0.6,0.2) -- (0,1);
\draw [postaction=decorate] (-1.85,1.5) -- (-1.75,1.5) ;
\draw [postaction=decorate] (-1.9,0.5) -- (-2,0.5) ;
\draw [thick] (-1.9,1) circle (0.50);
\node [above][ultra thick] at (-1.9,1.5) {$b$};
\node [above][ultra thick] at (-1.9,0) {$t$};
\filldraw (-2.5,0.9) rectangle (-2.3,1.1);
\draw     [decorate,decoration=snake]  (-1.4,1) -- (0,1);
\draw [postaction=decorate] (-2.4,1) -- (-3,1.8) ;
\draw [postaction=decorate] (-3,0.2) -- (-2.4,1) ;
\draw (-1.2,-0.4) node [align=center] {(b)};
\end{tikzpicture}
\begin{tikzpicture}
[
decoration={
	markings,
	mark=at position 0.65 with {\arrow[scale=1.5]{stealth'}};
}]
\hspace{0.5cm}
\draw     [decorate,decoration=snake]  (-1.5,0.4) -- (-0.51,0.4);
\draw     [decorate,decoration=snake]  (0.5,0.4) -- (1.5,0.4);
\draw [thick]  (0,0.4) circle (0.50);
\draw [postaction=decorate] (0.1,0.9) -- (0.15,0.9) ;
\draw [postaction=decorate] (-0.05,-0.1) -- (-0.15,-0.1) ;
\node [above][ultra thick] at (0,0.9) {$b$};
\node [above][ultra thick] at (0,-0.6) {$t$};
\draw (0,-1) node [align=center] {(c)};
\end{tikzpicture}
\begin{tikzpicture}
[cross/.style={path picture={
  \draw[black]
(path picture bounding box.south east) -- (path picture bounding box.north west) (path picture bounding box.south west) -- (path picture bounding box.north east);
}}]
\hspace{1.25cm}
\node [draw,circle,cross,minimum width=0.2 cm] at (0,0.5){};
\draw     [decorate,decoration=snake]  (-1,0.5) -- (-0.2,0.5);
\draw     [decorate,decoration=snake]  (0.2,0.5) -- (1,0.5);
\node [above][ultra thick] at (0,-0.4) {$\Delta v_T^2$};
\draw (0,-1) node [align=center] {(d)};
\end{tikzpicture}
\caption{One loop corrections in the large $y_t$ limit to the extraction of $\hat{G}_F$.
Diagrams a),b)  show the one loop contributions of Class 8 operators.
Diagram c) corresponds to the one loop correction to the $W$ mass in the SM in the large $y_t$ limit.
Diagram d) corresponds to the one loop correction to the vev.}
\label{GFfigs}
\end{figure}
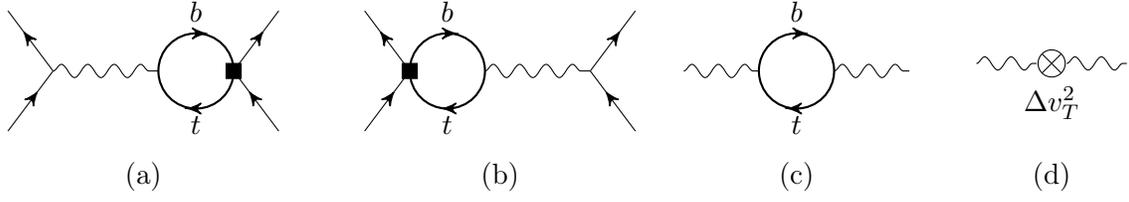

$G_F$ is extracted from the muon lifetime, which is dominated by the decay, $\mu^- \rightarrow e^- + \bar{\nu}_e + \nu_\mu$.
The local effective interaction for muon decay is defined as the $p^2 \ll \hat{m}_W^2$ limit of the SM calculation, so that the Effective Lagrangian used is
\begin{align}
\mathcal{L}_{G_F} \equiv  -\frac{4\mathcal{G}_F}{\sqrt{2}} \, \left(\bar{\nu}_\mu \, \gamma^\mu P_L \mu \right) \left(\bar{e} \, \gamma_\mu P_L \nu_e\right).
\end{align}
In the SMEFT, at LO the matching result \cite{Alonso:2013hga} onto this Lagrangian is
\begin{align}
-\frac{4\mathcal{G}_F}{\sqrt{2}} &=  -\frac{2}{\bar{v}_T^2} - 4 \, \hat{G}_F \, \delta G_F.
\label{gfermi}
\end{align}
where the leading order shift result is \cite{Alonso:2013hga,Berthier:2015oma}
\bea
\delta G_F &=& -\frac{1}{4 \, \hat{G}_F} \, \left(C_{\substack{ll \\ \mu ee \mu}} +  C_{\substack{ll \\ e \mu\mu e}}\right) + \frac{1}{2 \, \hat{G}_F} \left(C^{(3)}_{\substack{Hl \\ ee }} +  C^{(3)}_{\substack{Hl \\ \mu\mu }}\right).
\eea
The one loop correction to this result in the large $\lambda$ and  $y_t$ limit is straight forward to determine in the modified $\rm \overline{MS}$ scheme.
The one loop corrections to $\mathcal{L}_{G_F}$ in the matching result vanish when retaining only $\lambda,y_t$ corrections.
One loop corrections due to four fermion operators are shown in Fig.~\ref{GFfigs} a) and b). The finite terms for these contributions were reported in
Ref.~\cite{Gauld:2015lmb}. The two point function of the $W$ boson
is corrected as in Fig.~\ref{GFfigs} c).  In addition the diagrams in Fig.~\ref{WSMrenorm1} b) and c) contribute to the correction to the $W$ mass.
The vev  present in the $W$ mass is corrected as $\bar{v}_T^2 \rightarrow \bar{v}_T^2 + \Delta V^2$ as illustrated in
Fig.~\ref{GFfigs} d). The sum of these contributions lead to the result including one loop $\lambda$ and $y_t$ corrections
\begin{align}
-\frac{4\mathcal{G}_F}{\sqrt{2}} &=  -\frac{2}{\bar{v}_T^2} \left(1 - \frac{\Delta V^2}{\bar{v}_T^2} - \frac{\Delta m_W^2}{\bar{m}_W^2} \right) - 4 \, \hat{G}_F \, \delta G_F - \Delta \psi^4.
\label{gfermiNLO}
\end{align}
with a normalization change compared to Ref.~\cite{Berthier:2015oma} on the coefficient of $C^{(r)}_{HWB}$.
The finite terms reported in Ref.~\cite{Gauld:2015lmb} combined with the results derived from Ref.~\cite{Jenkins:2013wua} give
\bea
\Delta \psi^4 = \frac{N_c \, y_t^2}{16 \, \pi^2} \left[- \left(C^{(3)}_{\substack{lq \\ eett}} +  C^{(3)}_{\substack{lq \\ \mu \mu tt}}\right)
+ 2 \left(C^{(3)}_{\substack{lq \\ eett}} +  C^{(3)}_{\substack{lq \\ \mu \mu tt}} - C^{(3)}_{\substack{Hl \\ ee }} - C^{(3)}_{\substack{Hl \\ \mu\mu }} \right)\log \left(\frac{\Lambda^2}{\hat{m}_t^2}\right) \right],
\eea
where the operators are renormalized and defined at the matching scale $\sim \Lambda$ to infer the limits on the high energy theory matched onto the SMEFT.
We define a short hand notation to capture the one loop correction to $\mathcal{G}_F$ in reporting results. The one loop correction $\Delta G_F$ is introduced with a normalization
\bea
\hat{G}_F = \frac{1}{\sqrt{2} \, \bar{v}_T^2} + \sqrt{2} \, \hat{G}_F \, \delta G_F + \sqrt{2} \, \hat{G}_F \, \Delta G_F,
\eea
where
\bea
\Delta G_F =  -  \hat{G}_F \, \Delta V^2 \, (1- 2 \, \sqrt{2} \, \delta G_F) - \frac{\Delta m_W^2}{\sqrt{2} \, \hat{m}_W^2} + \frac{\Delta \psi^4}{4 \, \hat{G}_F} - \frac{\Delta m_W^2}{\sqrt{2} \, \hat{m}_W^2} \frac{\delta m_W^2}{\hat{m}_W^2}.
\eea
We use the definition of $\hat{m}_W^2$ in Ref.~\cite{Berthier:2015oma} so that at tree level $\hat{m}_W^2 = c_{\hat{\theta}}^2 \,  \hat{m}_Z^2$
and the leading order SMEFT corrections result is
\bea
\delta m_W^2= -\hat{m}_W^2 \left(\frac{\delta s_{\that}^2}{s_{\that}^2}+ 4 \, \hat{m}_W^2 \, C^{(r)}_{HWB} + \sqrt{2} \, \delta G_F\right),
\eea
\section{External state wavefunction finite terms to one loop}
In addition to $ \Delta \mathcal{A}_{SMEFT}$ one loop finite terms also
result from the renormalization conditions adopted. As we use a modified $\rm \overline{MS}$ scheme, these finite terms
include one loop corrections to the vev ($\Delta v$), the Weinberg angle to rotate the fields to the mass eigenstate form (in $\Delta \bar{s}_\theta^2$) and the shift in the pole mass Lagrangian term $\Delta \bar{m}_Z$. These finite terms are specified in the previous sections.

It remains to define the one loop terms related to fixing the position and residue of the pole for the external states in the LSZ formula \cite{Lehmann:1954rq}.
To distinguish these finite terms from those appearing in renormalization factors when an on-shell subtraction scheme is used
these terms are referred to as R factors in Refs.~\cite{Chiu:2009mg,Hartmann:2015oia,Hartmann:2015aia},
a notation we adopt here. For $Z \rightarrow \bar{\psi} \, \psi$ decay these $R$ factors appear in the corresponding $S$ matrix element as
\bea
\langle Z |S | \bar{\psi_i} \, \psi_i \rangle = (1 + \frac{\Delta R_Z}{2}) (1 + \Delta R_{\psi_i}) \, i  \, \mathcal{A}_{Z\bar{\psi_i}\psi_i}.
\eea

\subsection{Finite terms for $R_Z$}
The two point function of the $Z$ boson is decomposed as
\bea
i \, \bar{\Gamma}_{\mu \, \nu}^Z(q) = - i \, g_{\mu \, \nu} (q^2 - \bar{m}_Z^2) - i \left( g_{\mu \, \nu} - \frac{q_{\mu} \, q_{\nu}}{q^2}\right)\, \Sigma_T^Z(q^2) - i \frac{q_{\mu} \, q_{\nu}}{q^2} \, \Sigma_L^Z(q^2).
\eea
This leads to $\Delta R_Z$ which is defined as
\bea
\Delta R_Z = - \left(\tilde{\rm Re} \frac{\partial  \, \Sigma_T^Z(q^2)}{\partial q^2} \right)_{q^2 = \bar{m}_{Z}^2}.
\eea
In the limit of retaining $y_t^2$ and $\lambda$ contributions in the vanishing gauge coupling limit $\Delta R_Z = 0$.
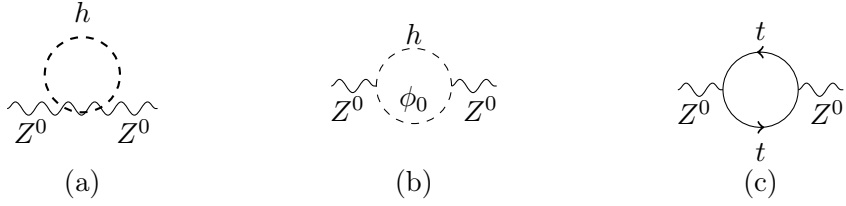
\begin{figure}
\begin{tikzpicture}
\hspace{-2cm}

\draw  [decorate,decoration=snake] (-1,0) -- (1,0);

\draw [thick] [dashed] (0,0.45) circle (0.50);

\draw (0,-1) node [align=center] {(a)};

\node [above][ultra thick] at (0,1) {$h$};
\node [left][ultra thick] at (-0.3,-0.3) {$Z^0$};
\node [right][ultra thick] at (0.3,-0.3) {$Z^0$};

\end{tikzpicture}
\begin{tikzpicture}
\draw [decorate,decoration=snake](-1.1,-0.3) -- (-0.5,-0.3);
\draw [decorate,decoration=snake](0.5,-0.3) -- (1.1,-0.3);

\draw  (0,-0.3) [dashed] circle (0.50);

\draw (0,-1.6) node [align=center] {(b)};

\node [above][ultra thick] at (0,+0.1) {$h$};
\node [above][ultra thick] at (0,-0.8) {$\phi_0$};
\node [left][ultra thick] at (-0.5,-0.6) {$Z^0$};
\node [right][ultra thick] at (0.5,-0.6) {$Z^0$};
\end{tikzpicture}
\begin{tikzpicture}
\hspace{2cm}
\draw [decorate,decoration=snake](-1.1,0) -- (-0.5,0);
\draw [decorate,decoration=snake](0.5,0) -- (1.1,0);

\draw  [->][thick]  (90:0.5)  -- + (0:-0.05) ;
\draw  [->][thick]  (270:0.5)  -- + (0:0.05) ;

\draw  (0,0) circle (0.50);

\draw (0,-1.25) node [align=center] {(c)};

\node [above][ultra thick] at (0,0.5) {$t$};
\node [above][ultra thick] at (0,-1.1) {$t$};
\node [left][ultra thick] at (-0.5,-0.3) {$Z^0$};
\node [right][ultra thick] at (0.5,-0.3) {$Z^0$};

\end{tikzpicture}

\caption{Diagrams determining $\delta R_Z$ using on-shell renormalization conditions.}
\label{WSMrenorm}
\end{figure}

\subsection{Finite terms for $R_{\psi_i}$}
Following Refs.~\cite{Denner:1994xt,Denner:1991kt,Gauld:2015lmb,Gauld:2016kuu} the two point function for the fermion fields is defined as
\bea
\bar{\Gamma}_{\psi_i}(p) = i ( \slashed{p} - m_{\psi_i}) + i \, \Sigma_{\psi_i}(\slashed{p}),
\eea
and in addition
\begin{align}
\Sigma_{\psi_i}(\slashed{p}) & = \Big(\slashed{p} \left(P_L \Sigma_{\psi_i}^L(p^2) + P_R \Sigma_{\psi_i}^R(p^2)\right) + m_{\psi_i} \, \left( \Sigma_{\psi_i}^S(p^2) P_L + \Sigma_{\psi_i}^{S*} (p^2) P_R \right) \Big),
\end{align}
and the R factors for the left-handed and right-handed fermion field fields respectively are
\begin{align}
\Delta R_{\psi_i}^{L} & = - \tilde{\rm Re} \, \Sigma_{\psi_i}^{L} (m_{\psi_i}^2) + \Sigma_{\psi_i}^S(m_{\psi_i}^2) - \Sigma_{\psi_i}^{S*}(m_{\psi_i}^2)
\nn \\
& - m_{\psi_i}^2 \frac{\partial}{\partial p^2} \tilde{\rm Re} \Big( \Sigma_{\psi_i}^L(p^2) + \Sigma_{\psi_i}^R(p^2) + \Sigma_{\psi_i}^S(p^2) + \Sigma_{\psi_i}^{S*}(p^2) \Big) |_{p^2 = m_{\psi_i}^2},
\nn \\
\Delta R_{\psi_i}^{R} & = - \tilde{\rm Re} \, \Sigma_{\psi_i}^{R} (m_{\psi_i}^2)
\nn \\
& - m_{\psi_i}^2 \frac{\partial}{\partial p^2} \tilde{\rm Re} \Big( \Sigma_{\psi_i}^L(p^2) + \Sigma_{\psi_i}^R(p^2) + \Sigma_{\psi_i}^S(p^2) + \Sigma_{\psi_i}^{S*}(p^2) \Big) |_{p^2 = m_{\psi_i}^2}.
 \end{align}
To find the fermion $R$ factors for large $y_t$ corrections we calculate the finite part of Fig.~\ref{wavefunction} for external $b$ quarks, giving
\begin{align}\label{fermionfiniteR}
\Delta R_b^L & = - \tilde{\rm Re} \, \Sigma_b^L(m_b^2), \\
&=  \frac{\hat{m}_t^2}{16 \, \pi^2} \, \left(\sqrt{2} \hat{G}_F(1- \sqrt{2} \delta G_F) +  C_{\substack{uH\\ 33}}^\star - 2 \, C_{Hq}^{(3)}\right) \left[-\frac{3}{4} - \frac{1}{2} \log \left[\frac{\Lambda^2}{\hat{m}_t^2}\right] \right].\nn
\end{align}
All other fermion $R$ factors relevant for $Z$ decay vanish in this limit.
\section{One loop results for the observables $\bar{\Gamma}_Z, \, \bar{\Gamma}_{Z \rightarrow \bar{\psi}_i \, \psi}, \, \bar{\Gamma}_{Z \rightarrow Had}, \, \bar{R}_\ell^0, \, \bar{R}_b^0$}\label{fullresults}
We report partial results for $\bar{\Gamma}_{Z \rightarrow \bar{\psi}_i \, \psi_i}$ and the ratios $\bar{R}_\ell^0$, $\bar{R}_b^0$ in this section. It is possible to report forward-backward or other asymmetries in this limit as well, however, the neglect of gauge corrections in that case, where IR radiation and box diagram effects arising from gauge coupling corrections could significantly affect those quantities is problematic.\footnote{See Ref.\cite{Elvang:2016qvq} for recent related discussion.}
Note that $\bar{\Gamma}_{Z \rightarrow Had} = 2 \, \bar{\Gamma}_{Z \rightarrow \bar{u} \, u} + 2 \, \bar{\Gamma}_{Z \rightarrow \bar{d} \, d} + \bar{\Gamma}_{Z \rightarrow \bar{b} \, b}$, $\bar{R}_{\ell}^0 = \bar{\Gamma}_{Z \rightarrow Had}/\bar{\Gamma}_{Z \rightarrow \bar{\ell} \, \ell}$, and $\bar{R}_{b}^0 = \bar{\Gamma}_{Z \rightarrow \bar{b} \, b}/\bar{\Gamma}_{Z \rightarrow Had}$.
Results on the tree level redefinition of $\bar{\Gamma}_Z$ in the SMEFT that build up to these results, were reported in Refs.~\cite{Alonso:2013hga,Berthier:2015oma,Berthier:2015gja,Berthier:2016tkq}.
In Ref.~\cite{Berthier:2015oma} $\delta$ shifts to effective vector and axial couplings
are reported for the $Z$, for all final state fermion pairs. Working with axial and vector couplings is advantageous due to the important impact of the
accidental numerical suppression of the leptonic vectorial coupling (in the SM) in studies of this form \cite{Altarelli:116932,Berthier:2016tkq}.
Conversely, calculating the interference with SMEFT corrections in the massless final state limit can be efficiently studied using effective chiral couplings and fields. To fix our notation  when using such chiral couplings and fields we note
\bea
\mathcal{L}_{Z,eff}  =  g_{Z,eff}  \,   \left(J_\mu^{Z \ell} Z^\mu + J_\mu^{Z \nu} Z^\mu + J_\mu^{Z u} Z^\mu +  J_\mu^{Z d} Z^\mu \right),
\eea
where $g_{Z,eff} = - \, 2 \, 2^{1/4} \, \sqrt{\hat{G}_F} \, \hat{m}_Z$ and $(J_\mu^{Z \psi})^{pr} = \bar{\psi}_p \, \gamma_\mu \left[(\bar{g}^{\psi}_L)_{eff}^{pr} \, P_L +  (\bar{g}^{\psi}_R)_{eff}^{pr} \, P_R \right] \psi_r$ for $\psi = \{u,d,\ell,\nu \}$. In a minimal linear MFV scenario $(J_\mu^{Z \psi})_{pr} \simeq (J_\mu^{Z \psi}) \delta_{pr}$ when quark mass dependence in flavour changing interactions is neglected. The coupling shifts are defined as
\bea
\delta (g^{\psi}_{L,R})_{pr} = (\bar{g}^{\psi}_{L,R})^{eff}_{pr} - (g^{\psi}_{L,R})^{SM}_{pr},
\eea
and
\bea
\delta (g^{\psi}_{L})_{pr} = \delta (g^{\psi}_{V})_{pr}+ \delta (g^{\psi}_{A})_{pr}, \quad \delta (g^{\psi}_{R})_{pr} = \delta (g^{\psi}_{V})_{pr} - \delta (g^{\psi}_{A})_{pr}.
\eea
Our normalization convention is $(g^{\psi}_{V})^{SM} = T_3/2 - \mathcal{Q}^\psi \, \bar{s}_\theta^2$ and $(g^{\psi}_{A})^{SM} = T_3/2$ where $T_3 = 1/2$ for $u_i,\nu_i$ and $T_3 = -1/2$ for $d_i,\ell_i$
and $\mathcal{Q}^\psi = \{-1,2/3,-1/3 \}$ for $\psi = \{\ell,u,d\}$. The leading order shift results for $\delta (g^{\psi}_{V})_{pr}$, $\delta (g^{\psi}_{A})_{pr}$ are reported in Ref.~\cite{Berthier:2015oma}.\footnote{Again we stress the different normalization convention for the operator $Q_{HWB}$ when comparing to past results.}
With this convention
\bea
\bar{\Gamma}_{Z \rightarrow \bar{\psi}^{p} \, \psi^r} = \frac{\sqrt{2} \, \hat{G}_F \, \hat{m}_Z^3 \, N_c}{6 \, \pi} \, \left(\left|(\bar{g}^\psi_L)_{pr}^{eff}\right|^2 + \left|(\bar{g}^\psi_R)_{pr}^{eff}\right|^2 \right),
\eea
and $ \bar{\Gamma}_Z = \sum_\psi \, \sum_{p,r} \, \bar{\Gamma}_{Z \rightarrow \bar{\psi}^p \, \psi^r}$.

\subsection{One loop corrections in the SMEFT}\label{oneloopsection}
\subsubsection{Charged Lepton effective couplings}
For charged lepton final states the leading order (flavour symmetric) SMEFT effective coupling shifts are \cite{Berthier:2015oma}
\bea\label{leptonleadingshift}
\delta (g^{\ell}_L)_{ss}&=&\delta \bar{g}_Z \, (g^{\ell}_{L})^{SM}_{ss} - \frac{1}{2 \sqrt{2} \hat{G}_F} \left(C_{\substack{H \ell \\ ss}}^{(1)} + C_{\substack{H \ell \\ ss}}^{(3)} \right) - \delta s_\theta^2, \\
\delta(g^{\ell}_R)_{ss}&=&\delta \bar{g}_Z \, (g^{\ell}_{R})^{SM}_{ss}  - \frac{1}{2 \, \sqrt{2} \, \hat{G}_F}
C_{\substack{H e \\ss}} - \delta s_\theta^2,
\eea
where
\bea
\delta \bar{g}_Z =- \frac{\delta G_F}{\sqrt{2}} - \frac{\delta m_Z^2}{2\hat{m}_Z^2} + s^2_{\hat{\theta}} \, c^2_{\hat{\theta}} \, 4 \, \hat{m}_Z^2 \,  C_{HWB},
\eea
while the one loop corrections are
\bea\label{leptonloopshift}
\Delta (g^{\ell}_L)_{ss}&=&\Delta \bar{g}_Z \, (g^{\ell}_{L})^{SM}_{ss} + \frac{N_c \, \hat{m}_t^2}{16 \, \pi^2} \,  \log \left[\frac{\Lambda^2}{\hat{m}_t^2} \right] \, \left[C_{\substack{\ell q\\ ss33}}^{(1)} - C_{\substack{\ell q\\ ss33}}^{(3)} - C_{\substack{\ell u\\ ss33}} \right] - \Delta s_\theta^2,  \\&-&\frac{1}{2}\left(\frac{\Delta G_F}{\hat{G}_F} + \Delta V^2 \right)\left(C_{\substack{H \ell \\ ss}}^{(1)} + C_{\substack{H \ell \\ ss}}^{(3)} \right),\nn
\\
\Delta(g^{\ell}_R)_{ss}&=&\Delta \bar{g}_Z \, (g^{\ell}_{R})^{SM}_{ss} + \frac{N_c \, \hat{m}_t^2}{16 \, \pi^2} \,  \log \left[\frac{\Lambda^2}{\hat{m}_t^2} \right] \,
\left[-C_{\substack{eu\\ ss33}}+ C_{\substack{q e \\ 33ss}} \right]  - \Delta s_\theta^2,\\ &-&\frac{1}{2}\left(\frac{\Delta G_F}{\hat{G}_F} + \Delta V^2 \right)C_{\substack{H e \\ss}},\nn
\eea
while
\bea
\Delta \bar{g}_Z &=&- \frac{\Delta G_F}{\sqrt{2}} - \frac{\Delta m_Z^2}{2\hat{m}_Z^2} +\frac{\delta G_F\Delta m_Z^2}{2\sqrt{2}\hat{m}_Z^2}+\frac{\delta m_Z^2\Delta G_F}{2\sqrt{2}\hat{m}_Z^2} +4\hat{m}_Z^2s^2_{\hat{\theta}}c^2_{\hat{\theta}}C_{HWB}\left[\frac{\Delta m_Z^2}{2\hat{m}_Z^2}+\frac{\Delta G_F}{\sqrt{2}}\right], \nn\\
&\,&\hspace{-0.5cm} + \left[\left(\frac{\Delta G_F}{\hat{G}_F} + \Delta V^2 \right) C_{HWB}+ \Delta^{yt}_{HWB} \right]  4 \, \sqrt{2} \, s^2_{\hat{\theta}} \, c^2_{\hat{\theta}} \, \hat{m}_Z^2 \, \hat{G}_F.
\eea
\subsubsection{Neutrino  effective couplings}
For neutrino final states the leading order SMEFT effective coupling shifts are \cite{Berthier:2015oma}
\bea\label{neutrinoleadingshift}
\delta (g^{\nu}_L)_{tt}&=&\delta \bar{g}_Z \, (g^{\nu}_{L})^{SM}_{tt} - \frac{1}{2 \, \sqrt{2} \, \hat{G}_F} \left( C_{\substack{H \ell \\ tt}}^{(1)} - C_{\substack{H \ell \\ tt}}^{(3)} \right),
\eea
while the one loop corrections are
\bea\label{leptonloopshift}
\Delta (g^{\nu}_L)_{tt}&=&\Delta \bar{g}_Z \, (g^{\nu}_{L})^{SM}_{tt} + \frac{N_c \, \hat{m}_t^2}{16 \, \pi^2} \,  \log \left[\frac{\Lambda^2}{\hat{m}_t^2} \right] \, \left[C_{\substack{\ell q\\ tt33}}^{(1)} + C_{\substack{\ell q\\ tt33}}^{(3)} - C_{\substack{\ell u\\ tt33}} \right],\\&-&\frac{1}{2}\left(\frac{\Delta G_F}{\hat{G}_F} + \Delta V^2 \right)\left( C_{\substack{H \ell \\ tt}}^{(1)} - C_{\substack{H \ell \\ tt}}^{(3)} \right)\nn.
\eea
\subsubsection{Up quark effective couplings}
For up quark final states the leading order SMEFT effective coupling shifts are \cite{Berthier:2015oma}
\bea\label{upleadingshift}
\delta (g^{u}_L)_{pp}&=&\delta \bar{g}_Z \, (g^{u}_{L})^{SM}_{pp}  + \frac{2}{3} \delta s_\theta^2   +
\frac{1}{2 \, \sqrt{2} \, \hat{G}_F} \left(- C_{\substack{H q \\ pp}}^{(1)} + \, C_{\substack{H q \\ pp}}^{(3)} \right),\\
\delta(g^{u}_R)_{pp}&=&\delta \bar{g}_Z \, (g^{u}_{R})^{SM}_{pp} + \frac{2}{3} \delta s_\theta^2
- \frac{1}{2 \, \sqrt{2} \, \hat{G}_F} C_{\substack{H u \\ pp}},
\eea
while the one loop corrections are
\bea\label{uploopshift}
\Delta (g^{u}_L)_{pp}&=&\Delta \bar{g}_Z \, (g^{u}_{L})^{SM}_{pp} + \frac{2}{3} \Delta s_\theta^2
+ \frac{1}{2}\left(\frac{\Delta G_F}{\hat{G}_F} + \Delta V^2 \right)\left(- C_{\substack{H q \\ pp}}^{(1)} + \, C_{\substack{H q \\ pp}}^{(3)} \right),\\&+& \frac{N_c \, \hat{m}_t^2}{16 \, \pi^2} \,  \log \left[\frac{\Lambda^2}{\hat{m}_t^2} \right] \, \left[C_{\substack{qq\\ 33pp}}^{(1)} + C_{\substack{qq\\ pp33}}^{(1)} + C_{\substack{qq\\ 33pp}}^{(3)} + C_{\substack{qq\\ pp33}}^{(3)} -C_{\substack{qu\\ pp33}}^{(1)} \right],\nn\\
\Delta(g^{u}_R)_{pp}&=&\Delta \bar{g}_Z \, (g^{u}_{R})^{SM}_{pp}  + \frac{2}{3} \Delta s_\theta^2
+ \frac{N_c \, \hat{m}_t^2}{16 \, \pi^2} \,  \log \left[\frac{\Lambda^2}{\hat{m}_t^2} \right] \, \left[C_{\substack{qu\\ 33pp}}^{(1)} - C_{\substack{uu\\ 33pp}} - C_{\substack{uu\\ pp33}}\right]\\&-&\frac{1}{2}\left(\frac{\Delta G_F}{\hat{G}_F} + \Delta V^2 \right)C_{\substack{H u \\ pp}}.\nn
\eea
\subsubsection{Down quark effective couplings}
For down quark final states, the leading order SMEFT effective coupling shifts are \cite{Berthier:2015oma}
\bea\label{downleadingshift}
\delta (g^{d}_L)_{rr}&=&\delta \bar{g}_Z \, (g^{d}_{L})^{SM}_{rr}  -  \frac{1}{3} \delta s_\theta^2
- \frac{1}{2 \, \sqrt{2} \, \hat{G}_F} \left(C_{\substack{H q \\ rr}}^{(1)}  +  \, C_{\substack{H q \\ rr}}^{(3)} \right), \\
\delta(g^{d}_R)_{rr}&=&\delta \bar{g}_Z \, (g^{d}_{R})^{SM}_{rr} -  \frac{1}{3} \delta s_\theta^2 \,
- \,  \frac{1}{2 \, \sqrt{2} \, \hat{G}_F} C_{\substack{H d \\ rr}},
\eea
while the one loop corrections are
\bea\label{downloopshift}
\Delta (g^{d}_L)_{rr}&=&\Delta \bar{g}_Z (g^{d}_{L})^{SM}_{rr}  +
\frac{N_c \hat{m}_t^2}{16 \pi^2}  \log \left[\frac{\Lambda^2}{\hat{m}_t^2} \right]  \left[C_{\substack{qq\\ 33rr}}^{(1)} + C_{\substack{qq\\ rr33}}^{(1)} - C_{\substack{qq\\ 33rr}}^{(3)}
- C_{\substack{qq\\ rr33}}^{(3)} - C_{\substack{qu\\ rr33}}^{(1)}\right], \nn \\&-&\frac{1}{2}\left(\frac{\Delta G_F}{\hat{G}_F} + \Delta V^2 \right)\left(C_{\substack{H q \\ rr}}^{(1)}  +  \, C_{\substack{H q \\ rr}}^{(3)} \right)
+ \delta_{br} \frac{\hat{m}_t^2}{4 \, \pi^2} \, \left[C_{\substack{qq\\ 3333}}^{(3)} \left(- 1 +  \log \left[\frac{\Lambda^2}{\hat{m}_t^2} \right] \right) \right] -  \frac{1}{3} \Delta s_\theta^2, \nn\\
&-& \delta_{br} \, \frac{\hat{m}_t^2}{16 \, \pi^2}  \, \left[\left(\frac{1}{4} - \frac{1}{2} \log \left[\frac{\Lambda^2}{\hat{m}_t^2} \right]\right) C_{Hu} + C_{Hq}^{(1)} \right]
- \delta_{br} \, \Delta R_b^L \, \left((g^{d}_{L})^{SM}_{rr} + \delta (g^{d}_L)_{rr}\right), \nn \\
&-&  \delta_{br} \, \frac{\hat{m}_t^2}{16 \, \pi^2}  \, C_{Hq}^{(3)}  \, \left[\frac{1}{2} -  \mathcal{Q}_b \, s_{\hat{\theta}}^2 + (3 - 2 \, \mathcal{Q}_b \, s_{\hat{\theta}}^2 ) \log \left[\frac{\Lambda^2}{\hat{m}_t^2} \right]\right],  \\
&-&  \delta_{br} \, \frac{\hat{m}_t^2}{4 \, \pi} \, \tilde{\alpha} \, (c_{\hat{\theta}}^2 - s_{\hat{\theta}}^2) \, C_{HWB} \, (\mathcal{Q}_u-1)\left[\frac{3}{2} + \log \left[\frac{\Lambda^2}{\hat{m}_t^2} \right]\right], \nn \\
\Delta(g^{d}_R)_{rr}&=&\Delta \bar{g}_Z \, (g^{d}_{R})^{SM}_{rr}  -  \frac{1}{3} \Delta s_\theta^2 +
\frac{N_c \, \hat{m}_t^2}{16 \, \pi^2} \,  \log \left[\frac{\Lambda^2}{\hat{m}_t^2} \right] \, \left[C_{\substack{qd\\ 33rr}}^{(1)}  - C_{\substack{ud\\ 33rr}}^{(1)}  \right],\\&-&\frac{1}{2}\left(\frac{\Delta G_F}{\hat{G}_F} + \Delta V^2 \right)C_{\substack{H d \\ rr}}\nn.
\eea
It is interesting to note that the left handed bottom quark coupling in the SMEFT is perturbed by a number of effects
that are not present in the leptonic and up quark couplings. This observation becomes
even more interesting when considering the (statistically insignificant) indication that the left handed
bottom quark coupling has a greater than $\sim 2 \sigma$ preference for a non-SM value in global analyses (at tree level)
in the SMEFT \cite{Berthier:2015oma,Berthier:2015gja,Berthier:2016tkq,deBlas:2016ojx,Falkowski:2014tna}. Although we caution such tree level analyses
are subject to significant theoretical uncertainties in the SMEFT \cite{Berthier:2015oma,Berthier:2015gja,Berthier:2016tkq},
performing the one loop calculation reported in this paper provides a more solid theoretical framework that makes
such a tentative indication of a deviation even more intriguing.

\section{Phenomenology and Numerics for $\bar{\Gamma}_Z, \, \bar{\Gamma}_{Z \rightarrow \bar{\psi}_i \, \psi}, \, \bar{\Gamma}_{Z \rightarrow Had}, \, \bar{R}_\ell^0, \, \bar{R}_b^0$}
\label{phenosection}
In the SMEFT, at tree level, one has  \cite{Berthier:2015oma}
\bea
 \bar{\Gamma} \left(Z \rightarrow \psi \bar{\psi} \right) &=& \frac{ \, \sqrt{2} \, \hat{G}_F \hat{m}_Z^3 \, N_c}{6 \pi} \left( |\bar{g}^{\psi}_L|^2 + |\bar{g}^{\psi}_R|^2 \right), \\
 \bar{\Gamma} \left(Z \rightarrow {\rm Had} \right) &=& 2 \, \bar{\Gamma} \left(Z \rightarrow u \bar{u} \right)+ 2  \, \bar{\Gamma} \left(Z \rightarrow d \bar{d} \right)
+ \bar{\Gamma} \left(Z \rightarrow b \bar{b} \right).
 \eea
The modification of the decay widths in the SMEFT compared to the SM at leading order in the power counting (and tree level) is given as:
\bea
\delta \bar{\Gamma}_{Z \rightarrow \ell \bar{\ell}}&=& \frac{\sqrt{2} \, \hat{G}_F \hat{m}_Z^3}{6 \pi} \, \left[2 \, g^{\ell}_R \, \delta g^{\ell}_R + 2 \, g^{\ell}_L \, \delta g^{\ell}_L \right] + \delta\bar{\Gamma}_{Z \rightarrow \bar{\ell} \, \ell, \psi^4}, \\
\delta \bar{\Gamma}_{Z \rightarrow \nu \bar{\nu}}&=& \frac{\sqrt{2} \, \hat{G}_F \hat{m}_Z^3}{6 \pi} \, \left[ 2 \, g^{\nu}_L \, \delta g^{\nu}_L \right] + \delta \bar{\Gamma}_{Z \rightarrow \nu \bar{\nu},\psi^4},
\eea
\bea
\delta \bar{\Gamma}_{Z \rightarrow {\rm Had}}&=& 2 \, \delta \bar{\Gamma}_{Z \rightarrow \bar{u} u} + 2 \, \delta \bar{\Gamma}_{Z \rightarrow \bar{d} d}+\delta \bar{\Gamma}_{Z \rightarrow \bar{b} b}, \\ &=& \frac{3 \, \sqrt{2} \, \hat{G}_F \hat{m}_Z^3}{6 \, \pi} \left[4 \, g^{u}_R \, \delta g^{u}_R + 4 \, g^{u}_L \, \delta g^{u}_L + 4 \, g^{d}_R \, \delta g^{d}_R + 4 \, g^{d}_L \, \delta g^{d}_L
+ 2 \, g^{b}_R \, \delta g^{b}_R + 2 \, g^{b}_L \, \delta g^{b}_L\right], \nonumber \\ &+& \delta \bar{\Gamma}_{Z \rightarrow {\rm Had}, \psi^4},  \\
\delta \bar{\Gamma}_{Z} &=& 3 \, \delta \bar{\Gamma}_{Z \rightarrow \ell \bar{\ell}} + 3 \, \delta \bar{\Gamma}_{Z \rightarrow \nu \bar{\nu}} + \delta \bar{\Gamma}_{\rm Had}.
\eea
The shift in the ratios of decay rates follows as $R^0_{\ell}=\frac{\bar{\Gamma}_{\rm Had} + \delta \bar{\Gamma}_{\rm Had}}{\bar{\Gamma}_{Z \rightarrow \bar{\ell} \ell} + \delta \bar{\Gamma}_{Z \rightarrow \bar{\ell} \ell}}$  and  $R^0_{b}=\frac{\bar{\Gamma}_{Z \rightarrow \bar{b} b} + \delta \bar{\Gamma}_{Z \rightarrow \bar{b} b}}{\bar{\Gamma}_{\rm Had} + \delta \bar{\Gamma}_{\rm Had}}$. These expressions directly define $\bar{R}^0_{\psi}= R^0_{\psi} + \delta R^0_{\psi}$. Note that we also indicate the
dependence on interference with four fermion operators, denoted as $\delta \bar{\Gamma}_{Z \rightarrow \psi \bar{\psi},\psi^4}$ in the SMEFT reported in Ref.\cite{Berthier:2015oma}.
These tree level, numerically suppressed terms should be considered when fits to LEPI data including loop corrections are developed in more detail.

The corrections at one loop follow a similar pattern, and are given as
\bea
\Delta \bar{\Gamma}_{Z \rightarrow \ell \bar{\ell}}&=& \frac{\sqrt{2} \, \hat{G}_F \hat{m}_Z^3}{6 \pi} \, \left[2 \, g^{\ell}_R \, \Delta g^{\ell}_R + 2 \, g^{\ell}_L \, \Delta g^{\ell}_L
+2 \, \delta g^{\ell}_R \, \Delta g^{\ell}_R + 2 \, \delta g^{\ell}_L \, \Delta g^{\ell}_L \right], \\
\Delta \bar{\Gamma}_{Z \rightarrow \nu \bar{\nu}}&=& \frac{\sqrt{2} \, \hat{G}_F \hat{m}_Z^3}{6 \pi} \, \left[ 2 \, g^{\nu}_L \, \Delta g^{\nu}_L+ 2 \, \delta g^{\nu}_L \, \Delta g^{\nu}_L \right], \\
\Delta \bar{\Gamma}_{Z \rightarrow Had}&=& 2 \, \Delta \bar{\Gamma}_{Z \rightarrow \bar{u} u} + 2 \, \Delta \bar{\Gamma}_{Z \rightarrow \bar{d} d}+\Delta \bar{\Gamma}_{Z \rightarrow \bar{b} b}, \\ &=& \frac{3 \, \sqrt{2} \, \hat{G}_F \hat{m}_Z^3}{6 \, \pi} \left[4 \, (g^{u}_R + \delta g^{u}_R) \, \Delta g^{u}_R + 4 \, (g^{u}_L+ \delta g^{u}_L ) \, \Delta g^{u}_L + 4 \, (g^{d}_R+ \delta g^{d}_R) \, \Delta g^{d}_R \right], \nn \\
&+& \frac{3 \, \sqrt{2} \, \hat{G}_F \hat{m}_Z^3}{6 \, \pi} \left[4 \, (g^{d}_L
+ \delta g^{d}_L) \, \Delta g^{d}_L
+ 2 \, (g^{b}_R+ \delta g^{b}_R )\, \Delta g^{b}_R + 2 \, (g^{b}_L+ \delta g^{b}_L) \, \Delta g^{b}_L\right], \nonumber  \\
\Delta \bar{\Gamma}_{Z} &=& 3 \, \Delta \bar{\Gamma}_{Z \rightarrow \ell \bar{\ell}} + 3 \Delta \bar{\Gamma}_{Z \rightarrow \nu \bar{\nu}} +\Delta \bar{\Gamma}_{\rm Had}.
\eea
In the cross terms of order $\delta g^{\psi}_{L/R} \, \Delta \, g^{\psi}_{L/R}$ only the terms leading order
in the SMEFT power counting expansion in $\Delta \, g^{\psi}_{L/R}$ are retained in the interference term.
The shift in the ratios of decay rates follows directly.
In presenting numerical results, we factor the $\Lambda$ dependence out of the Wilson coefficient and scale the suppression scale(s) to $1 \, {\rm TeV}$.
As a result all the numerical expressions reported should be understood to be implicitly multiplied by a factor of $(1 \, {\rm TeV})^2/\Lambda^2$.

To clarify our notational conventions, we note that we denote corrections which are linearly suppressed by a dimension-6 operator coefficient as order $\delta$, and corrections which are present at 1-loop as order $\Delta$. Thus, we will present results that are at order $\delta$, corresponding to known, tree-level SMEFT effects, and new loop-level results at order $\delta\,\Delta$. The results are listed in the Appendix.

\section{Conclusions}\label{conclude}
In this article, we have calculated a set of one loop corrections to the observables
$\bar{\Gamma}_Z, \, \bar{\Gamma}_{Z \rightarrow \bar{\psi}_i \, \psi}$, $\bar{\Gamma}_{Z \rightarrow {\rm Had}}, \, \bar{R}_\ell^0, \, \bar{R}_b^0$.
We have developed results where $\lambda$ and $y_t$ dependent corrections are retained at one loop, while relative $\bar{g}_i$ dependence
in the loop corrections are dropped. Our results incorporate previously known Renormalization Group terms, but also include finite terms.
The numerical version of the results are given in Section \ref{phenosection}. The phenomenological conclusions and implications are extensive.
We postpone a detailed discussion of some of these issues to future publications, but summarize here the most important conclusions.

\begin{itemize}
\item{How large are the corrections? We have presented our numerical results scaled in $\rm TeV$ units, i.e. with the implicit multiplication of $(1 {\rm TeV})^2/\Lambda^2$
being understood. The relative size of the one loop SMEFT corrections to the leading order SMEFT corrections varies dramatically and is UV dependent.
The results can be conservatively estimated as being a relative correction on the order of $\mathcal{O}(10 \%)$ up to the ratio of the unknown Wilson coefficients.
Smaller and larger corrections are also possible. This estimate does not rely on a large log enhancement and is somewhat larger than a naive expectation even so. The importance of these corrections strongly depends
on the unknown Wilson coefficient matching pattern, but for this very reason, strong model independent conclusions are particularly sensitive to these
effects. The logarithmic terms are not particularly dominant numerically when the cut off scale is in the $\rm TeV$ range.}
\item{The most basic point is that at one loop a large number of new parameters in the SMEFT contribute, that are not present at tree level.
It is not the case that a choice of the renormalization scale $\mu^2 = \hat{m}_Z^2$ can remove all of the new parameters that enter.
The reason is that the SMEFT (and the SM) is a multi-scale theory and SMEFT loop corrections also result from interference with the pure SM loop corrections.
At tree level, the ten SMEFT parameters that enter LEPI data in the Warsaw basis are
\bea
\{C_{He}, C_{Hu}, C_{Hd}, C_{H \ell}^{(1)}, C_{H \ell}^{(3)},C_{Hq}^{(1)},C_{Hq}^{(3)},C_{\ell \ell},C_{HWB},C_{HD} \}.
\eea
At one loop, even in our chosen limit of $\rm U(3)^5$ symmetry, $\rm CP$ symmetry, and only retaining $y_t$ and $\lambda$ corrections,
the additional SMEFT parameters present in the Warsaw basis are
\bea
\{C_{qq}^{(1)}, C_{qq}^{(3)}, C_{qu}^{(1)},C_{uu}, C_{qd}^{(1)},C_{ud}^{(1)},C_{\ell q}^{(1)},C_{\ell q}^{(3)},C_{\ell u},C_{qe},C_{eu},C_{Hu}, C_{HB}+ C_{HW},C_{uB},C_{uW},C_{uH} \}.
\nn \eea
With an additional sixteen parameters in our partial calculation it is clear that once the size of loop corrections are reached
in the SMEFT, the parameters contributing at tree level to precise LEP data become unbounded. The size of these corrections is UV dependent,
but is not robustly below the percent level \cite{Berthier:2015oma,Berthier:2015gja,Berthier:2016tkq}, and can be greater. As a result, it is difficult to take seriously
claims that LEP data constrains the parameters appearing at tree level to the per-mille level in a {\it completely model independent fashion}.
This point has already been made in general terms in the literature but is strongly reinforced by the explicit results reported here. This once again insists that such claims do not lead to SMEFT parameters being set to zero in LHC analyses and data reporting to avoid UV bias.}
\item{As has been noted in Section \ref{oneloopsection}, the contributions to the left handed effective bottom coupling at one loop,
compared to the remaining one loop corrections to the other effective couplings, shows an interesting pattern. It is reasonable to expect the left handed effective bottom coupling to be relatively
more perturbed in the SMEFT in this limit of calculation. Whether this points to an underlying explanation of some
partially decoupled new physics effects with important one loop corrections in the SMEFT
leading to the (statistically insignificant) anomaly in LEPI data in the left handed bottom coupling is inconclusive,
but very interesting to consider.}
\item{We have reported numerical results in Section \ref{phenosection} making the tadpole finite term $\Delta \bar{v}_T$ explicit.
The purpose of presenting our results in this manner is to emphasize that the tadpole contributions do affect the result in our $\overline{\rm MS}$ scheme.
The naive expectation is that a pure on-shell renormalization scheme will lead to a cancelation of tadpole effects. This expectation is based on the assumption that the manner in which the on-shell scheme will fix Lagrangian parameters in the SM, will carry over to the SMEFT.
There is ample reason to question this naive expectation. One way to understand this is to consider the "evanescent" scheme dependence
(see Appendix \ref{evanescentappendix})
present at one loop until one considers experimentally extracting the Wilson coefficients, to then use such measured quantities to predict a deviation in a
related process.
This implies that a systematic set of renormalization conditions is required in the SMEFT to unambiguously fix all parameters in the
Lagrangian including the SMEFT Wilson coefficients, beyond those usually employed in an on-shell scheme for the SM. Clearly a full one loop calculation, including gauge coupling dependence is required
to conclusively examine the tadpole issue in the SMEFT. Related to this point, we stress that we make no particular claim that the
modified $\rm \overline{MS}$ scheme we use is in any way preferred, this is simply a subtraction scheme choice with its own benefits and challenges.
Further, the demonstration of the robustness of interpreting $W$ mass measurements in the SMEFT \cite{Bjorn:2016zlr} supports the on-shell scheme employed
and developed for the SMEFT in Refs.~\cite{Gauld:2015lmb,Gauld:2016kuu} being of at least equal interest.}
\item{We have presented results with the theory renormalized at the scale $\mu^2 = \Lambda^2$ in order to make it more transparent to infer constraints
on underlying models integrated out and matched onto the SMEFT. To form model independent conclusions
it is necessary to seriously engage with these loop corrections and consider the further loop effects
that are still uncalculated, when interpreting strong LEP experimental constraints. In much recent literature, SMEFT loop corrections have generally been (implicitly) assumed to be zero and ignored.
Alternatively, some works have argued these effects are not important by essentially invoking UV bias and assumptions. These approaches do not lead to model
independent interpretations of precise LEP data.
We encourage the reader to use these results to decide for themselves the size of these corrections in various models of interest, and
what is a reasonable model independent degree of constraint to assert due to LEP data in the SMEFT.\footnote{The results of this calculation can also be compared
to arguments and claims in YR4 \cite{deFlorian:2016spz} and Ref.\cite{Passarino:2016pzb}.}}
\end{itemize}

Our results are positive for the hopes of physics beyond the SM being discoverable
in the long term LHC experimental program, despite LEP constraints.
We also believe our results strongly encourage the further development of the SMEFT as a meaningful paradigm for
interpreting precise experimental measurements. Loop corrections can be directly calculated
when treating the SMEFT as a well defined field theory. This reduces significant theoretical errors and avoids implicit UV bias in naive leading order analyses.
Such calculations are simply required for a precise and serious interpretation of the most precise experimental measurements
considering the global data set. Fortunately, these calculations can be directly performed using standard EFT methods.
There is still an enormous amount of work to do to project the data consistently into the SMEFT.

\section*{Acknowledgements}
MT acknowledges generous support from the Villum Fonden and partial support by the Danish National Research Foundation (DNRF91).
The project leading to this application has received funding from the European Union's Horizon 2020 research
and innovation programme under the Marie Sklodowska-Curie grant agreement No 660876, HIGGS-BSM-EFT.
WS acknowledges generous support from the Danish Council for Independent Research and the Alexander von Humboldt Foundation, in the framework of the Sofja Kovalevskaja Award 2016, endowed by the German Federal Ministry of Education and Research.
MT and CH thank Joaquin Blasco for discussions in a very early stage of this work.
MT also thanks  Ilaria Brivio, Aneesh Manohar, Giampiero Passarino and Benjamin Pecjak for
discussions and (in a sense) the Higgs XS working group for a motivating environment to develop these results. We also thank Ilaria Brivio and Aneesh Manohar for comments on the manuscript.

\newpage
\appendix

\section{Appendix}
\subsection{One
loop vev finite terms}
In the $\lambda$, $y_t$ dominance limit of interest in this work, the tadpole correction to the one point
function of the Higgs field has the form \cite{Hartmann:2015oia,Hartmann:2015aia}
\bea
\frac{\Delta \bar{v}_T}{\bar{v}_T} = \frac{3 \, \lambda}{16 \, \pi^2} \left(1+ \log \left[\frac{\mu^2}{m_h^2} \right] \right) - \frac{2 \, N_c}{16 \, \pi^2} \, \frac{m_t^2}{m_h^2} \, y_t^2
\left(1+ \log \left[\frac{\mu^2}{m_t^2} \right] \right) + \mathcal{O} \left(\frac{\bar{v}_T^2}{16 \, \pi^2 \, \Lambda^2} \right) + \cdots
\eea
Additional corrections are also present in the one point function of the h field in the SMEFT, indicated here as $\mathcal{O} \left(\frac{\bar{v}_T^2}{16 \, \pi^2 \, \Lambda^2} \right)$, proportional to $1/\Lambda^2$ at one loop. We postpone a detailed
discussion of these terms to a future publication.
\subsection{Class 8 Scheme dependence in $d$ dimensions.}\label{evanescentappendix}
A subtlety is present in defining the one loop finite terms in the SMEFT for Class 8 operators, when calculating $\bar{\Gamma}_Z$.
As is well known, see for example related discussion in Refs.~\cite{Buras:1989xd,Dugan:1990df,Jack:1994bn,Chanowitz:1979zu,Harlander:2006rj,Bern:2015xsa}, $d \neq 4$ scheme dependence can exist in a calculation of a particular observable at one loop, when considering finite terms in amplitudes. This occurs in the case of interest here, due to an odd parity fermion loop in $d$ dimensions due to a number of Class 8 operators. We refer to this scheme dependence as "evanescent", due to its relationship to the
issues of $d$ dimensional effects and evanescent operators discussed in Refs.~\cite{Buras:1989xd,Dugan:1990df}.

Various schemes can be chosen to define $\gamma_5$ in $d$ dimensions. Here we consider Naive dimensional regularization, which assumes that $\gamma_5$ is anti-commuting with the other $\gamma$ matrices; and the t'Hooft- Veltman scheme (HV) \cite{'tHooft:1972fi}, in which $\gamma_5$ is anti-commuting with $\gamma$ matrices in $d=4$ dimensions and commuting in $d = -2 \, \epsilon$
dimensions. Dimensional reduction,
in which $d=4$ is assumed for $\gamma_5$, is another possibility.  We use naive dimensional regularization in our main results including finite terms. This requires some clarification of the cancelation of this $d \neq 4$ "evanescent" scheme dependence when considering relations between measurable quantities, which in this case formally includes the renormalized SMEFT Wilson coefficients. If these Wilson coefficients are non-zero, the extraction of these parameters corresponds to another set of input parameters in general. The expectation based on known examples is that the scheme dependence is unphysical, and will cancel in physical relations between measured quantities \cite{Buras:1989xd,Dugan:1990df}. We find that this is indeed the case as follows.

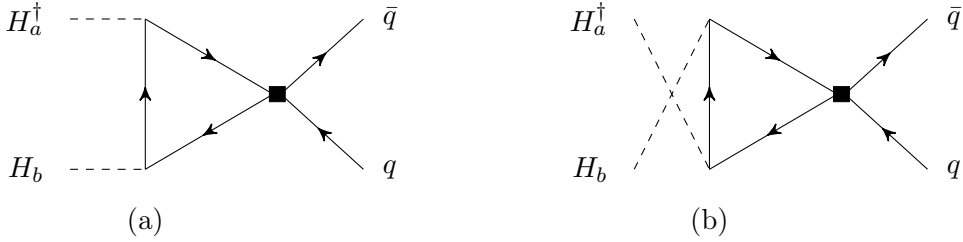
\begin{figure}[!h]
\begin{tikzpicture}
[
decoration={
	markings,
	mark=at position 0.55 with {\arrow[scale=1.5]{stealth'}};
}]
\hspace{-4cm}
\draw  [dashed]  (-1,1) -- (0,1);
\draw  [dashed]  (-1,-1) -- (0,-1);

\draw [postaction=decorate] (0,1) -- (1.7,0);
\draw [postaction=decorate] (1.7,0) -- (0,-1);
\draw [postaction=decorate] (0,-1) -- (0,1);
\draw [postaction=decorate] (1.8,0) -- (2.9,1);
\draw [postaction=decorate] (2.9,-1) -- (1.8,0);

\filldraw (1.65,-0.1) rectangle (1.86,0.1);

\draw (0,-1.7) node [align=center] {(a)};

\node [left][ultra thick] at (-1.2,1) {$H^{\dagger}_{a}$};
\node [left][ultra thick] at (-1.2,-1) {$H_b$};
\node [right][ultra thick] at (3,1) {$\bar{q}$};
\node [right][ultra thick] at (3,-1) {$q$};

\hspace{7.5cm}

\draw  [dashed]  (-1,1) -- (0,-1);
\draw  [dashed]  (-1,-1) -- (0,1);

\draw [postaction=decorate] (0,1) -- (1.7,0);
\draw [postaction=decorate] (1.7,0) -- (0,-1);
\draw [postaction=decorate] (0,-1) -- (0,1);
\draw [postaction=decorate] (1.8,0) -- (2.9,1);
\draw [postaction=decorate] (2.9,-1) -- (1.8,0);

\filldraw (1.65,-0.1) rectangle (1.86,0.1);

\draw (0,-1.7) node [align=center] {(b)};

\node [left][ultra thick] at (-1.2,1) {$H^{\dagger}_{a}$};
\node [left][ultra thick] at (-1.2,-1) {$H_b$};
\node [right][ultra thick] at (3,1) {$\bar{q}$};
\node [right][ultra thick] at (3,-1) {$q$};
\end{tikzpicture}
\caption{Evanescent one loop matching correction onto $C_{Hq}^{(1)}$.}
\label{mix}
\end{figure}

Consider the contribution of the operator $\mathcal{Q}_{qq}^{(1)}$ to the naive amplitude for the decay $Z \rightarrow \bar{\psi}_p \, \psi_p$.
The finite terms proportional to $y_t^2$ in the HV scheme are found to be
\bea
i \mathcal{A}^{HV} =  -  \frac{i \delta_{pr}}{16 \, \pi^2}  \, (C_{\substack{qq\\ 33pr}}^{(1)}+C_{\substack{qq\\ pr33}}^{(1)} ) \, m_z \, v \, y_t^2 \, \bar{u}_p \, \tilde{\gamma}_{\alpha} \, P_L \, u_r, \label{evanescent}
\eea
for the $p=r$ flavour indices summed over in this decay. Here the $\sim
$ superscript indicates an explicitly 4 dimensional $\gamma$ matrix. This result differs from the result obtained in a naive dimensional regularization calculation by a factor
\begin{align}\label{schemedep}
i \mathcal{A}^{HV-NDR}  =  -  \frac{i \, \delta_{pr}}{16 \, \pi^2} \, (C_{\substack{qq\\ 33pr}}^{(1)}+C_{\substack{qq\\ pr33}}^{(1)} ) \, m_z \, v \, y_t^2 \, \bar{u}_p \, \tilde{\gamma}_{\alpha} \, P_L \, u_r.
\end{align}
For the two schemes to lead to the same result, we expect the above extra contribution in the HV scheme to be canceled by a finite matching correction at one loop. This occurs for the Wilson coefficient $C_{\substack{Hq \\pr}}^{(1)}$ contributing to the decay, due to an evanescent $d \neq 4$ effect. Calculating the two diagrams shown in Fig.\ref{mix} in the HV scheme with $d \neq 4$ in the unbroken phase of the theory,
one finds the required finite matching correction
\begin{align}
C_{\substack{Hq \\ pr}}^{(1)} & = C_{\substack{Hq \\ pr}}^{(1)} +  \frac{1}{48 \pi^2} \left(C_{\substack{qq \\ prst}}^{(1)} + C_{\substack{qq \\ stpr}}^{(1)}\right) (2 [Y_u^{\dagger}Y_u]^{st} + [Y_uY_u^{\dagger}]^{st} ).
\end{align}
This exactly cancels the scheme dependence in Eqn.~\ref{schemedep}, when inserted in the tree level matrix element for $C_{\substack{Hq \\ pr}}^{(1)}$. In this manner,
the evanescent scheme dependence for the Class 8 operators cancels out of relations between observables, as expected.

This result is equivalent to that achieved by the usual method of introducing explicit counter terms to regain gauge invariance, but makes clear that this class of scheme choices is no different from any other. Ultimately, the consistent application of a renormalization scheme to both input parameters and predicted observables will cancel the scheme choices made.

\subsection{Numerical results for $\bar{\Gamma}_{Z \rightarrow \bar{\ell} \, \ell}$}
The $\delta$ correction to $\bar{\Gamma}_{Z \rightarrow \bar{\ell} \, \ell}$ is given by
\bea
\frac{\delta \bar{\Gamma}_{Z \rightarrow \bar{\ell} \, \ell}}{10^{-2}}
&=& \left[1.07 C_{H\ell}^{(1)}-0.939 C_{He} - 0.298 \,C_{HD} -0.117 \, C_{H\ell}^{(3)}  - 0.049 \, C_{HWB}  + 0.595 \, C_{\ell \ell} \right]. \nn \\
\eea
The $\delta \, \Delta$ correction to $\bar{\Gamma}_{Z \rightarrow \bar{\ell} \, \ell}$ has the contributions
\bea
\frac{\delta \Delta \bar{\Gamma}_{Z \rightarrow \bar{\ell} \, \ell}}{10^{-3}} &=& \left[ \left(0.071  \Delta \bar{v}_T + 0.201 \right) C_{He} + \left(0.153  \Delta \bar{v}_T + 0.065 \right)C_{H\ell}^{(1)} - \left(0.094 \,  \Delta \bar{v}_T + 0.118\right) C_{HD}, \right. \nn \\
&\,& \hspace{-0.15cm} - \left(0.390 \Delta \bar{v}_T + 0.476 \right) C_{H\ell}^{(3)}
-\left(0.085 \, \Delta \bar{v}_T + 0.117 \right) \, C_{HWB}  + 0.271\left(\Delta \bar{v}_T + 1\right) \, C_{\ell \ell}, \nn  \\
&\,& \hspace{-0.15cm} +0.112 \, C_{\ell q}^{(3)}
+ 0.007 \left(C_{HB}+C_{HW}\right) + 0.797\, \Delta \bar{v}_T \Big],
\eea
and $(\delta \Delta \bar{\Gamma}_{Z \rightarrow \bar{\ell} \, \ell}/10^{-3}) \times (\Lambda^2/(1 \, {\rm TeV})^2)$ also has the logarithmic terms
\bea
\frac{\delta \Delta \bar{\Gamma}_{Z \rightarrow \bar{\ell} \, \ell}}{10^{-3}} &=& \left[0.342 \, C_{He} + 0.153 \, C_{H\ell}^{(1)} -0.218 \, C_{HD} - 0.657 \, C_{H\ell}^{(3)}
 - 0.213 \, C_{HWB} + 0.517 \, C_{\ell \ell},\nn\right.\\
&\,&-0.022 \, C_{\ell q}^{(3)} +0.202 (C_{\ell u}-C_{\ell q}^{(1)})+ 0.176 C_{q e} -0.027\, C_{uW}-0.016\, C_{uB} \Big]
\, \log \left[\frac{\Lambda^2}{\hat{m}_t^2}\right] \nn\\
 &+& \Big[6.21 \times 10^{-6} \, C_{HD}  + 0.015 \, C_{He} - 0.018 \, C_{H\ell}^{(1)} + 0.002 \, C_{H\ell}^{(3)}\nn\\
&-& (1.39 \, C_{HWB} + 9.73 \, C_{\ell \ell}) 10^{-3} \Big]
\, \log \left[\frac{\Lambda^2}{\hat{m}_h^2}\right].
\eea
\subsection{Numerical results for $\bar{\Gamma}_{Z \rightarrow \bar{\nu} \, \nu}$}
The $\delta$ correction to $\bar{\Gamma}_{Z \rightarrow \bar{\nu} \, \nu}$ is given by
\bea
\frac{\delta \bar{\Gamma}_{Z \rightarrow \bar{\nu} \, \nu}}{10^{-2}}
&=& \left[ - 2.01 \,C_{Hl}^{(1)} - 0.503 \,C_{HD}  + 1.01 \, C_{\ell \ell} \,\right].
\eea
The $\delta \, \Delta$ correction to $\bar{\Gamma}_{Z \rightarrow \bar{\nu} \, \nu}$  has the contributions
\bea
\frac{\delta \Delta \bar{\Gamma}_{Z \rightarrow \bar{\nu} \, \nu}}{10^{-3}} &=& \left[-\left(0.082 \,  \Delta \bar{v}_T - 0.136 \right) C_{H\ell}^{(1)}
 - \left(0.041 \,  \Delta \bar{v}_T + 0.051 \right) C_{HD}  +0.189 \, C_{\ell q}^{(3)} + 1.35 \, \Delta \bar{v}_T \right. \nn \\
&\,& \hspace{-1.75cm} - \left(0.327 \, \Delta \bar{v}_T + 0.411 \right) C_{H\ell}^{(3)}  - \left(0.016\,  \Delta \bar{v}_T + 0.044 \right)C_{HWB}
+ \left(0.204 \, \Delta \bar{v}_T + 0.138 \right) \, C_{\ell \ell}\Big],
\eea
and the $\delta \, \Delta$ corrections to $\bar{\Gamma}_{Z \rightarrow \bar{\nu} \, \nu}$ also has the logarithmic terms
\bea
\frac{\delta \Delta \bar{\Gamma}_{Z \rightarrow \bar{\nu} \, \nu}}{10^{-3}} &=& \left[ 0.189 \, C_{H\ell}^{(1)}  -0.095 \, C_{HD}-0.378 \, C_{H\ell}^{(3)}
 - 0.074 \, C_{HWB} + 0.284 \, C_{\ell \ell},\right.\nn\\
&\,& +0.378 (C_{\ell q}^{(1)} - C_{\ell u}) -0.062 \, C_{uW} - 0.037 \, C_{uB} \Big]
\, \log \left[\frac{\Lambda^2}{\hat{m}_t^2}\right], \\
&\,& \hspace{-0.75cm}  +\Big[ \left(5.93 \, C_{H\ell}^{(3)} + 7.41 \, C_{HD}\right) \times 10^{-6}
 + 0.033 \, C_{H\ell}^{(1)}  -0.003 \, C_{HWB}-  0.016 \, C_{\ell \ell}\Big]
\, \log \left[\frac{\Lambda^2}{\hat{m}_h^2}\right]. \nn
\eea
\subsection{Numerical results for $\bar{\Gamma}_{Z \rightarrow \bar{u} \, u}$}
Similarly the $\delta$ correction to $\bar{\Gamma}_{Z \rightarrow \bar{u} \, u}$ is given by
\bea
\frac{\delta \bar{\Gamma}_{Z \rightarrow \bar{u} \, u}}{10^{-2}}
&=&  - 1.37 \, C_{HD} - 5.49 \, C_{H\ell}^{(3)} - 4.16 \left(C_{Hq}^{(1)} - C_{Hq}^{(3)}\right) +1.88 \, C_{Hu}
 - 0.559 \, C_{HWB} + 2.74 \, C_{\ell \ell}. \nn \\
\eea
The $\delta \, \Delta$ correction to $\bar{\Gamma}_{Z \rightarrow \bar{u} \, u}$ has the contributions
\bea
\frac{\delta \Delta \bar{\Gamma}_{Z \rightarrow \bar{u} \, u}}{10^{-3}} &=& \left[ - \left(0.168 \, \Delta \bar{v}_T + 0.211 \right) C_{HD}
- \left(1.59 \, \Delta \bar{v}_T + 1.35 \right) C_{H\ell}^{(3)} - \left(0.143 \, \Delta \bar{v}_T + 0.402 \right) C_{Hu}  \right. \nn \\
&\,& \hspace{2.5mm} - \left(0.388 \, \Delta \bar{v}_T - 0.005 \right) (C_{Hq}^{(1)}-C_{Hq}^{(3)}) - \left(0.093 \, \Delta \bar{v}_T + 0.138 \right)C_{HWB}
+3.68 \, \Delta \bar{v}_T, \nn  \\
&\,& \hspace{2.5mm}+0.516 \, C_{\ell q}^{(3)}
+ \left(0.794 \, \Delta \bar{v}_T +0.674\right) \, C_{\ell \ell}
+ 0.079 \left(C_{HB}+C_{HW}\right)  \Big],
\eea
and the $\delta \, \Delta$ correction to $\bar{\Gamma}_{Z \rightarrow \bar{u} \, u}$ also has the logarithmic terms
\bea
\frac{\delta \Delta \bar{\Gamma}_{Z \rightarrow \bar{u} \, u}}{10^{-3}} &=& \left[ -0.389 \, C_{HD} -1.62\, C_{H\ell}^{(3)}
-0.117 \left(C_{Hq}^{(1)}-C_{Hq}^{(3)}\right) - 0.684\,C_{Hu}
 - 0.248 \, C_{HWB}, \right. \nn \\
&\,& \hspace{0.25cm}   + 1.32 \, C_{\ell \ell} - 1.03 \, C_{\ell q}^{(3)} +  1.56 \left(C_{qq}^{(3)}+C_{qq}^{(1)}\right) -1.13\, C_{qu}^{(1)}+0.706\,C_{uu}, \\
&\,& \hspace{0.25cm}-0.056\, C_{uW}-0.034\, C_{uB} \Big]
\, \log \left[\frac{\Lambda^2}{\hat{m}_t^2}\right]+\Big[ 2.31\times10^{-5} \, C_{HD} - 0.03\, C_{Hu},\nn \\
&\,& \hspace{0.25cm}+ 0.068 \left(C_{Hq}^{(1)}-C_{Hq}^{(3)}\right)-0.003 \, C_{HWB}
 + 0.090 \, C_{H\ell}^{(3)} - 0.045 \, C_{\ell \ell}\Big]
\, \log \left[\frac{\Lambda^2}{\hat{m}_h^2}\right].\nn
\eea
\subsection{Numerical results for $\bar{\Gamma}_{Z \rightarrow \bar{d} \, d}$}
The $\delta$ correction to $\bar{\Gamma}_{Z \rightarrow \bar{d} \, d}$ (where $d = \{d,s,b\}$) is given by
\bea
\frac{\delta \bar{\Gamma}_{Z \rightarrow \bar{d} \, d}}{10^{-2}}
&=&- 0.939\, C_{Hd} - 1.58\, C_{HD} - 6.31\, C_{H\ell}^{(3)} + 5.10 \left(C_{Hq}^{(1)} + C_{Hq}^{(3)}\right) - 0.510 \, C_{HWB} + 3.15 \, C_{\ell \ell} . \nn \\
\eea
The $\delta \, \Delta$ correction to $\bar{\Gamma}_{Z \rightarrow \bar{d} \, d}$ (where $d = \{d,s\}$) has the contributions
\bea
\frac{\delta \Delta \bar{\Gamma}_{Z \rightarrow \bar{d} \, d}}{10^{-3}} &=& \left[\left(0.071 \,  \Delta \bar{v}_T + 0.201 \right) C_{Hd}
 - \left(0.115 \,  \Delta \bar{v}_T + 0.144 \right) C_{HD},
- \left(1.45 \, \Delta \bar{v}_T + 1.08 \right) C_{H\ell}^{(3)} \right. \nn \\
&\,& \hspace{0.25cm} + \left(0.316\, \Delta \bar{v}_T - 0.206 \right) \left(C_{Hq}^{(1)}+C_{Hq}^{(3)}\right)  - \left(0.024 \, \Delta \bar{v}_T + 0.064 \right)C_{HWB}
+4.23 \, \Delta \bar{v}_T ,\nn   \\
&\,& \hspace{0.25cm} + \left(0.727 \, \Delta \bar{v}_T +0.541\right)  C_{\ell \ell} +0.593 \, C_{\ell q}^{(3)}
+ 0.072 \left(C_{HB}+C_{HW}\right) \Big],
\eea
and the $\delta \, \Delta$ corrections to $\bar{\Gamma}_{Z \rightarrow \bar{d} \, d}$ (where $d = \{d,s\}$) also has the logarithmic terms
\bea
\frac{\delta \Delta \bar{\Gamma}_{Z \rightarrow \bar{d} \, d}}{10^{-3}} &=& \left[ 0.342 \, C_{Hd}  -0.266 \, C_{HD} - 0.995\, C_{H\ell}^{(3)}
-0.225 \left(C_{Hq}^{(1)}+C_{Hq}^{(3)}\right)
 - 0.110 \, C_{HWB}, \right. \nn \\
&\,& \hspace{0.25cm}   + 1.09 \, C_{\ell \ell} - 1.19 \, C_{\ell q}^{(3)}+0.176 \left( C_{qd}^{(1)}-C_{ud}^{(1)}\right) +  1.92\left( C_{qq}^{(3)}-C_{qq}^{(1)}\right) +0.958\, C_{qu}^{(1)}, \nn\\
&\,&\hspace{2.5mm}-0.091\, C_{uW}-0.055\, C_{uB}  \Big]
\, \log \left[\frac{\Lambda^2}{\hat{m}_t^2}\right]
+ \Big[ (2.43\times10^{-5}\, C_{HD}+0.015\, C_{Hd}, \nn \\
&\,& \hspace{0.25cm} +0.103\, C_{Hl}^{(3)}-0.083\left(C_{hq}^{(1)}+C_{hq}^{(3)}\right)- 0.005 \, C_{HWB} - 0.052 \, C_{\ell \ell}\Big] \, \log \left[\frac{\Lambda^2}{\hat{m}_h^2}\right].
\eea
\subsection{Numerical results for $\bar{\Gamma}_{Z \rightarrow \bar{b} \, b}$}
The $\delta$ correction to $\bar{\Gamma}_{Z \rightarrow \bar{b} \, b}$ is identical to that for other down-type quarks given in the previous subsection.
The $\delta \, \Delta$ correction to $\bar{\Gamma}_{Z \rightarrow \bar{b} \, b}$  has the contributions
\bea
\frac{\delta \Delta \bar{\Gamma}_{Z \rightarrow \bar{b} \, b}}{10^{-3}} &=& \left[\left(0.071 \,  \Delta \bar{v}_T + 0.201 \right) C_{Hd}
 - \left(0.115 \,  \Delta \bar{v}_T + 0.221 \right) C_{HD}
- \left(1.45\, \Delta \bar{v}_T + 1.59 \right) C_{H\ell}^{(3)} \right. \nn \\
&\,& \hspace{-0.15cm} + \left(0.316 \,\Delta \bar{v}_T + 0.353 \right) C_{Hq}^{(1)}
+\left(0.316 \,\Delta \bar{v}_T + 0.016 \right) C_{Hq}^{(3)} + 0.080 \,C_{Hu}, \nn  \\
&\,& \hspace{-0.15cm} - \left(0.024\, \Delta \bar{v}_T + 0.102 \right)C_{HWB}
+ \left(0.727 \, \Delta \bar{v}_T + 0.796 \right) \, C_{\ell \ell}
+0.593 \, C_{\ell q}^{(3)},  \\
&\,& \hspace{-0.15cm} + 1.28 \, C_{qq}^{(3)} + 0.101 \, C_{uH}
+ 0.072 \left(C_{HB}+C_{HW}\right) +4.23 \,\Delta \bar{v}_T \Big], \nn
\eea
and the $\delta \, \Delta$ corrections to $\bar{\Gamma}_{Z \rightarrow \bar{b} \, b}$ also has the logarithmic terms
\bea
\frac{\delta \Delta \bar{\Gamma}_{Z \rightarrow \bar{b} \, b}}{10^{-3}} &=& \left[ 0.342 \, C_{Hd}  -0.317 \, C_{HD} - 1.33\, C_{H\ell}^{(3)}
-0.066\, C_{Hq}^{(1)} + 0.807\, C_{Hq}^{(3)} - 0.160\, C_{Hu}, \right. \nn \\
&\,& \hspace{0.25cm}  -1.92\, C_{qq}^{(1)} - 0.135 \, C_{HWB}+ 1.26 \, C_{\ell \ell} - 1.19 \, C_{\ell q}^{(3)} + 0.176 \left(C_{qd}^{(1)}-C_{ud}^{(1)}\right), \nn\\
&\,& \hspace{2.5mm} +0.639 \, C_{qq}^{(3)} +0.958\, C_{qu}^{(1)} +0.067\, C_{uH} -0.091\, C_{uW}-0.055\, C_{uB}\Big]
\, \log \left[\frac{\Lambda^2}{\hat{m}_t^2}\right], \nn \\
&+&\Big[ 0.015\, C_{Hd} + 2.43\times10^{-5} \, C_{HD} - 0.083\left(C_{Hq}^{(1)}+C_{Hq}^{(3)}\right)
 + 0.103 \, C_{H\ell}^{(3)} -0.005 \, C_{HWB}  \nn \\
  &\,& \hspace{0.25cm}  - 0.052 \, C_{\ell \ell}\Big]
\, \log \left[\frac{\Lambda^2}{\hat{m}_h^2}\right].
\eea
\subsection{Numerical results for $\bar{\Gamma}_{Z \rightarrow {\rm Had}}$}
The $\delta$ correction to $\bar{\Gamma}_{Z \rightarrow {\rm Had}}$ is given by
\bea
\frac{\delta \bar{\Gamma}_{Z \rightarrow {\rm Had}}}{10^{-2}}
&=& \left[ - 2.82 \,C_{Hd} - 7.47 \,C_{HD} - 29.9 \,C_{H\ell}^{(3)}
+6.97 \,C_{Hq}^{(1)} + 23.6 \,C_{Hq}^{(3)} + 3.75 \,C_{Hu} \right. \nn \\
&\,& \hspace{0.25cm} - 2.65  \,C_{HWB}
+ 14.9 \, C_{\ell \ell} \, \Big].
\eea
The $\delta \, \Delta$ correction to $\bar{\Gamma}_{Z \rightarrow {\rm Had}}$ has the contributions
\bea
\frac{\delta \Delta \bar{\Gamma}_{Z \rightarrow {\rm Had}}}{10^{-3}} &=& \left[\left(0.214 \,  \Delta \bar{v}_T + 0.603 \right) C_{Hd}
 - \left(0.681 \,  \Delta \bar{v}_T + 0.932 \right) C_{HD}
- \left(7.54\, \Delta \bar{v}_T + 6.45 \right) C_{H\ell}^{(3)}, \right. \nn \\
&\,& \hspace{-0.75cm} +\left(0.174\, \Delta \bar{v}_T -0.049 \right) C_{Hq}^{(1)}
+\left(1.73 \,\Delta \bar{v}_T - 0.406 \right) C_{Hq}^{(3)}  -\left(0.286\, \Delta \bar{v}_T + 0.725 \right) C_{Hu}, \nn \\
&\,& \hspace{-0.75cm}
- \left(0.256 \,\Delta \bar{v}_T + 0.507 \right)C_{HWB}
+ \left(3.77 \, \Delta \bar{v}_T + 3.22 \right) \, C_{\ell \ell}  +1.28 \, C_{qq}^{(3)}+ 20.0 \,\Delta \bar{v}_T , \nn  \\
&\,& \hspace{-0.75cm}
+2.81 \, C_{\ell q}^{(3)}
 +0.101 \, C_{uH}
  + 0.374 \, \left(C_{HB}+C_{HW}\right)\Big],
\eea
and the logarithmic terms
\bea
\frac{\delta \Delta \bar{\Gamma}_{Z \rightarrow {\rm Had}}}{10^{-3}}&=& \left[
1.03 \, C_{Hd} - 1.63\, C_{HD} - 6.56\, C_{H\ell}^{(3)} - 0.750 \, C_{Hq}^{(1)}+ 0.590\, C_{Hq}^{(3)}
- 1.53\, C_{Hu} , \right. \nn \\
&\,& \hspace{0.25cm} - 0.85 \,C_{HWB}  + 6.09\, C_{\ell \, \ell} - 5.62\, C_{\ell q}^{(3)} +0.529 \left(C_{qd}^{(1)}-C_{ud}^{(1)}\right)+ 7.60 \,C_{qq}^{(3)} + 0.605\, C_{qu}^{(1)},\nn\\
&\,& \hspace{0.25cm}  -2.62\, C_{qq}^{(1)} + 0.067 \, C_{uH}+1.41\, C_{uu}-0.386\, C_{uW}-0.232\, C_{uB} \Big]
\, \log \left[\frac{\Lambda^2}{\hat{m}_t^2}\right], \nn \\
&\, &\hspace{-0.35cm}  + \Big[ 0.046 \,C_{Hd} + 1.19\times10^{-4}\, C_{HD} -0.114\, C_{Hq}^{(1)} -0.386\, C_{Hq}^{(3)} - 0.061\, C_{Hu}
 + 0.489 \, C_{H\ell}^{(3)}, \nn \\
&\,& \hspace{0.25cm}  - 0.020 \, C_{HWB} - 0.244\, C_{\ell \ell} \Big]
\, \log \left[\frac{\Lambda^2}{\hat{m}_h^2}\right].
\eea
\subsection{Numerical results for $\bar{\Gamma}_{Z}$}
The $\delta$ correction to $\bar{\Gamma}_{Z}$ is given by
\bea
\frac{\delta \bar{\Gamma}_Z}{10^{-2}} &=& \left[-2.82 \left(C_{Hd} +C_{He}+ C_{H\ell}^{(1)}\right) - 9.87 \,C_{HD} -30.2 \, C_{H\ell}^{(3)} + 6.97 \,C_{Hq}^{(1)} + 23.6 \,C_{Hq}^{(3)} , \right. \nn \\
&\,& + 3.75 \, C_{Hu} - 2.80 \, C_{HWB} + 19.7 \, C_{\ell \ell} \,\Big].
\eea
Similarly, the $\delta \, \Delta$ correction to $\bar{\Gamma}_Z$ has the contributions
\bea
\frac{\delta \Delta \bar{\Gamma}_Z}{10^{-3}} &=& \left[ \left(0.214 \, \Delta \bar{v}_T + 0.603 \right) \left(C_{Hd}+ C_{He}+ C_{H\ell}^{(1)}\right) - \left(1.09 \,  \Delta \bar{v}_T + 1.44\right) C_{HD}, \right. \nn \\
&\,& \hspace{0.25cm} - \left(9.69\, \Delta \bar{v}_T + 9.11 \right) C_{H\ell}^{(3)}
+ \left(0.174 \, \Delta \bar{v}_T  - 0.049 \right) \, C_{Hq}^{(1)} + \left( 1.73 \, \Delta \bar{v}_T - 0.406 \,\right) C_{Hq}^{(3)}, \nn \\
&\,& \hspace{0.25cm}  - \left(0.286 \, \Delta \bar{v}_T  + 0.725 \right)\, C_{Hu}  - \left(0.560 \, \Delta \bar{v}_T + 1.00 \right) \, C_{HWB},  \\
&\,& \hspace{0.25cm}  + \left(5.20 \, \Delta \bar{v}_T +4.45\right) \, C_{\ell \ell} +3.71 \, C_{\ell q}^{(3)}
 +1.28 \, C_{qq}^{(3)},  \nn \\
&\,&\hspace{0.25cm} + 0.101 \, C_{uH} + 0.395 \left(C_{HB}+C_{HW}\right) +26.5 \, \Delta \bar{v}_T \Big], \nn
\eea
and the $\delta \, \Delta$ correction to $\bar{\Gamma}_Z$ also has the logarithmic terms
\bea
\frac{\delta \Delta \bar{\Gamma}_Z}{10^{-3}} &=& \left[1.03 \left(C_{Hd}+ C_{He}+ C_{H\ell}^{(1)}\right) -2.56 \, C_{HD} - 9.66 \, C_{H\ell}^{(3)}
- 0.749 \, C_{Hq}^{(1)} + 0.590\, C_{Hq}^{(3)}, \right.  \\
&\,& \hspace{0.2cm}  - 1.53\, C_{Hu}  - 1.71 \, C_{HWB} + 8.49 \, C_{\ell \ell} -5.69 \, C_{\ell q}^{(3)}+ 7.60 \, C_{qq}^{(3)} , \nn \\
&\,& \hspace{0.2cm} +0.529 \left(C_{\ell q}^{(1)}+C_{qd}^{(1)}+C_{qe}+C_{qd}^{(1)}- C_{\ell u}- C_{ud}^{(1)}-C_{eu}\right) \nn \\
&\,&\hspace{0.2cm} -2.62\, C_{qq}^{(1)} + 0.605 \, C_{qu}^{(1)} + 0.067 \, C_{uH} +1.41 \, C_{uu}-0.651 \, C_{uW}-0.391\, C_{uB}\Big]
 \log \left[\frac{\Lambda^2}{\hat{m}_t^2}\right],\nn\\
&+& \left[0.046 \left(C_{Hd}+ C_{He}+ C_{H\ell}^{(1)}\right) +1.60 \times 10^{-4}\,C_{HD},\right.-0.114\,C_{Hq}^{(1)} -0.386 \,C_{Hq}^{(3)},   \nn \\
&\,&\hspace{0.2cm} - 0.061\, C_{Hu}+  0.495 \, C_{H\ell}^{(3)}  -0.323 \, C_{\ell \ell} - 0.034 \, C_{HWB} \Big]
\, \log \left[\frac{\Lambda^2}{\hat{m}_h^2}\right]. \nn
\eea
\subsection{Numerical results for $\bar{R}_\ell^0$}
The $\delta$ correction to $\bar{R}_\ell^0$ is given by
\bea
\frac{\delta \bar{R}_\ell^0}{10^{-2}}
&=& - 33.8 \,C_{Hd} + 226 \,C_{He} - 258 \,C_{H\ell}^{(1)} - 18.2 \,C_{HD} - 331 \,C_{H\ell}^{(3)}
+83.7 \,C_{Hq}^{(1)}  \nn \\
&\,& \hspace{0.25cm}+ 283 \,C_{Hq}^{(3)} + 45.1 \,C_{Hu} - 19.9  \,C_{HWB}
+ 36.3 \, C_{\ell \ell}.
\eea
Similarly, the $\delta \, \Delta$ correction to  $\bar{R}_\ell^0$  has the contributions
\bea
\frac{\delta \Delta \bar{R}_\ell^0}{10^{-3}} &=& \left[ - \left(33.3 \,\Delta \bar{v}_T + 66.4 \right) C_{He}
- \left(18.4 \,\Delta \bar{v}_T - 4.85 \right) C_{H\ell}^{(1)}  +\left(5.81 \,\Delta \bar{v}_T + 11.3 \right) C_{Hd}, \right. \\
&\,&\hspace{0.2cm} +\left(17.9 \,\Delta \bar{v}_T + 22.3 \right) C_{HD} +\left(35.6\, \Delta \bar{v}_T + 78.1 \right) C_{H\ell}^{(3)}
-\left(5.93\, \Delta \bar{v}_T + 10.7 \right) C_{Hq}^{(1)}, \nn \\
&\,&\hspace{0.2cm} -\left(6.39 \,\Delta \bar{v}_T + 40 \right) C_{Hq}^{(3)} - \left(7.74 \,\Delta \bar{v}_T + 14.1 \right) C_{Hu}
+ \left(19.6 \,\Delta \bar{v}_T + 24.9 \right) C_{HWB}, \nn \\
&\,&\hspace{-1.6cm} -\left(27.0\, \Delta \bar{v}_T + 36.6 \right) C_{\ell \ell} + 6.83 \,C_{\ell q}^{(3)}  + 15.3\, C_{qq}^{(3)} + 1.21\, C_{uH} + 2.81 \left(C_{HB}+C_{HW}\right) + 48.6 \,\Delta v \Big],\nn
\eea
and the $\delta \, \Delta$ correction to $\bar{R}_\ell^0$  also has the logarithmic terms
\bea
\frac{\delta \Delta \bar{R}_\ell^0}{10^{-3}}&= & \left[19.8 \,C_{Hd} - 118 \,C_{He} +4.18\, C_{H\ell}^{(1)} +41.3 \, C_{HD} + 154 \, C_{H\ell}^{(3)}
-27.5 \, C_{Hq}^{(1)} - 55.7\, C_{Hq}^{(3)}, \right. \nn \\
&\,& \hspace{0.2cm}  - 28.3\, C_{Hu}  + 46.0 \, C_{HWB} -68.2 \, C_{\ell \ell} +48.5 \, (C_{\ell q}^{(1)} -C_{\ell u})-62.2 \, C_{\ell q}^{(3)}+6.36 C_{qd}^{(1)}\nn \\
&\,& \hspace{0.2cm} -42.4 \, (C_{qe}-C_{eu}) - 31.5 C_{qq}^{(1)} + 91.2 \, C_{qq}^{(3)} + 7.26 \, C_{qu}^{(1)}\nn\\
&\,&\hspace{0.2cm} + 0.810 \, C_{uH}  - 6.36 C_{ud}^{(1)} +16.9 \, C_{uu}+1.78 \, C_{uW}+1.07 \, C_{uB}\Big]
 \log \left[\frac{\Lambda^2}{\hat{m}_t^2}\right], \nn  \\
&\, &\hspace{0.2cm} + \Big[0.552 \,C_{Hd} - 1.48 \times10^{-4}\,C_{HD}- 3.69\, C_{He}+ 4.22\, C_{H\ell}^{(1)}+5.40 \, C_{H\ell}^{(3)}, \nn \\
&\,&\hspace{0.2cm} -  1.37\, C_{Hq}^{(1)}- 4.63\, C_{Hq}^{(3)} - 0.736\, C_{Hu} - 0.593\, C_{\ell \ell} + 0.092\, C_{HWB} \Big]
\, \log \left[\frac{\Lambda^2}{\hat{m}_h^2}\right].
\eea
\subsection{Numerical results for $\bar{R}_b^0$}
The $\delta$ correction to $\bar{R}_\ell^b$ is given by
\bea
\frac{\delta R_b^0}{10^{-2}}
&=& - 0.192 \,C_{Hd}+0.039 \,C_{HD} + 0.158 \,C_{H\ell}^{(3)}
+2.13 \,C_{Hq}^{(1)} -0.055 \,C_{Hq}^{(3)}, \nn \\
 &\,&\hspace{0.2cm} -0.494 \,C_{Hu} + 0.043 \,C_{HWB}
- 0.079 \, C_{\ell \ell}.
\eea
Similarly, the $\delta \, \Delta$ correction to  $\bar{R}_b^0$  has the contributions
\bea
\frac{\delta \Delta R_b^0}{10^{-3}} &=& \left[ \left(0.036 \,\Delta \bar{v}_T + 0.083 \right) C_{Hd}
+ \left(0.011\, \Delta \bar{v}_T + 0.013 \right) C_{HD}  +\left(0.084 \,\Delta \bar{v}_T - 0.014 \right) C_{H\ell}^{(3)}, \right. \nn \\
&\,&\hspace{0.2cm} -\left(0.085\, \Delta \bar{v}_T +0.152 \right) C_{Hq}^{(1)} -\left(0.016 \,\Delta \bar{v}_T + 0.019 \right) C_{Hq}^{(3)}
+\left(0.099\, \Delta \bar{v}_T + 0.208 \right) C_{Hu}, \nn \\
&\,&\hspace{0.2cm} -\left(0.042\, \Delta \bar{v}_T - 0.007 \right) C_{\ell \ell}
+ \left(0.013\, \Delta \bar{v}_T + 0.009 \right) C_{HWB} -0.015\, C_{\ell q}^{(3)} , \nn \\
&\,&\hspace{0.2cm}+ 0.597 \,C_{qq}^{(3)}  + 0.047\, C_{uH} -0.006 \left(C_{HB}+C_{HW}\right) -0.106 \, \Delta v \Big],
\eea
and the $\delta \, \Delta$ correction to $\bar{R}_b^0$  also has the logarithmic terms
\bea
\frac{\delta \Delta R_b^0}{10^{-3}} &= & \left[0.129\, C_{Hd} +0.025\, C_{HD} +0.067\, C_{H\ell}^{(3)}  - 0.559 \, C_{Hq}^{(1)}
+ 0.383 \,C_{Hq}^{(3)} +0.240\, C_{Hu} , \right. \nn \\
&\,& \hspace{0.2cm}  + 0.023 \, C_{HWB} -0.049 \, C_{\ell \ell} +0.030 \, C_{\ell q}^{(3)}+0.036\left(C_{qd}^{(1)}-C_{ud}^{(1)}\right)
 - 0.618 \, C_{qq}^{(3)},  \\
&\,&\hspace{0.2cm}-0.803\, C_{qq}^{(1)}+0.494 \, C_{qu}^{(1)} -0.002 \, C_{uB}
+0.032 \, C_{uH} -0.004 \, C_{uW}-0.186 \, C_{uu} \Big]
 \log \left[\frac{\Lambda^2}{\hat{m}_t^2}\right] \nn\\
&\,&\hspace{0.2cm} + \left[- 8.94 \times10^{-7}\,C_{HD}+\Big(0.313\, C_{Hd} - 3.49\, C_{Hq}^{(1)}+0.090\, C_{Hq}^{(3)}- 0.258\, C_{H\ell}^{(3)}, \right. \nn \\
&\,&\hspace{0.2cm} +0.808\, C_{Hu}  + 0.129\, C_{\ell \ell}- 0.020 \, C_{HWB}\Big)10^{-2} \Big]
\, \log \left[\frac{\Lambda^2}{\hat{m}_h^2}\right]. \nn
\eea

\begin{table}
\begin{center}
\small
\begin{minipage}[t]{4.45cm}
\renewcommand{\arraystretch}{1.5}
\begin{tabular}[t]{c|c}
\multicolumn{2}{c}{$1:X^3$} \\
\hline
$Q_G$                & $f^{ABC} G_\mu^{A\nu} G_\nu^{B\rho} G_\rho^{C\mu} $ \\
$Q_{\widetilde G}$          & $f^{ABC} \widetilde G_\mu^{A\nu} G_\nu^{B\rho} G_\rho^{C\mu} $ \\
$Q_W$                & $\epsilon^{IJK} W_\mu^{I\nu} W_\nu^{J\rho} W_\rho^{K\mu}$ \\
$Q_{\widetilde W}$          & $\epsilon^{IJK} \widetilde W_\mu^{I\nu} W_\nu^{J\rho} W_\rho^{K\mu}$ \\
\end{tabular}
\end{minipage}
\begin{minipage}[t]{2.7cm}
\renewcommand{\arraystretch}{1.5}
\begin{tabular}[t]{c|c}
\multicolumn{2}{c}{$2:H^6$} \\
\hline
$Q_H$       & $(H^\dag H)^3$
\end{tabular}
\end{minipage}
\begin{minipage}[t]{5.1cm}
\renewcommand{\arraystretch}{1.5}
\begin{tabular}[t]{c|c}
\multicolumn{2}{c}{$3:H^4 D^2$} \\
\hline
$Q_{H\Box}$ & $(H^\dag H)\Box(H^\dag H)$ \\
$Q_{H D}$   & $\ \left(H^\dag D_\mu H\right)^* \left(H^\dag D_\mu H\right)$
\end{tabular}
\end{minipage}
\begin{minipage}[t]{2.7cm}

\renewcommand{\arraystretch}{1.5}
\begin{tabular}[t]{c|c}
\multicolumn{2}{c}{$5: \psi^2H^3 + \hbox{h.c.}$} \\
\hline
$Q_{eH}$           & $(H^\dag H)(\bar l_p e_r H)$ \\
$Q_{uH}$          & $(H^\dag H)(\bar q_p u_r \widetilde H )$ \\
$Q_{dH}$           & $(H^\dag H)(\bar q_p d_r H)$\\
\end{tabular}
\end{minipage}

\vspace{0.25cm}

\begin{minipage}[t]{4.7cm}
\renewcommand{\arraystretch}{1.5}
\begin{tabular}[t]{c|c}
\multicolumn{2}{c}{$4:X^2H^2$} \\
\hline
$Q_{H G}$     & $H^\dag H\, G^A_{\mu\nu} G^{A\mu\nu}$ \\
$Q_{H\widetilde G}$         & $H^\dag H\, \widetilde G^A_{\mu\nu} G^{A\mu\nu}$ \\
$Q_{H W}$     & $H^\dag H\, W^I_{\mu\nu} W^{I\mu\nu}$ \\
$Q_{H\widetilde W}$         & $H^\dag H\, \widetilde W^I_{\mu\nu} W^{I\mu\nu}$ \\
$Q_{H B}$     & $ H^\dag H\, B_{\mu\nu} B^{\mu\nu}$ \\
$Q_{H\widetilde B}$         & $H^\dag H\, \widetilde B_{\mu\nu} B^{\mu\nu}$ \\
$Q_{H WB}$     & $ H^\dag \tau^I H\, W^I_{\mu\nu} B^{\mu\nu}$ \\
$Q_{H\widetilde W B}$         & $H^\dag \tau^I H\, \widetilde W^I_{\mu\nu} B^{\mu\nu}$
\end{tabular}
\end{minipage}
\begin{minipage}[t]{5.2cm}
\renewcommand{\arraystretch}{1.5}
\begin{tabular}[t]{c|c}
\multicolumn{2}{c}{$6:\psi^2 XH+\hbox{h.c.}$} \\
\hline
$Q_{eW}$      & $(\bar l_p \sigma^{\mu\nu} e_r) \tau^I H W_{\mu\nu}^I$ \\
$Q_{eB}$        & $(\bar l_p \sigma^{\mu\nu} e_r) H B_{\mu\nu}$ \\
$Q_{uG}$        & $(\bar q_p \sigma^{\mu\nu} T^A u_r) \widetilde H \, G_{\mu\nu}^A$ \\
$Q_{uW}$        & $(\bar q_p \sigma^{\mu\nu} u_r) \tau^I \widetilde H \, W_{\mu\nu}^I$ \\
$Q_{uB}$        & $(\bar q_p \sigma^{\mu\nu} u_r) \widetilde H \, B_{\mu\nu}$ \\
$Q_{dG}$        & $(\bar q_p \sigma^{\mu\nu} T^A d_r) H\, G_{\mu\nu}^A$ \\
$Q_{dW}$         & $(\bar q_p \sigma^{\mu\nu} d_r) \tau^I H\, W_{\mu\nu}^I$ \\
$Q_{dB}$        & $(\bar q_p \sigma^{\mu\nu} d_r) H\, B_{\mu\nu}$
\end{tabular}
\end{minipage}
\begin{minipage}[t]{5.4cm}
\renewcommand{\arraystretch}{1.5}
\begin{tabular}[t]{c|c}
\multicolumn{2}{c}{$7:\psi^2H^2 D$} \\
\hline
$Q_{H l}^{(1)}$      & $(H^\dag i\overleftrightarrow{D}_\mu H)(\bar l_p \gamma^\mu l_r)$\\
$Q_{H l}^{(3)}$      & $(H^\dag i\overleftrightarrow{D}^I_\mu H)(\bar l_p \tau^I \gamma^\mu l_r)$\\
$Q_{H e}$            & $(H^\dag i\overleftrightarrow{D}_\mu H)(\bar e_p \gamma^\mu e_r)$\\
$Q_{H q}^{(1)}$      & $(H^\dag i\overleftrightarrow{D}_\mu H)(\bar q_p \gamma^\mu q_r)$\\
$Q_{H q}^{(3)}$      & $(H^\dag i\overleftrightarrow{D}^I_\mu H)(\bar q_p \tau^I \gamma^\mu q_r)$\\
$Q_{H u}$            & $(H^\dag i\overleftrightarrow{D}_\mu H)(\bar u_p \gamma^\mu u_r)$\\
$Q_{H d}$            & $(H^\dag i\overleftrightarrow{D}_\mu H)(\bar d_p \gamma^\mu d_r)$\\
$Q_{H u d}$ + h.c.   & $i(\widetilde H ^\dag D_\mu H)(\bar u_p \gamma^\mu d_r)$\\
\end{tabular}
\end{minipage}

\vspace{0.25cm}

\begin{minipage}[t]{4.75cm}
\renewcommand{\arraystretch}{1.5}
\begin{tabular}[t]{c|c}
\multicolumn{2}{c}{$8:(\bar LL)(\bar LL)$} \\
\hline
$Q_{\ell \ell}$        & $(\bar l_p \gamma_\mu l_r)(\bar l_s \gamma^\mu l_t)$ \\
$Q_{qq}^{(1)}$  & $(\bar q_p \gamma_\mu q_r)(\bar q_s \gamma^\mu q_t)$ \\
$Q_{qq}^{(3)}$  & $(\bar q_p \gamma_\mu \tau^I q_r)(\bar q_s \gamma^\mu \tau^I q_t)$ \\
$Q_{\ell q}^{(1)}$                & $(\bar l_p \gamma_\mu l_r)(\bar q_s \gamma^\mu q_t)$ \\
$Q_{\ell q}^{(3)}$                & $(\bar l_p \gamma_\mu \tau^I l_r)(\bar q_s \gamma^\mu \tau^I q_t)$
\end{tabular}
\end{minipage}
\begin{minipage}[t]{5.25cm}
\renewcommand{\arraystretch}{1.5}
\begin{tabular}[t]{c|c}
\multicolumn{2}{c}{$8:(\bar RR)(\bar RR)$} \\
\hline
$Q_{ee}$               & $(\bar e_p \gamma_\mu e_r)(\bar e_s \gamma^\mu e_t)$ \\
$Q_{uu}$        & $(\bar u_p \gamma_\mu u_r)(\bar u_s \gamma^\mu u_t)$ \\
$Q_{dd}$        & $(\bar d_p \gamma_\mu d_r)(\bar d_s \gamma^\mu d_t)$ \\
$Q_{eu}$                      & $(\bar e_p \gamma_\mu e_r)(\bar u_s \gamma^\mu u_t)$ \\
$Q_{ed}$                      & $(\bar e_p \gamma_\mu e_r)(\bar d_s\gamma^\mu d_t)$ \\
$Q_{ud}^{(1)}$                & $(\bar u_p \gamma_\mu u_r)(\bar d_s \gamma^\mu d_t)$ \\
$Q_{ud}^{(8)}$                & $(\bar u_p \gamma_\mu T^A u_r)(\bar d_s \gamma^\mu T^A d_t)$ \\
\end{tabular}
\end{minipage}
\begin{minipage}[t]{4.75cm}
\renewcommand{\arraystretch}{1.5}
\begin{tabular}[t]{c|c}
\multicolumn{2}{c}{$8:(\bar LL)(\bar RR)$} \\
\hline
$Q_{le}$               & $(\bar l_p \gamma_\mu l_r)(\bar e_s \gamma^\mu e_t)$ \\
$Q_{lu}$               & $(\bar l_p \gamma_\mu l_r)(\bar u_s \gamma^\mu u_t)$ \\
$Q_{ld}$               & $(\bar l_p \gamma_\mu l_r)(\bar d_s \gamma^\mu d_t)$ \\
$Q_{qe}$               & $(\bar q_p \gamma_\mu q_r)(\bar e_s \gamma^\mu e_t)$ \\
$Q_{qu}^{(1)}$         & $(\bar q_p \gamma_\mu q_r)(\bar u_s \gamma^\mu u_t)$ \\
$Q_{qu}^{(8)}$         & $(\bar q_p \gamma_\mu T^A q_r)(\bar u_s \gamma^\mu T^A u_t)$ \\
$Q_{qd}^{(1)}$ & $(\bar q_p \gamma_\mu q_r)(\bar d_s \gamma^\mu d_t)$ \\
$Q_{qd}^{(8)}$ & $(\bar q_p \gamma_\mu T^A q_r)(\bar d_s \gamma^\mu T^A d_t)$\\
\end{tabular}
\end{minipage}

\vspace{0.25cm}

\begin{minipage}[t]{3.75cm}
\renewcommand{\arraystretch}{1.5}
\begin{tabular}[t]{c|c}
\multicolumn{2}{c}{$8:(\bar LR)(\bar RL)+\hbox{h.c.}$} \\
\hline
$Q_{ledq}$ & $(\bar l_p^j e_r)(\bar d_s q_{tj})$
\end{tabular}
\end{minipage}
\begin{minipage}[t]{5.5cm}
\renewcommand{\arraystretch}{1.5}
\begin{tabular}[t]{c|c}
\multicolumn{2}{c}{$8:(\bar LR)(\bar L R)+\hbox{h.c.}$} \\
\hline
$Q_{quqd}^{(1)}$ & $(\bar q_p^j u_r) \epsilon_{jk} (\bar q_s^k d_t)$ \\
$Q_{quqd}^{(8)}$ & $(\bar q_p^j T^A u_r) \epsilon_{jk} (\bar q_s^k T^A d_t)$ \\
$Q_{lequ}^{(1)}$ & $(\bar l_p^j e_r) \epsilon_{jk} (\bar q_s^k u_t)$ \\
$Q_{lequ}^{(3)}$ & $(\bar l_p^j \sigma_{\mu\nu} e_r) \epsilon_{jk} (\bar q_s^k \sigma^{\mu\nu} u_t)$
\end{tabular}
\end{minipage}
\end{center}
\caption{\label{op59}
The independent dimension-six operators built from Standard Model fields which conserve baryon number, as given in
Ref.~\cite{Grzadkowski:2010es}. The flavour labels $p,r,s,t$ on the $Q$ operators are suppressed on the left hand side of
the tables. This table is taken from Ref.\cite{Alonso:2013hga}.}
\end{table}

\bibliographystyle{JHEP}
\bibliography{bibliography2}

\end{document}